\documentclass[numberedappendix]{emulateapj}
\usepackage{apjfonts}
\usepackage{multirow}
\usepackage{longtable}
\usepackage{subfig}
\usepackage[countmax]{subfloat}
\usepackage{amsmath}
\usepackage{color}

\newcommand{\eqnref}[1]{Equation (\ref{#1})}

\newcommand{\tabref}[1]{Table~\ref{#1}}
\newcommand{\figref}[1]{Figure~\ref{#1}}
\newcommand{\secref}[1]{Section~\ref{#1}}
\newcommand{\appref}[1]{Appendix~\ref{#1}}

\newcommand{\C}[1]{\ensuremath{{}^{#1}{\rm C}}}
\newcommand{\0}[1]{\ensuremath{{}^{#1}{\rm O}}}
\newcommand{\Ne}[1]{\ensuremath{{}^{#1}{\rm Ne}}}
\newcommand{\Ni}[1]{\ensuremath{{}^{#1}{\rm Ni}}}
\newcommand{\cdens}{\ensuremath{\rho_{c,0}}}
\newcommand{\tcool}{\ensuremath{\tau_{\rm cool}}}
\newcommand{\tIGE}{\ensuremath{t_{\rm IGE}}}
\newcommand{\tDDT}{\ensuremath{t_{\rm DDT}}}
\newcommand{\rhoDDT}{\ensuremath{\rho_{\rm DDT}}}

\newcommand{\Msol}{\ensuremath{{\rm M}_\odot}}
\newcommand{\chisq}{\ensuremath{\chi^2_\nu}}
\newcommand{\proxy}{\ensuremath{M_{\rho_7 > 2}^{\rm \,DDT}}}
\newcommand{\MIGE}{\ensuremath{M_{\rm IGE}}}
\shorttitle{SNIa Dependence on Central Density}

\begin{document}

\submitted{Submitted to the Astrophysical Journal 2012 February 13}

\title{Evaluating Systematic Dependencies of Type Ia Supernovae:\\
The Influence of Central Density}

\author{
Brendan K.\ Krueger\altaffilmark{1,2},
Aaron P.\ Jackson\altaffilmark{2,3},
Alan C.\ Calder\altaffilmark{2,4},
Dean M.\ Townsley\altaffilmark{5}
Edward F.\ Brown\altaffilmark{6,7},
Francis X.\ Timmes\altaffilmark{7,8}
}

\altaffiltext{1}{
email: \texttt{brendan.krueger@stonybrook.edu}
}
\altaffiltext{2}{
Department of Physics \& Astronomy,
The State University of New York - Stony Brook, Stony Brook, NY, USA
}
\altaffiltext{3}{
present address: Laboratory for Computational Physics and Fluid Dynamics,
Naval Research Laboratory, Washington, DC, USA
}
\altaffiltext{4}{
New York Center for Computational Sciences,
The State University of New York - Stony Brook, Stony Brook, NY, USA
}
\altaffiltext{5}{
Department of Physics and Astronomy 
The University of Alabama, Tuscaloosa, AL, USA
}
\altaffiltext{6}{
Department of Physics and Astronomy,
Michigan State University, East Lansing, MI, USA
}
\altaffiltext{7}{
The Joint Institute for Nuclear Astrophysics,
Notre Dame, IN, USA
}
\altaffiltext{8}{
School of Earth and Space Exploration,
Arizona State University, Tempe, AZ, USA
}

\begin{abstract}
We present a study exploring a systematic effect on the brightness of type Ia
supernovae using numerical models that assume the single-degenerate paradigm.
Our investigation  varied the central density of the progenitor white dwarf at
flame ignition, and considered its impact on the explosion yield, particularly
the production and distribution of radioactive \Ni{56}, which powers the light
curve.  We performed a suite of two-dimensional simulations with randomized
initial conditions, allowing us to characterize the statistical trends that we
present. 
The simulations indicate that production of Fe-group material is statistically
independent of progenitor central density, but the mass of stable Fe-group
isotopes is tightly correlated with central density, with a decrease in the
production of \Ni{56} at higher central densities. These results imply
progenitors with higher central densities produce dimmer events.
We provide details of the post-explosion distribution of \Ni{56} in the models,
including the lack of a consistent centrally-located deficit of \Ni{56}, which
may be compared to observed remnants.
By performing a self-consistent extrapolation of our model yields and
considering the main-sequence lifetime of the progenitor star and the elapsed
time between the formation of the white dwarf and the onset of accretion, we
develop a brightness-age relation that improves our prediction of the expected
trend for single degenerates and we compare this relation with observations.
\end{abstract}

\keywords{hydrodynamics --- nuclear reactions, nucleosynthesis, abundances
--- supernovae: general --- white dwarfs}

\section{Introduction}
\label{sec:intro}

Type Ia supernovae (SNeIa; singular SNIa) are bright, transient astronomical
events identified by a peak-light spectrum showing no evidence of hydrogen but
absorption lines of singly-ionized silicon~\citep{Mink41, Fili97}.  These
events follow from explosive thermonuclear burning of degenerate stellar
material composed principally of C and O, which synthesizes $\sim\!\!0.6 \Msol$
of radioactive \Ni{56}.  The decay of this \Ni{56} powers the light
curve~\citep{truranetal67, colgatemckee69, arnett:type, pinto.eastman:physics}.

The progenitor systems of these explosions remain the subject of considerable
debate and active research.  Observations, however, indicate these events
largely form a homogeneous class.  \citet{PhillipsRelation} identified a
relationship between the maximum B-band magnitude of an event and its rate of
decline.  This ``brighter equals broader'' relationship has been extended to
additional bands with templates from nearby events, allowing these events to be
calibrated as an extension of the astronomical distance ladder (see
\citealt{jha2007} for a description).  This property, along with the brightness
of SNeIa, which makes them visible over great distances, enables the use of
SNeIa to probe the structure and expansion history of the universe, allowing
studies of various cosmological models' parameters~\citep{riessetal1998,
perlmutter.aldering.ea:measurements, albetal2006, Kirshner09}, with recent work
constraining cosmological parameters to within a few
percent~\citep{riessetal11, sullivanetal11}.  Recent observational studies of
SNeIa have begun to correct for correlations of the brightness of a SNIa with
properties of the host galaxy \citep{ConleyEtAl11}.  Many SNIa observations are
restricted to broadband photometry, so knowledge of host galaxy properties is
correlated.  The inability to deconvolve these properties from each other is
among the larger sources of uncertainty in cosmological constraints from SNeIa,
so advancing the understanding of how brightness correlates with host galaxy
properties may contribute significantly to reducing the uncertainties of
cosmological parameters.

The brightness, and therefore ``broadness'', of a SNIa is determined
principally by the amount of \Ni{56} synthesized during the explosion.
Observations report that SNeIa appear to have an intrinsic scatter of a few
tenths of a magnitude after calibration, forcing a minimum uncertainty in any
distances measured by using SNeIa as standardizable
candles~\citep{JacobyEtAl92, Kirshner09}.  An important goal of theoretical
research into SNeIa, from the standpoint of cosmology, is to understand the
sources of scatter and to identify potential systematic biases by studying the
effects of various properties on the mechanism and nucleosynthetic yield of the
SNIa.  The surrounding stellar population, the metallicity and mass of the
progenitor, the thermodynamic state of the progenitor, the cooling and
accretion history of the progenitor, and other parameters are known to affect
the lightcurves of SNeIa; the role of these ``secondary'' parameters is the
subject of considerable study~\citep[e.g.,][]{Roepetal06_2, hoeetal2010}.
Additionally, many of these effects may be interconnected in complex
ways~\citep{DomiHoefStra01, LesaffreEtAl06, townetal09}.

Observational campaigns are gathering information about SNeIa at an
unprecedented rate.  \citet{Scannapieco2005The-Type-Ia-Sup} and
\citet{MannucciEtAl06} showed that the delay time (elapsed time between star
formation and the supernova event) data are best fit by a bimodal delay time
distribution (DTD) with a prompt component that tracks less than 1 Gyr after
star formation and a tardy component that occurs several Gyr later.
\citet{GallagherEtAl08} demonstrate a correlation between brighter SNeIa and
shorter delay times, which they state is consistent with the bimodality
described by~\citeauthor{MannucciEtAl06}, but also with a continuous relation.
\citet{howelletal+09}, \citet{NeillEtAl09} and \citet{BrandtEtAl10} also find
such a correlation between the delay time and brightness of a SNIa.  While the
degeneracy of age and metallicity in observations could obscure these
correlations, \cite{howelletal+09} note that the scatter in brightness of this
observed relation is unlikely to be explained by the effect of metallicity.

For this theoretical study, we adopt the model known as the single-degenerate
paradigm.  This model assumes that a SNIa is the result of a thermonuclear
disruption of a white dwarf (WD) in a mass-transferring binary system with
either a main-sequence or red-giant companion star~\citep[see] [and references
therein]{branchetal1995, Fili97, hillebrandt.niemeyer:type, livio2000,
roepke2006, LiEtAl2011, NugentEtAl2011, BloomEtAl2011}.  Recent observational
evidence, however, suggests other progenitors such as the merging of two white
dwarfs may explain many events~\citep{scalzo:2010, Yuan:2010}.  In the
single-degenerate scenario, the WD is formed when the primary star goes through
a giant phase and expels a planetary nebula.  Once the primary becomes a WD, it
is initially not in contact with the companion star, and it slowly cools as
thermal energy is radiated away.  Once the companion star evolves and fills its
Roche lobe, mass-transfer begins to carry low-mass elements from the envelope
of the companion to the surface of the WD.  If the accretion rate exceeds
$\sim\!\!  10^{-7}$~\Msol~yr$^{-1}$, the H-rich material can steadily
burn~\citep{NomotoEtAl07} and the WD gains mass, which heats and compresses
the WD, driving up both the temperature and density in the core.  Once the
temperature rises enough for carbon burning to begin, the core of the WD begins
to convect; this is known as the ``simmering'' phase.  This simmering phase
lasts on order of 10$^3$~yr, and ends when a flame is ignited, which occurs
approximately when the eddy turnover time becomes shorter than the local
nuclear runaway time.  Our initial models attempt to parameterize the WD at the
end of the simmering phase, just at the beginning of the thermonuclear
deflagration, which will in turn cause an explosion that will disrupt the
entire WD in a SNIa.

The explosion mechanism we use (within the single-degenerate
paradigm) is that of a deflagration to detonation transition (DDT).
After ignition, the flame propagates as a subsonic deflagration for a while
and then transitions to a supersonic detonation that rapidly consumes
the star~\citep{1986SvAL, woosley90, Khokhlov1991Delayed-detonat, hokowh95,
HoefKhok96, KhokhlovEtAl97, NiemWoos97, hwt98, Niem99}. We describe the
details of our implementation of this explosion mechanism below.

In the single-degenerate paradigm, a longer delay time can be explained by a
longer elapsed time between the formation of the WD and the onset of accretion.
During this period, the WD is in isolation and cools, hence the moniker the
``WD cooling time'' (\tcool).  Following the cooling time is a period of
accretion, during which the WD is compressed and heats, approaching the
conditions for ignition of the thermonuclear runaway.  The decrease in
temperature during the cooling time, which is determined by \tcool, influences
the density structure of the WD just prior to ignition, with a longer \tcool\ 
resulting in a higher central density when the core reaches the ignition
temperature~\citep{LesaffreEtAl06}.  Thus, a correlation between central
density and the brightness of an event would suggest a correlation between
delay time and the brightness of an event.

In this manuscript, we expand on our earlier investigation on the effect of
\tcool\ on the brightness of the explosion.  In~\citet{KruegerEtAl10} we
reported that as the central density of the progenitor WD increases, the
production of radioactive \Ni{56} decreases due to increased neutronization
rates, producing a dimmer event.  Using the results of \citet{LesaffreEtAl06},
we related the WD central density to \tcool\ and were able to compare our
results to the observations of \citet{NeillEtAl09}.  Here we present additional
details of our models; a statistical analysis of the results including the
assessment of intrinsic scatter; the distribution of Fe-group
elements within the remnant; and a potentially-observable effect to demonstrate
the connection between age, progenitor central density, and brightness. We also
revised our previously-reported trend in brightness with age to account for
the main sequence evolution of the WD progenitor.

In \secref{sec:method} we discuss the methodology of our suite of simulations,
followed by details of the code we used to perform our simulations in
\secref{sec:code}.  We present the results of our simulations in
\secref{sec:results}, and discuss how these results compare with previous
studies in \secref{sec:discussion}.  \secref{sec:conclusions} contains a brief
summary and final conclusions.

\section{Methodology}
\label{sec:method}

As described above, our explosion models assume the DDT mechanism in which a 
flame ignited in the core propagates as a subsonic deflagration and then 
transitions to a detonation. We simulate an explosion from ignition through 
the detonation phase until burning effectively ceases. Our models are by 
necessity incomplete, however, in that we do not produce light curves and 
spectra with which we could compare to actual observations. Instead, we rely 
on the mass of \Ni{56} synthesized in our models and compare this result 
from our simulations to \Ni{56} masses inferred from 
observations~\citep{howelletal+09}.

For the study, we adopted the theoretical framework first applied in
\citet{townetal09} for a statistical study of a suite of simulations performed
with randomized initial conditions.  We constructed a set of five progenitor
models with different central densities (\cdens), and therefore slightly
different masses, and from these performed suites
of two-dimensional simulations.  For each progenitor model 
we applied thirty sets of initial conditions consisting of randomized 
perturbations on an initially-burned region.  We refer to each of these 
thirty as a realization, each of which is seeded by a random
number used to generate a unique power spectrum of spherical harmonics (see
\appref{app:init} for details of the realizations).  The spectra are used as
initial perturbations to a spherical ``match head'' in the center of the
progenitor star.  Each progenitor WD had the same seed values applied,
resulting in the same thirty perturbations.  This procedure allows us to
characterize the intrinsic scatter in the models and check for systematic
biases in the realizations across different progenitors, such as how the
morphology of the initial conditions may influence the final result.

Complete details of our models and simulations follow, but we preface
the description by mentioning that one limitation of our study is the 
use of two-dimensional models and the parameterization of inherently 
three-dimensional phenomena. In particular, two-dimensional models lack 
any meaningful consideration of turbulence and its effect on the flame 
because the turbulence found in the interior of a simmering white dwarf 
and its interaction with the flame are inherently three-dimensional. 
The problem is compounded by the fact that this interaction occurs
partially on unresolvable scales, necessitating
use of sub-grid-scale models~\citep[see][for an example]{Schmetal06a}. 

There are two critical ways that omitting consideration of turbulence influences
this work. The first is in the calculation of flame speeds during the deflagration
phase. The burning model (described below) relies on an input flame speed to propagate
a model flame during the deflagration. Turbulence will interact with this flame,
stretching it and thereby boosting the burning rate \citep[see][and references
therein]{Schmetal06a,Schmetal06b}.  Our present models boost the
input flame speed from tabulated laminar values to compensate for buoyancy 
effects~\citep{townetal09} but do not include the effect of turbulence-flame
interactions for reasons discussed above. Effectively, we assume the
increase in flame surface (and hence the burning rate) is dominated
by stretching due to buoyancy rather than turbulence.

The second critical way that omitting consideration of turbulence influences this 
work is in the criteria for the DDT~\citep[see][and references 
therein]{seitenzahletal11}. For this study, we parameterized the DDT criterion 
as a threshold density, \rhoDDT, with a detonation initiated when the 
top of a rising plume of burned material reaches this density. As we 
describe below, this threshold density determines the duration of the 
deflagration and thus strongly influences the outcome of an explosion. 
Our {\em a priori} choice for this parameter led to higher than expected
yields of radioactive \Ni{56}, necessitating a rescaling of our results
for comparison to observations. Confirmation of the trends we present 
from extrapolated results awaits a future study with a more consistent 
treatment of these issues.

\subsection{Initial White Dwarf Models}
\label{sec:init}

As demonstrated by \citet{LesaffreEtAl06}, for a given zero-age main-sequence
mass, the properties of a progenitor WD such as central temperature and density
at ignition of the deflagration can be constructed as functions of \tcool.  We
take the leading-order effect from varying \tcool; that is, vary \cdens\ while
holding all other parameters constant.  This choice allows us to disentangle
the effects of \cdens\ from the effects of other parameters.  For our \cdens,
we chose $1-5 \times 10^9$~g~cm$^{-3}$ in steps of $1 \times 10^9$~g~cm$^{-3}$.
We then constructed a series of five parameterized WD progenitor models in
hydrostatic equilibrium.

\begin{figure}
   \centering
   \includegraphics[angle=270, width=\columnwidth]{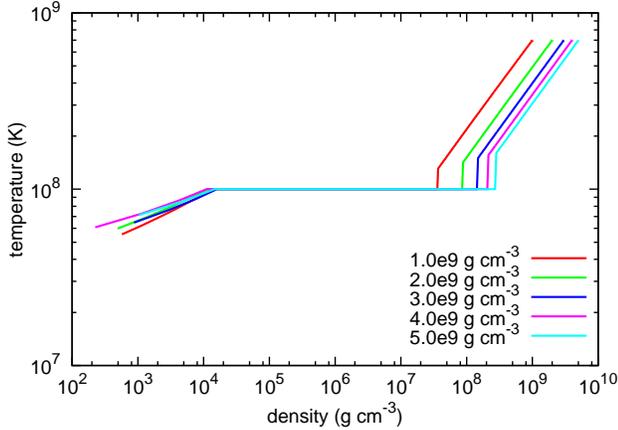}
   \caption{\label{fig:prog_TvRho}The structure of our progenitor WDs in
      $\rho$-$T$ space.  Below $\sim\!10^4$~g~cm$^{-3}$ is the power-law
      atmosphere.  In the center of the figure is the isothermal envelope, with
      the same temperature for all progenitors.  In the upper right region of
      the figure is the adiabatic core; all 5 progenitors are isentropic in
      this region, but the value of the entropy varies between progenitors.}
\end{figure}

\figref{fig:prog_TvRho} presents the profiles of the progenitor WDs in the
$\rho$-$T$ plane.  The core of each WD is isentropic due to convection, the
intermediate (``envelope'') region is isothermal due to the high conductivity
of degenerate matter, and the outer (``atmosphere'') region has a power-law
temperature dependence that was chosen to mimic a radiative atmosphere.  For a
model to explode, the central temperature must be in the range where the carbon
burning begins a runaway, which is approximately $7 - 8 \times 10^8$~K.
Varying the isothermal envelope temperature would have a relatively
insignificant effect on the mass, as the envelope contains only a small
fraction of the mass; primarily the mass is set by \cdens.  Thus we have chosen
a central temperature at the low end of the carbon ignition range and allow the
total mass to vary as a function of \cdens.  \tabref{tab:DensityMass} shows the
total mass ($M_{\rm tot}$) and mass of the isentropic core ($M_{\rm core}$) for
each progenitor.

\begin{table}
   \caption{Central densities, masses, and radii of progenitor WDs.}
   \centering
   \begin{tabular}{c c c c}
   \hline \hline
   \cdens\ (g~cm$^{-3}$) & $M_{\rm tot}$ (\Msol)
                         & $M_{\rm core}$ (\Msol) & $R$ (km) \\
   \hline
   $1 \times 10^9$       & 1.345 & 1.180 & 2500 \\
   $2 \times 10^9$       & 1.368 & 1.162 & 2076 \\
   $3 \times 10^9$       & 1.379 & 1.144 & 1852 \\
   $4 \times 10^9$       & 1.385 & 1.131 & 1716 \\
   $5 \times 10^9$       & 1.389 & 1.121 & 1604 \\
   \hline
   \end{tabular}
   \label{tab:DensityMass}
\end{table}

The core of each WD has a lower C/O ratio than the envelope; this is primarily
due to the composition of different regions of the star at the end of the
asymptotic giant branch (AGB) phase and the subsequent mixing of these regions,
with additional contributions from the consumption of C during the simmering
phase~\citep{Straetal03, PiroBild07, Chametal08, PiroChan08}.  Our
parameterized models for this study assume a fixed carbon abundance.  As we
plan to also study the dependence on central carbon abundance, we have
maintained a clear separation between the central-density and the
carbon-abundance studies by varying only central density and not the core
carbon fraction.  The composition discontinuity between the core and the
envelope causes a temperature discontinuity~\citep{PiroChan08}, shown by the
short vertical line segments between the core and the envelope in
\figref{fig:prog_TvRho}.  The composition is listed in
\tabref{tab:ProgComposition}.  We use \Ne{22} as a placeholder to represent the
neutron-rich isotopes present in a SNIa.  The abundance of \Ne{22} is
calibrated to achieve the electron-to-baryon ratio of the material present in a
SNIa, but sedimentation effects are not included.  See \citet{jacketal10},
specifically Section~2, for a more detailed discussion.

\begin{table}
   \centering
   \caption{Composition of the progenitor WDs.}
   \begin{tabular}{c c c}
   \hline \hline
   \multirow{2}{*}{isotope} & \multicolumn{2}{c}{mass fraction} \\
                            & core & envelope \\
   \hline
   \C{12}                   & 40\% & 50\% \\
   \0{16}                   & 57\% & 48\% \\
   \Ne{22}                  &  3\% &  2\% \\
   \hline
   \end{tabular}
   \label{tab:ProgComposition}
\end{table}

\subsection{Ensemble of Simulations}
\label{sec:ensemble}

\begin{figure*}
   \centering
   \begin{tabular}{cc}
   \subfloat{\label{fig:MorphCompA}
      \includegraphics[width=\columnwidth]{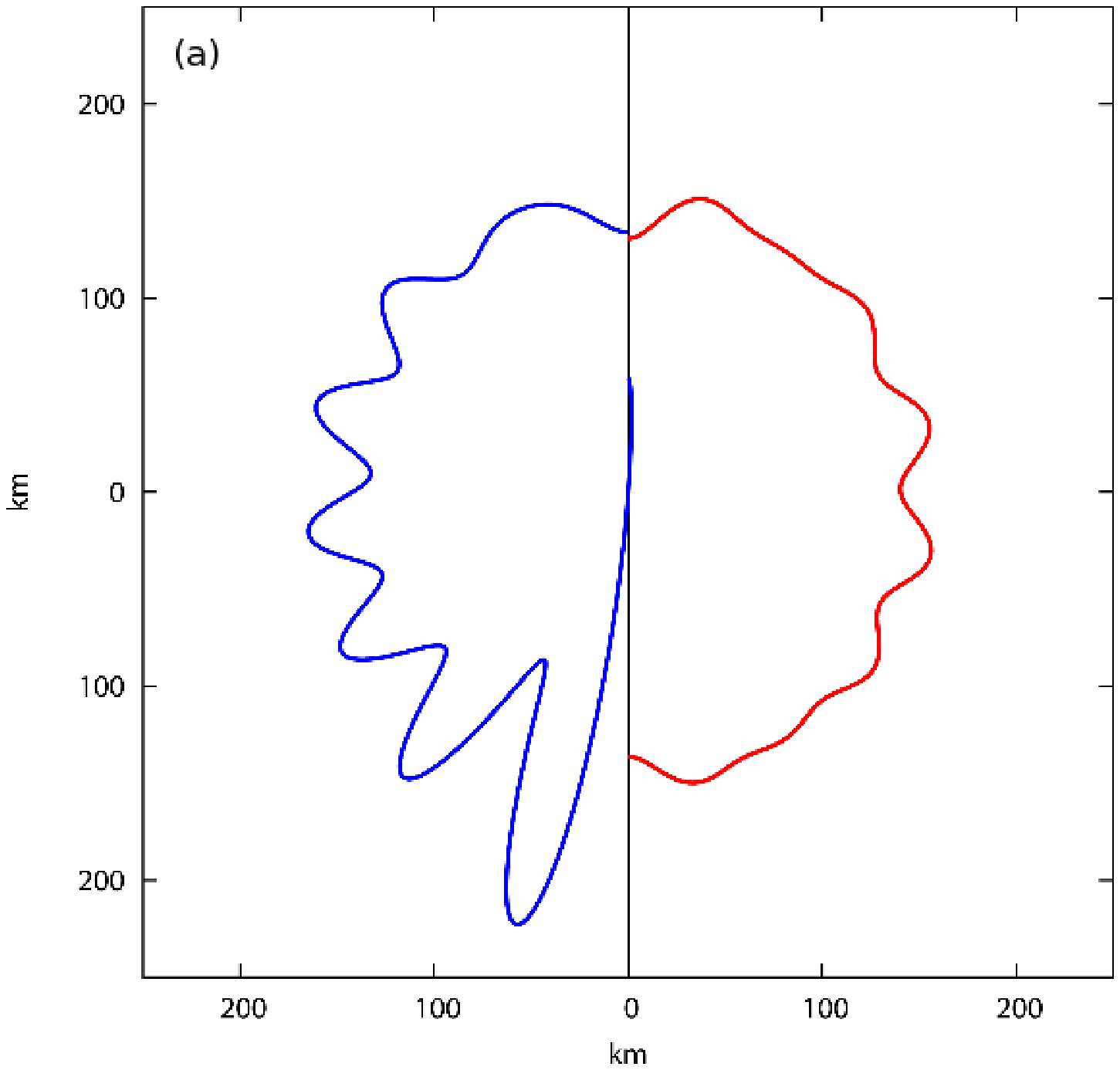}
   } &
   \subfloat{\label{fig:MorphCompB}
      \includegraphics[width=\columnwidth]{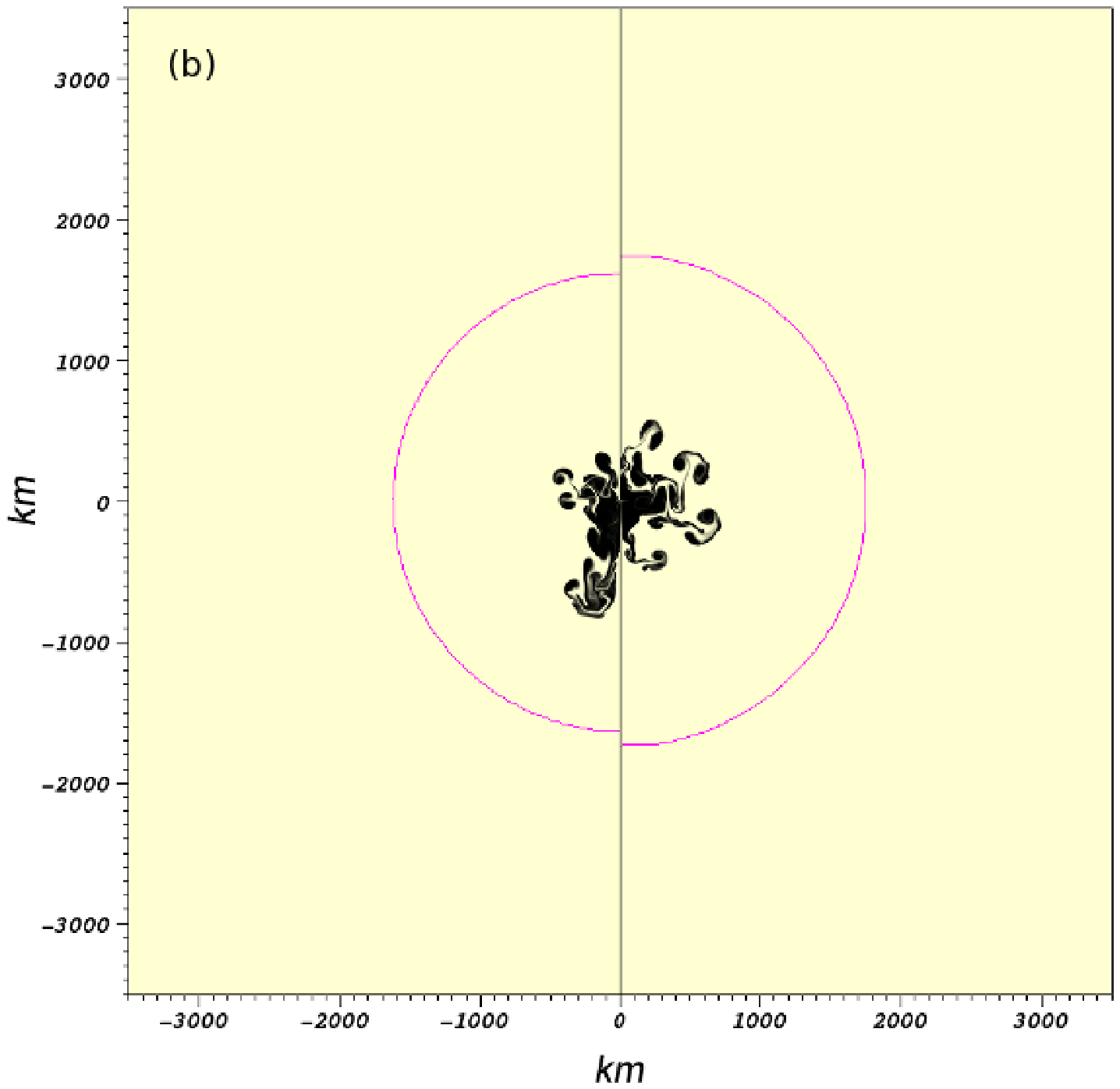}
   } \\
   \subfloat{\label{fig:MorphCompC}
      \includegraphics[width=\columnwidth]{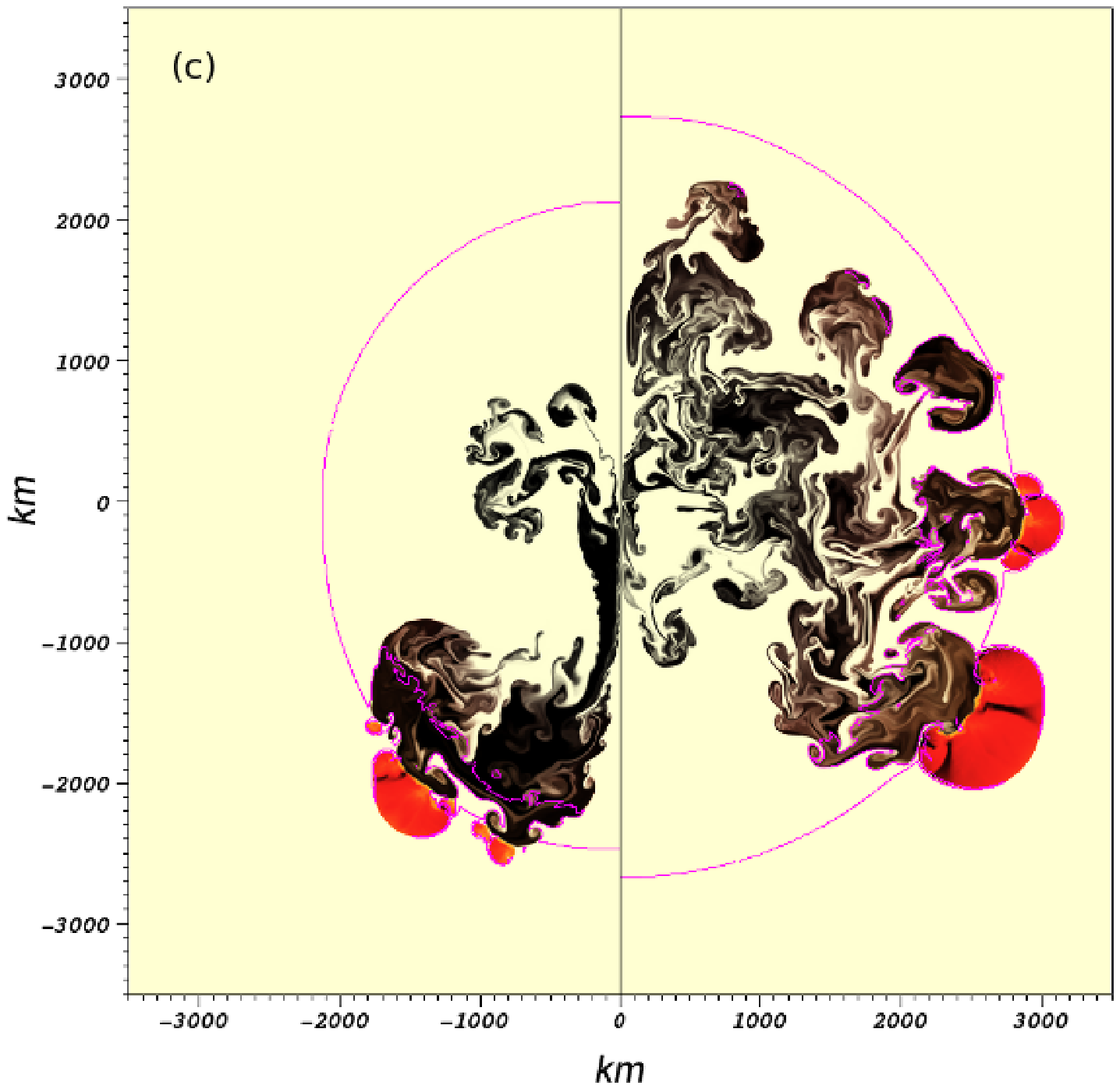}
   } &
   \subfloat{\label{fig:MorphCompD}
      \includegraphics[width=\columnwidth]{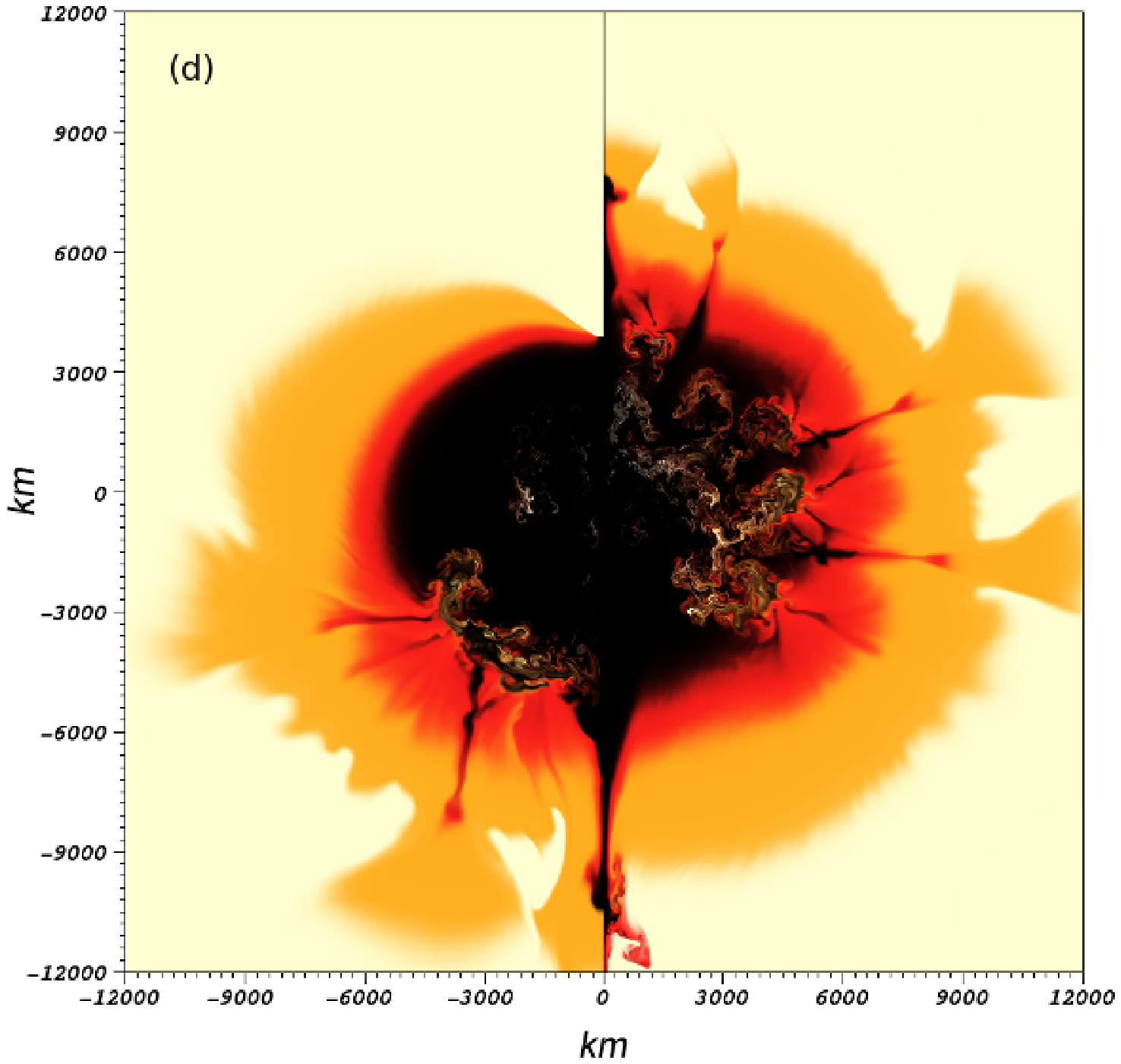}
   }
   \end{tabular}
   \caption{Illustrations of the phases of
      the DDT model of a SNIa;
      each panel displays two of our realizations: 21 (left) and 10
      (right).  Panel~\ref{fig:MorphCompA} shows the initial flame surfaces;
      realization 21 has the greatest deviation from the mean radius, and
      realization 10 has the least deviation from the mean radius.
      Panels~\ref{fig:MorphCompB}~--~\ref{fig:MorphCompD} are snapshots from
      our simulations, with $\cdens = 3 \times 10^9$~g~cm$^{-3}$;
      Panel~\ref{fig:MorphCompB} shows the early deflagration phase,
      Panel~\ref{fig:MorphCompC} shows the first DDT events, and
      Panel~\ref{fig:MorphCompD} shows the later detonation phase.  The colors
      represent the four stages of the burning model discussed in
      \secref{sec:burning}: cream represents unburned fuel, gold represents ash
      from carbon burning, red represents material in NSQE, and black
      represents material in NSE.  Only in the detonation stage do the three
      burning processes separate out spatially; they are co-located during the
      deflagration.  Our simulations extend to $\sim\!6.5\times10^8$~km, but
      these images show only the inner regions; the spatial scale varies
      between panels, with only panels~\ref{fig:MorphCompB} and
      \ref{fig:MorphCompC} having the same spatial extent in order to
      illustrate the expansion that occurs during the deflagration phase.}
   \label{fig:MorphComp}
\end{figure*}

The thirty unique realizations for each of our five progenitor models,
consisting of a set of perturbations on the initially-burned region, allowed
us to perform a suite of 150 two-dimensional, axisymmetric simulations of
SNeIa.  A simulation begins with a region of burned material at the center of
the star with the perturbation from sphericity given as spherical harmonics
with a prescribed range of $12 \leq \ell \leq 16$ and random amplitudes; the
amplitudes for each realization are shown in \appref{app:init}.  Each
realization has a unique seed (for the random number generator), allowing each
realization to be applied to the five progenitor models.
\figref{fig:MorphCompA} shows two example perturbations that span the space of
the random perturbations.  Realization 21, on the left, is representative of
the ``spikiest'' initial conditions; i.e., the greatest deviation from the mean
radius.  Realization 10, on the right, is representative of the ``smoothest''
initial conditions; i.e., the smallest deviation from the mean radius.

\section{The Simulation Code}
\label{sec:code}

The simulations were performed using a customized version of the FLASH
code\footnote{available from http://flash.uchicago.edu}, an Eulerian
adaptive-mesh compressible hydrodynamics code developed by the ASC/Alliances
Center for Astrophysical Thermonuclear Flashes at the University of
Chicago~\citep{fryxell_2000_aa, calder.fryxell.ea:on}.  The equation of state
(EoS) that we use is the fully-ionized electron-ion plasma
EoS~\citep{timmes.swesty:accuracy, fryxell_2000_aa}.  The version of FLASH used
here is the same as that used in \citet{jacketal10}.  Customizations perform
two main functions: First to implement the energy release due to explosive
carbon-oxygen fusion, in both deflagration and detonation propagation modes, as
well as a provision for a transition from deflagration to detonation.  Second,
criteria for mesh refinement that capture the important physics with a suitable
degree of efficiency.  Details of various components are given by
\citet{townsley.calder.ea:flame}, \citet{townetal09}, and \citet{jacketal10},
with important additional and supporting information on implementation of some
pieces of physics in \citet{Caldetal07} and \citet{SeitTownetal09}.

As mentioned in the introductory material, the simulations 
implement a multi-dimensional version of the deflagration-to-detonation 
transition model, in which the flame is born
as a subsonic deflagration and later transitions to a supersonic
detonation~\citep{1986SvAL, woosley90, Khokhlov1991Delayed-detonat, hokowh95,
HoefKhok96, KhokhlovEtAl97, NiemWoos97, hwt98, Niem99}.  The initial conditions
prescribe the deflagration at the start of a simulation.  As the deflagration
proceeds, the flame is subject to fluid instabilities
(\figref{fig:MorphCompB}), and when the top of a rising bubble reaches the
threshold density, it is assumed to transition into a supersonic detonation
(Figures~\ref{fig:MorphCompC} and \ref{fig:MorphCompD}; see \secref{sec:ddt}
for details of the transition).  See \citet{townetal09},
\citet{maedaetal2010}, \citet{jacketal10},
\citet{roepkeetal11}, \citet{seitenzahletal11}, and references therein for
examples of recent work assuming this explosion mechanism.

\subsection{Burning Model}
\label{sec:burning}

Since the carbon-oxygen fusion occurring in SNeIa proceeds to nuclear
statistical equilibrium (NSE), in which all reactions among all nuclides are,
to good approximation, fast relative to the hydrodynamic timescales, the
nuclear processing necessarily involves a large number of reactions and
nuclides.  This is especially true when calculating $e^{-}$ capture on Fe-group
elements (IGEs), as the overall effective $e^-$ capture rate is a combination
of contributions from captures on a wide variety of nuclides with comparable
individual rates.  As explored in our earlier work (\citealp{Caldetal07,
townsley.calder.ea:flame}; see also \citealp{khokhlov+00}), it is possible to
abstract the burning process from hundreds of nuclides to just a few fluid
state variables with appropriately chosen reaction dynamics and energetics
computed using a large set of nuclides.  This enables tremendous gains in
computational efficiency, making 3-d simulations and extensive 2-d studies such
as this feasible.

Our burning model consists of three reaction progress variables that describe
conversion between four states.  The first state is the initial, unburned
mixture of \C{12}, \0{16}, and \Ne{22} (representing general neutron excess, as
described in section \ref{sec:init}), which we call fuel.  The second state is
the result of the \C{12} fusing to roughly Si-group elements, which we call
ash.  The third state is the result of the remaining \0{16} burning to Si-group
elements, which we call nuclear statistical quasi-equilibrium (NSQE) material.
The fourth and final state is the result of NSQE material relaxing to nuclear
statistical equilibrium (NSE), containing mostly IGEs.  The three reaction
progress variables are
\begin{eqnarray*}
   \phi_{fa} & \quad & \text{Carbon consumption, fuel to ash}\\
   \phi_{aq} & \quad & \text{Oxygen consumption, ash to NSQE}\\
   \phi_{qn} & \quad & \text{Conversion of Si-group to Fe-group, NSQE to NSE.}
\end{eqnarray*}
Each variable evolves from 0 (unburned) to 1 (fully burned).  We also require
that $\phi_{fa} \ge \phi_{aq} \ge \phi_{qn}$ to enforce the time-ordering of
the four states.  A given cell will have mass fractions $1-\phi_{fa}$ of fuel,
$\phi_{fa} - \phi_{aq}$ of ash, $\phi_{aq} - \phi_{qn}$ of NSQE material, and
$\phi_{qn}$ of NSE material.

While a large portion of the energy release occurs in the consumption of
\C{12}, the final NSE state is not, during the explosion, energetically inert.
The NSE state is one in which the distribution of nuclides in the fluid, and
therefore the average nuclear binding energy, is determined by the fluid state.
That is, it participates in the EoS of the fluid, releasing or absorbing energy
as the pressure and density of the fluid change.  As a result, a significant
portion of the energy release for some fluid elements occurs well after the
``fully burned'' NSE state has been reached.  The reaction kinetics used are
given in \citet{jacketal10}.  We track several material properties in the fully
burned material, including the electron-to-baryon ratio ($Y_e$), ion-to-baryon
ratio ($Y_{\rm ion}$), and average nuclear binding energy per baryon
($\overline{q}$).

Our treatment of matter in NSE allows for the effects of weak reactions,
specifically electron captures, which serve to deleptonize the material.  Weak
processes (e.g.\  electron capture) are included in the calculation of the
energy input rate, as are neutrino losses, which are calculated by convolving
the NSE distribution with the weak interaction cross sections.  Both the NSE
state and the electron capture rates were calculated with a set of 443 nuclides
\citep{SeitTownetal09}.  Weak cross sections were taken from
\citet{FullerEtAl85}, \citet{OdaEtAl94}, and
\citet{langanke.martinez-pinedo:weak}, with newer rates superseding earlier
ones.

The treatment of electron capture is critical to the dynamics for three
reasons.  First, NSE is a dynamic equilibrium and the composition of material
in NSE evolves as the thermodynamic state evolves and/or $Y_e$ changes.
Binding energy can be released if the equilibrium evolves toward more tightly
bound nuclei, which changes the local temperature.  Second, the reduction in
$Y_e$ lowers the Fermi energy, reducing the primary pressure support of this
highly degenerate material and having an impact on the buoyancy of the
neutronized material.  Finally, neutrinos are emitted (since the star is
transparent to them) so that some energy is lost from the system.

By using the progress variables defined above and the local $Y_e$, we can
derive a local estimate for the abundance of \Ni{56}.  As was done in
\citet{townetal09}, we estimate the \Ni{56} abundance by assuming that the
first IGE material made as neutronization occurs is equal parts by mass
$^{54}$Fe and \Ni{58}.  The local mass fraction of \Ni{56} is estimated by
\begin{eqnarray}
\nonumber
Y_{e,n} &=& \frac{Y_e - (1-\phi_{qn})Y_{e,f}}{\phi_{qn}}\\
X_{\Ni{56}} &=& \max\left[\phi_{qn}\frac{Y_{e,n}-0.48212}{0.5-0.48212} ,
0\right]\ ,
\label{eqn:ni56est}
\end{eqnarray}
where $Y_{e,f}$ is the electron fraction in the unburned fuel and 0.48212 
is the electron fraction of material that is equal parts by mass $^{54}$Fe 
and \Ni{58}.  This should be a modestly accurate estimate of the
$^{56}$Ni production because the dynamics of the progress variable $\phi_{qn}$
have been calibrated to reproduce, in hydrodynamics, the production of IGEs
during incomplete silicon burning seen in direct calculations (Zeldovich, Von
Neumann, D\"oring; ZND; e.g.~\citealt{Khokhlov89}) of steady-state detonations.
When realization 2 from \citet{jacketal10} with $\rho_{\rm DDT} =
10^{7.1}$~g~cm$^{-3}$ is post-processed with a nuclear network, the $^{56}$Ni
yield from \eqnref{eqn:ni56est} is within 2\% of that determined by the
post-processing.  More detailed study of the accuracy of computed yields under
various conditions is the subject of separate, ongoing work.

\subsubsection{Deflagration}
\label{sec:flame}

Even at our highest resolution (4~km), the flame front is unresolved.  To
handle this we use an artificial, resolved reaction front that is governed by
the advection-diffusion-reaction (ADR) equation~\citep{Khok95, VladWeirRyzh06},
with special features to ensure that the front is stable and acoustically
quiet \citep{townsley.calder.ea:flame}.  Our ADR front is chosen to be resolved
over about 4 computational cells in order to obtain acceptably low acoustic
noise as it propagates across the grid and releases energy.  This creates an
extended ``partially burned'' region that requires some specialized treatment.
In such regions, particularly at high density, we say that the material is
well-separated into unburned and fully-burned material divided by a thin flame.
 However the spatial resolution cannot capture this and the average over a
mixture of fully burned and unburned results a ``partially-burned'' state.
Thus we have to make estimates of the correct thermodynamic state of the two
cases (unburned and fully burned) mixed together within the region.

Additionally, we enforce a minimum flame speed in order to prevent the flame
from being torn apart by Rayleigh-Taylor-induced turbulence.  The minimum flame
speed is
\begin{equation}
   S_{\rm min} = 0.5 \sqrt{Agm \Delta},
\end{equation}
where $A$ is the Atwood number, $g$ is the local acceleration of gravity,
$\Delta$ is the width of the grid cell, and $m$ is an adjustable parameter, set
to 0.04 for these simulations.  The flame speed is set by
\begin{equation}
   S = \max(S_{\rm min}, S_{\rm lam}).
\end{equation}
The Atwood number and the laminar flame speed, $S_{\rm lam}$, are both
functions of the local, unburned density estimate and the composition.  The
Atwood number varies by less than 0.01\% due to the amount of \Ne{22} present
so it is tabulated for a representative, constant \Ne{22} fraction.

\subsubsection{DDT}
\label{sec:ddt}

At present, the physical mechanism by which a DDT in degenerate supernova
material occurs is an area of current research~\citep[see][and references
therein]{roepke07, Seitetal09, Woosetal09, schmidtetal10,PoludnenkoEtAl2011}.
Simulations of supernovae involving a DDT assume it occurs via
the Zeldovich-gradient mechanism~(\citealp{KhokhlovEtAl97}, but see also
\citealp{Niem99}), 
in which a gradient in reactivity leads to a series of explosions that are in phase 
with the velocity of a steadily propagating detonation wave. 
Many authors suggest that when the flame reaches a state of distributed burning, which is
when turbulence on scales at or below the laminar flame width are fast enough to dominate
transport processes~\citep[see, e.g.,][]{pope87}, fuel and ash are mixed and the temperature of
the fuel is raised and "prepared" in such a way to produce the required reactivity
gradient. A requirement for distributed burning is that the ratio of turbulent intensity to
the laminar flame speed must exceed some unknown threshold, which is still actively
researched~\citep{NiemWoos97, KhokhlovEtAl97,GoloNiem05,
roepkeandhillebrandt2005,Aspdetal08,aspdenbellwoosley2010,poludnenkooran2010,poludnenkooran2011}.
Entrance into the distributed burning regime does not guarantee such a reactivity gradient
to form. ~\citet{Woosley07} and~\citet{Woosetal09} studied incorporating more stringent
requirements for these conditions to be met.

In the context of supernova models, 
the ratio of turbulent intensity to laminar flame speed changes most rapidly due to 
the change in laminar flame properties, which are strongly dependent on fuel density. 
Therefore, DDT is assumed to occur at a range of densities that vary somewhat but 
generally lie in the range of $10^{6.7}$ to $10^{7.7}$~g~cm$^{-3}$~\citep{KhokhlovEtAl97,
LisewskiEtAl00, Woosley07, RoepNiem07, bravGarc08, maedaetal2010, jacketal10}.  
We choose to ignite detonations where the flame reaches a
specific density, \rhoDDT  = $10^{7.1}$~g~cm$^{-3}$, which puts us somewhere 
in the middle of that range, consistent with a number of studies on this subject.  
We note that while parameterizing the DDT criteria by only density 
omits effects such as background turbulent intensity and density gradients, 
our choice of one threshold is intended to keep such variables constant so that
we can isolate and investigate the effect of varying the central density. 

The choice of \rhoDDT\ increases or decreases the duration of the deflagration 
phase, which increases or decreases the amount of expansion prior to the 
detonation and hence the yield of IGEs.  A companion study of the effects of 
varying \rhoDDT\ as a proxy for metallicity under the same statistical ensemble 
we use here indicated a slight over-production of \Ni{56} at this choice of
\rhoDDT~\citep{jacketal10}, a result borne out by this study.
The DDT transition is implemented by burning small regions ahead of rising
plumes.  When a plume reaches \rhoDDT, a circular region with a radius of 12~km
is selected 32~km radially outward from the point where the rising plume
reaches the transition density.  The reaction progress variables in this region
are then instantly increased to a fully-burned state.  This method conserves
energy, as the detonation is initiated by the sudden release of energy from the
conversion of fuel to NSE, not by an unphysical addition of extra energy.  Each
plume is limited to ignite no more than 2~--~3 detonations, with a minimum
separation distance of 200~km imposed between detonation points.  The full
details of this algorithm are given in Section~3.2 of \citet{jacketal10}.

Previous studies using this DDT mechanism found that DDT points with a 12~km
radius successfully generated detonations in all simulations performed.
However, our simulations showed that this size DDT point is not as robust for
this study.  One of our simulations ($\cdens = 1.0 \times 10^9$~g~cm$^{-3}$,
realization 8) deviated from the behavior of the other 149 simulations and
inspection showed that the first plume to reach \rhoDDT\ ignited several DDT
points that did not propagate as detonations; the first detonation to actually
propagate ignited when the second plume reached \rhoDDT\ significantly after
(and beneath) the first plume to reach \rhoDDT.  The failure of the detonation
of the first plume to reach \rhoDDT\ led to significantly more expansion of the
WD prior to the subsequent detonation, which led to some or all of the material
burning at a lower density than it should have, and the corresponding suspect
yields.  Thus we removed this simulation from our suite and performed the
analysis on the remaining 149 points.  Even if this ``failed DDT'' simulation
is included, the results presented in this paper change by no more than a few
percent.

\subsection{Mesh Refinement}
\label{sec:refinement}

Refinement is based on gradients in $\rho$ and $\phi_{fa}$, subject to the
limits imposed below.  We define three types of regions, subject to different
refinement limits:
\begin{enumerate}
   \item fluff (f): regions with $\rho < \rho_{\rm fluff}$
   \item star (*): non-energy-generating stellar material, $\rho >
      \rho_{\rm fluff}$
   \item energy generation (eg):\\ regions with $\epsilon_{\rm nuc} >
      \epsilon_{\rm eg}$ or $\dot\phi_{\rm fa}>\dot\phi_{\rm fa, eg}$
\end{enumerate}
where $\phi_{\rm fa}$ is the reaction progress variable from the ADR equation,
and $\epsilon_{\rm eg}$ and $\dot\phi_{\rm fa, eg}$ are parameters equal to
$10^{18}$~erg~g$^{-1}$~s$^{-1}$ and 0.2~s$^{-1}$ respectively.  These limits
are chosen so that all actively propagating flames or detonation fronts are
at the highest resolution.  We establish a minimum cell size for refinement of
each type of region such that $\Delta_f > \Delta_* > \Delta_{\rm eg}$.  We use
$\Delta_{\rm eg}=4$~km, $\Delta_{*}=16$~km and $\Delta_f$ to be as large as
allowable.  FLASH only allows adjacent sub-domains of the mesh to be of fixed
size (we use 16$\times$16 cells) and to differ by a single refinement level (a
factor of 2 in resolution).  These were found to be the lowest resolutions
which gave converged results in 1-d simulations \citep{townetal09}.

Fluff is the low-density area outside of the star; we choose
$\rho_{\rm fluff}=10^3$~g~cm$^{-3}$.  The FLASH code cannot properly handle
empty (zero-density) regions, so these regions are set to some low, but
non-zero, density so that they will not affect the dynamics of the star.  To
avoid rapidly cycling the refinement-derefinement of a region, a small amount
of hysteresis is introduced near the limits for refinement changes, so that,
for example, a refinement of grid resolution is not immediately derefined as a
result of slight changes due to the necessary interpolation.

\section{Results}
\label{sec:results}

The results presented here build on the initial results presented in
\citet{KruegerEtAl10} and extend the analysis beyond what was shown there.  The
explosions occur in two main phases, the deflagration phase (from ignition
until the first DDT occurs) and the subsequent detonation phase.  The principal
difference is that during the deflagration phase, the star has time to react to
the energy release.  Accordingly, the evolution is naturally divided by the
time at which the first DDT event occurs, which we define as \tDDT; this
duration also describes the time spent in the deflagration-dominated phase of
the SNIa evolution.  We also define \tIGE\ as the time when the production of
IGEs ceases.  This is the time at which burning ceases, and by this time the
NSE state is no longer evolving due to freezeout; thus energy release has
effectively ceased by \tIGE.  However, our models have not yet
entered into
free expansion by this time.  For the purpose of our simulations, \tIGE\ also
measures the duration of the entire SNIa event.  We found that the duration of
the deflagration phase (\tDDT) decreases with increasing density and is less
sensitive to density as the density increases, while the duration of the
detonation phase (equal to $\tIGE - \tDDT$) is very nearly constant for all
simulations, with a mean of 0.476~s and a standard deviation of 0.065~s.

The table in \appref{app:data} presents the masses of \Ni{56} and IGEs at
\tDDT\ and \tIGE.  Recall from \tabref{tab:DensityMass} that the total mass and
the mass of the convective core both change with density, but the variations
are only of a few percent.  Thus, we believe that effects from variations in
the total or convective core masses are negligible.  The results and trends we
describe follow from variations in the central density and the variations in
initial conditions from realization to realization.  Also, as discussed in
Section~3 of \citet{KruegerEtAl10}, the choice of the DDT transition density in
our simulations led to an overproduction of \Ni{56}.  Essentially, our models
are systematically too bright, but we believe that our trends are valid.
\citet{jacketal10} investigated the role of DDT density in our models and found
that the production of \Ni{56} is very sensitive to the choice of DDT density.
This choice determines the duration of the deflagration phase, which determines
the amount of expansion and, accordingly, the density profile of the star
during the detonation and the yield.  Future studies will be better calibrated
based on these results.

\subsection{Evolution}
\label{sec:Evolution}

\begin{figure}
   \includegraphics[angle=270, width=\columnwidth]{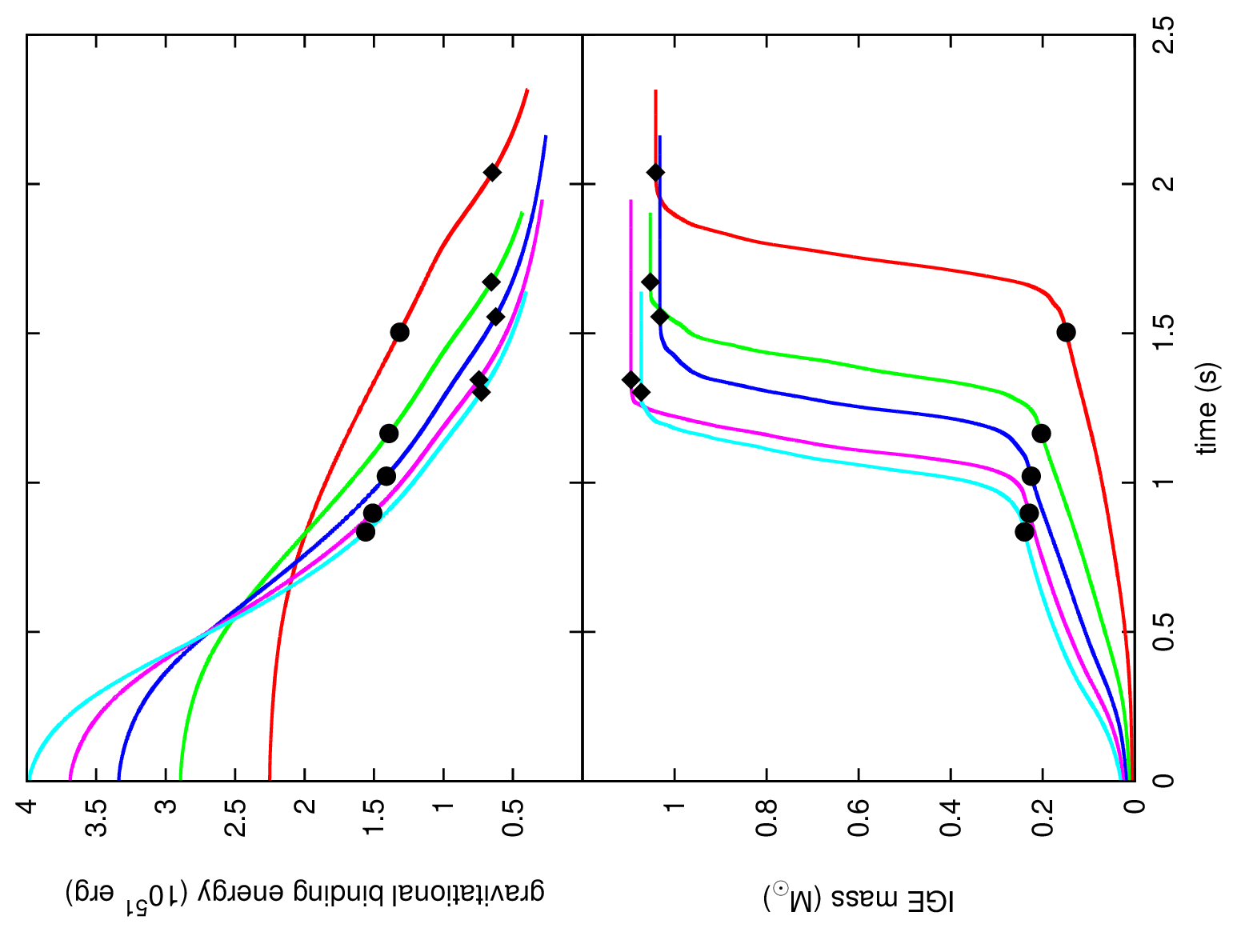}
   \caption{Sample curves showing the evolution of the gravitational binding
      energy and the mass of IGEs over the duration of a simulation.  These
      curves all use realization 5.  Curves are colored by \cdens: $1.0 \times
      10^9$~g~cm$^{-3}$ (red), $2.0 \times 10^9$~g~cm$^{-3}$ (green), $3.0
      \times 10^9$~g~cm$^{-3}$ (blue), $4.0 \times 10^9$~g~cm$^{-3}$ (magenta),
      $5.0 \times 10^9$~g~cm$^{-3}$ (cyan).  The times \tDDT\ and \tIGE\ are
      marked by black circles and black diamonds respectively.}
   \label{fig:evolution}
\end{figure}

\figref{fig:evolution} shows the evolution of the gravitational binding energy
and the mass of IGEs for the five simulations performed using realization 5,
each with a different \cdens.  During the deflagration-dominated
phase,
higher-\cdens\ progenitors experience faster burning, thus expanding the
star faster due to the faster energy release.  This effect can be seen by the
rapid drop of the binding energy for the higher-density simulations around
0.5~s in the upper panel.  The transition to a detonation in higher-\cdens\ 
progenitors occurs sooner.  Once the first DDT event occurs, the production of
IGEs proceeds much faster than in the deflagration phase, as may be seen by the
sudden increase of the slope in the lower panel after the detonation occurs.

During the detonation phase the drop in binding energy is similar for all
densities.  During this same period the curves of the IGE mass stay ordered,
with the differences between them following principally from the difference in
\tDDT.  The nonlinear morphological dependencies come into play as the
detonation slows and then stops; the total IGE yield plateaus to a constant
value, but that value is not (for a single realization) correlated with the
central density of the progenitor.  The leveling off of the mass of IGEs, due
to the cessation of burning, is apparent in the lower panel.  The time \tIGE\ 
was calculated for each simulation by finding the point at which the IGE mass
changes by less than 0.01\% over the preceding 0.01~s.

\subsection{Statistics}
\label{sec:Statistics}

\begin{figure}[t]
   \includegraphics[angle=270, width=\columnwidth]{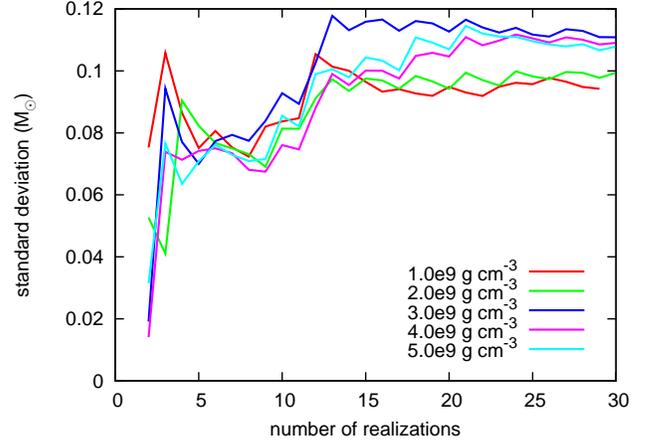}
   \caption{ Plot of the standard deviation of \Ni{56} yield as a function of
      the number of realizations, added in the order listed in
      \appref{app:init}, for each of the five central densities of the study.
      The standard deviation of the \Ni{56} mass converges around 15 total
      realizations.}
   \label{fig:Ni56StDev}
\end{figure}

We showed in \citet{KruegerEtAl10} (see especially Figure 2) that a single
initial morphology is, in general, insufficient to capture trends in SNeIa due
to the nonlinearities involved in the explosion process.  This observation
invites the question of how many initial morphologies are necessary to obtain
statistically-meaningful trends from the simulations.  \figref{fig:Ni56StDev}
presents the standard deviation of \Ni{56} yield as a function of the number of
realizations, with the realizations added in the order presented in
\appref{app:init}.  Shown are the standard deviations for each of the five
central densities of the study.  The obvious evolution of the standard
deviation until approximately 15 realizations are included in the average shows
that a statistically meaningful average requires approximately 15 realizations.
From this result, we conclude that our our sample of 30 realizations per
central density is sufficient to fully characterize the statistical trends we
present.  The choice of 30 realizations allows us to go somewhat beyond
characterizing the variation in the sample, allowing us to reduce our
uncertainty on the sample mean.

This result, the need for an average over an ensemble of simulations for
determining statistically-meaningful trends, is critical to the analysis of our
study.  Because the results discussed in this paper come from a statistical
analysis of our simulations, the results presented here generally cannot be
applied predictively to individual SNeIa, but only apply statistically to large
sets of SNeIa.  Exceptions to this rule will be noted explicitly.

\subsection{Initial Morphology Correlations}
\label{sec:init_corr}

\begin{figure}
   \centering
   \subfloat{\label{fig:init_final_corr_Rmax}
      \includegraphics[angle=270, width=\columnwidth]{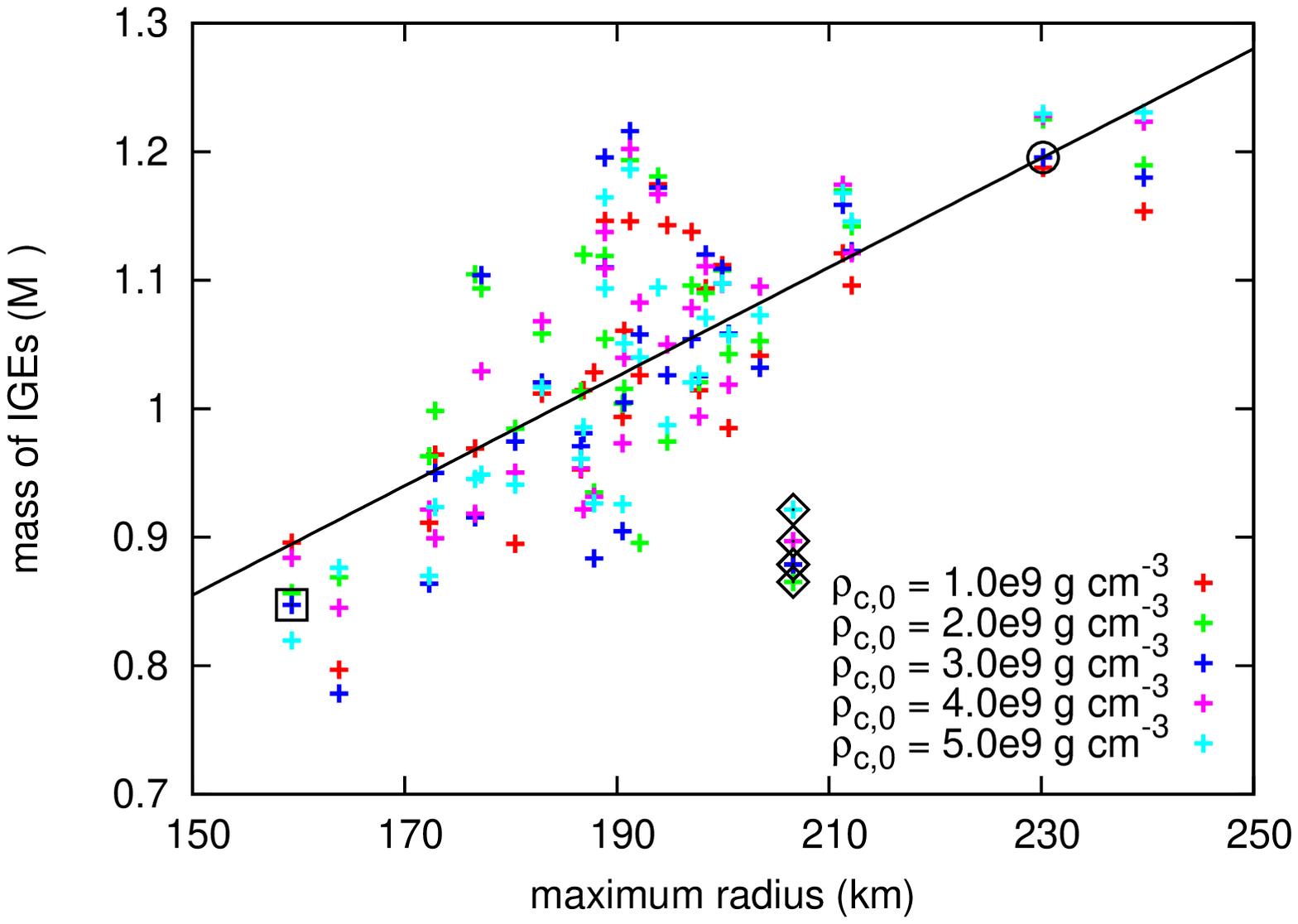}
   } \\
   \subfloat{\label{fig:init_final_corr_Rstdv}
      \includegraphics[angle=270,width=\columnwidth]{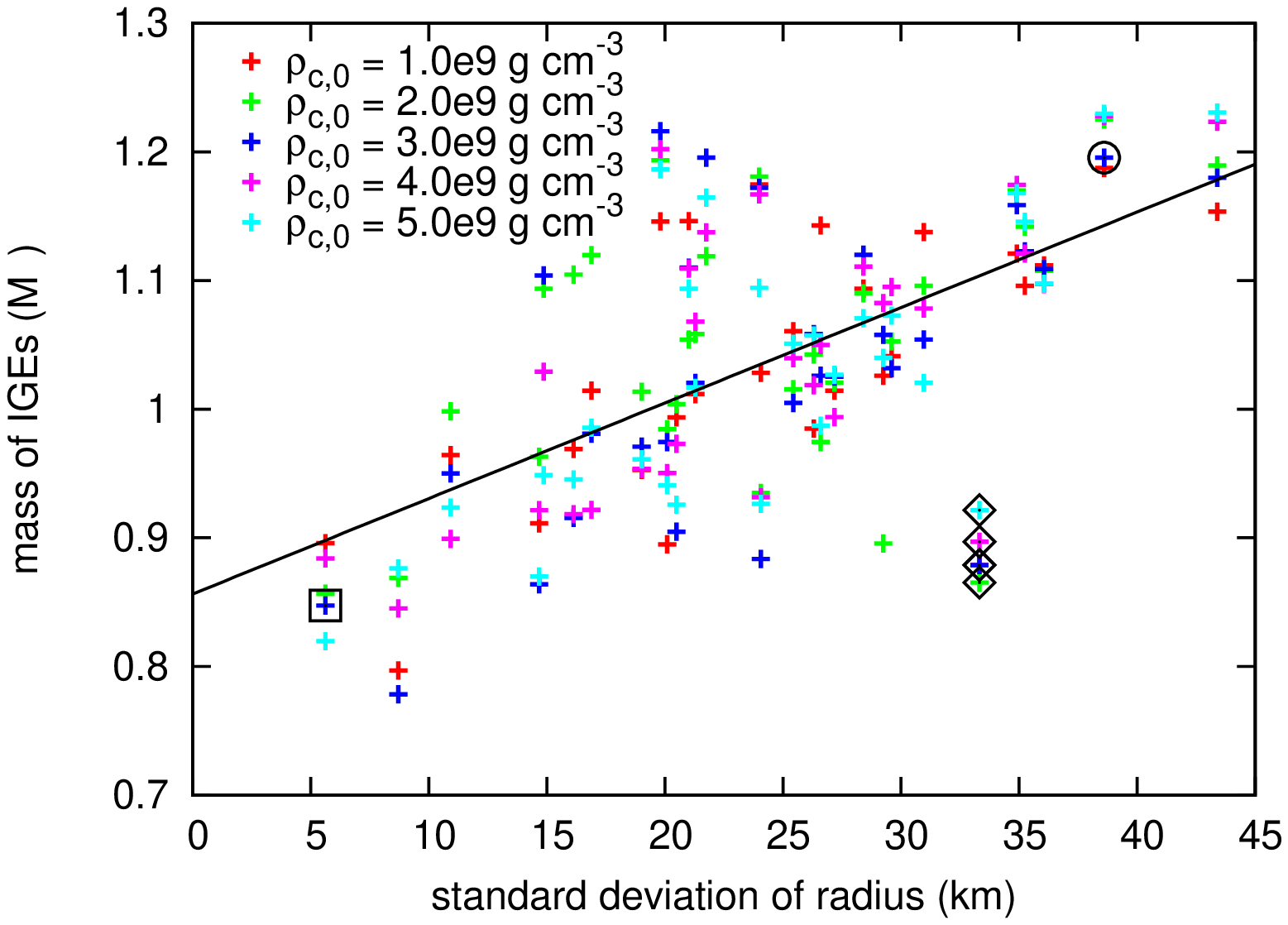}
   }
   \caption{Plots of the mass of IGEs as functions of the maximum
      initial flame radius and the standard deviation of the initial flame
      radius, both being treated as measures of how spiky the initial flame
      surface is, with simulations colored by \cdens.  The black square marks
      realization 10 and the black circle marks realization 21, both with
      $\cdens = 3\times10^9$~g~cm$^{-3}$.  These two simulations are shown in
      Figures~\ref{fig:MorphCompA} -- \ref{fig:MorphCompD}, and are examples of
      extreme cases: very smooth (realization 10) and very spiky (realization
      21).  Black diamonds mark realization 24, an example which runs counter
      to the general trend, and a good illustration of the ambiguity of the
      intermediate regions of these plots.}
   \label{fig:init_final_corr}
\end{figure}

We explored whether properties of the initial flame morphology
correlate with the final result.  We find that ``spikier'' initial flame
surfaces, that is initial morphologies that appear to have a small number of
large amplitude perturbations, give rise to faster plume growth and less
expansion, and, therefore, a higher yield of IGEs and \Ni{56}.  Several
quantities were tested as measures of the morphology of ``spikiness'', and each
gave the same qualitative result.  For example, \figref{fig:init_final_corr}
plots the final yield of IGEs as a function of the maximum radius of the
initial flame surface and standard deviation of the initial flame radius.

As illustrative cases, consider \figref{fig:MorphComp}.  The right panels shows
realization 10, $\cdens = 3\times10^9$~g~cm$^{-3}$, which is a very smooth
initial flame surface; this same realization is outlined by a black square in
\figref{fig:init_final_corr}.  The smooth configuration results in several
rising plumes approaching \rhoDDT, and these multiple plumes burn a larger
fraction of the star and release more energy than a single plume would during
its rise to \rhoDDT. The net effect is an increase in both \tDDT\ and in the
amount of expansion at \tDDT, leaving less mass at densities high enough to
burn to IGEs during the detonation phase.  Contrast this with the left panels
of \figref{fig:MorphComp}, showing realization 21, $\cdens =
3\times10^9$~g~cm$^{-3}$, which is a very spiky initial flame surface; this
same realization is outlined by a black circle in \figref{fig:init_final_corr}.
For this case, a single plume is dominant over all other features of the
initial flame surface and rapidly accelerates towards \rhoDDT\ with little
competition.  This gives a short deflagration phase, and burns a lower fraction
of the star prior to the detonation (see especially \figref{fig:MorphCompC}).
These two effects result in less expansion at \tDDT, leaving more mass at a
density high enough to burn to IGEs during the detonation, resulting in a
greater yield of IGEs and \Ni{56}.

In the center of each panel of \figref{fig:init_final_corr}, the trend is not
so clear.  Realization 24 (outlined by black diamonds) is a good example of the
lack of a clear trend in this intermediate range.  The initial flame surface
for realization 24 has multiple large spikes, so that it has a larger than
average maximum radius and radial standard deviation, and would be considered a
spiky case.  However, most of these spikes are of comparable maximum radius,
and none of these features develops into a dominant plume, as is typical of the
spiky cases; instead, the plumes grow together and the behavior is like that of
the smooth cases.  Thus realization 24 leads to high values of \tDDT\ and low
yields of IGEs and \Ni{56} relative to simulations of similar spikiness.  Thus
the intermediate range does not show a strong trend, while the extreme ranges
(multiple competing plumes starting from a smooth initial surface, or a single
dominant plume with no significant competition) show the trend more clearly.

\subsection{Yields}
\label{sec:Yields}

\begin{figure}
   \includegraphics[angle=270, width=\columnwidth]{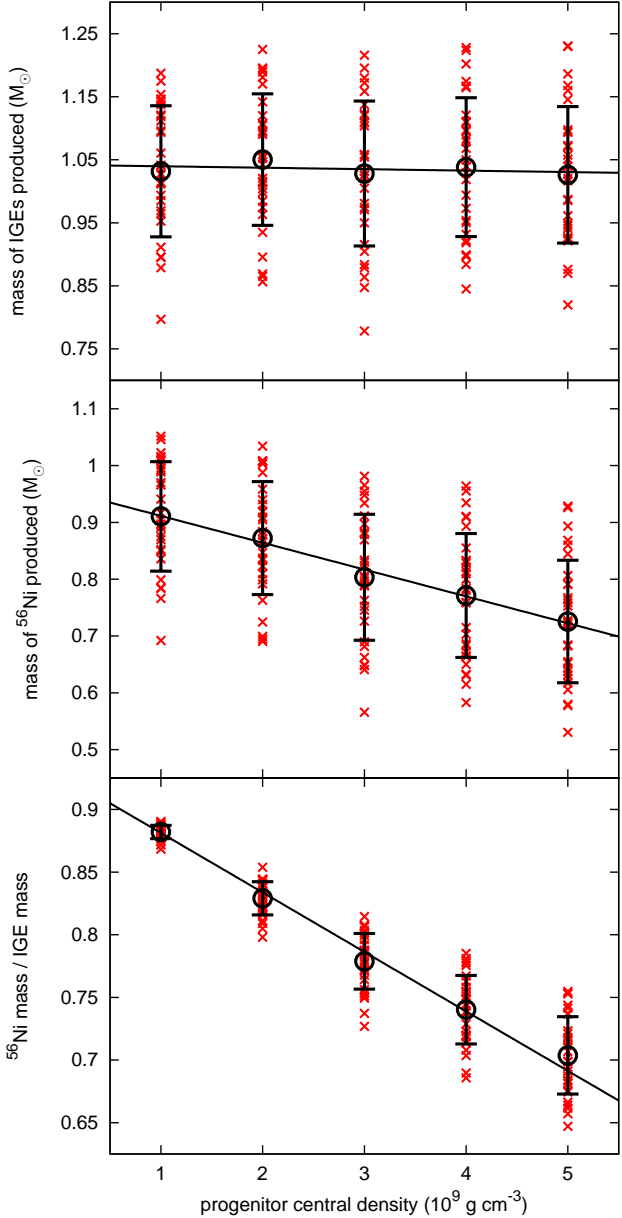}
   \caption{Final yields (at \tIGE) of our simulations plotted against \cdens.
      The black lines are the best-fit trend lines, with the averages and
      standard deviations marked by the circles and the vertical error bars.}
   \label{fig:yield_by_rho}
\end{figure}

\figref{fig:yield_by_rho} shows the masses of IGEs and \Ni{56} produced in the
149 simulations that were analyzed, along with the \Ni{56}-to-IGE mass ratio.
The figures plot yield vs.\ central density and also show average yields and
standard deviations at each central density, with best-fit trend lines.  The
mass of IGEs is consistent with a flat line; i.e., it is independent of \cdens.
However, the mass of \Ni{56} decreases with increasing \cdens.  The significant
scatter in the two masses is readily apparent; the mean standard deviations for
the IGEs and \Ni{56} masses are 0.108~$M_\odot$ and 0.105~$M_\odot$,
respectively.  As discussed in \citet{Mazzali2006The-54Fe58Ni/56} and
\citet{WoosleyEtAl07:LightCurves}, assuming a constant IGE mass and varying the
\Ni{56} mass produces SNeIa that lie approximately along the observed
width-luminosity relationship, while the width of the relationship allows the
IGE mass to vary somewhat (c.f.\ Figures 15 and 20 of
\citealt{WoosleyEtAl07:LightCurves}).  We note that the masses
of IGEs and in particular the masses of \Ni{56} from our simulations are
on average higher than accepted results for masses synthesized in actual
SNIa events~\citep[see][and references therein]{WoosleyEtAl07:LightCurves}.

The \Ni{56}-to-IGE mass ratio decreases with increasing \cdens, as would be
expected from a constant IGE mass and a decreasing mass of \Ni{56}.  However,
unlike the constant standard deviations of these two masses, the standard
deviation of the \Ni{56}-to-IGE mass ratio increases with \cdens.  The
variation in the \Ni{56}-to-IGE ratio is dominated by variation related to
neutronization: because neutronization can exaggerate differences that arise in
the hydrodynamics, we find that the standard deviation (variation) increases
with the total mass of stable IGEs synthesized

\begin{figure}[t]
   \includegraphics[angle=270, width=\columnwidth]{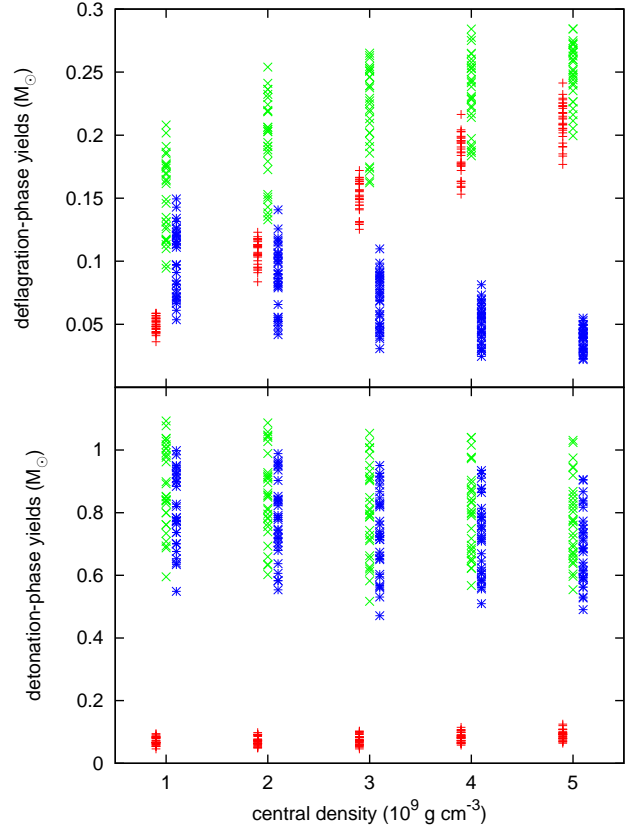}
   \caption{Plots of the masses of stable IGEs (red pluses), \Ni{56} (blue
      asterisks), and total IGEs (green crosses) produced during the
      deflagration phase (top panel) and the detonation phase (bottom panel).
      The symbols for each mass have been horizontally shifted slightly so the
      symbols do not overlap.}
   \label{fig:defdet}
\end{figure}

Additionally, we found that the yield from burning during the deflagration
phase is substantially different from the yield during the detonation phase.
\figref{fig:defdet} shows the yields of stable IGEs and \Ni{56} during the
deflagration and detonation phases.  The trend of increasing stable IGE yield
and decreasing \Ni{56} yield with increasing central density is most obvious in
the deflagration phase yield.  We interpret this result as following from the
fact that the detonation phase involves burning at densities that are typically
lower than those of the deflagration phase due to expansion of the star during
the deflagration phase.  The neutronization rate increases with density, so it
shows up most strongly in the deflagration phase before significant expansion
occurs.

\subsection{Distribution of \Ni{56}}
\label{sec:NiDistribution}

\begin{figure}
   \centering
   \subfloat{\label{fig:Ni56_profile1}
      \includegraphics[angle=270,width=\columnwidth]{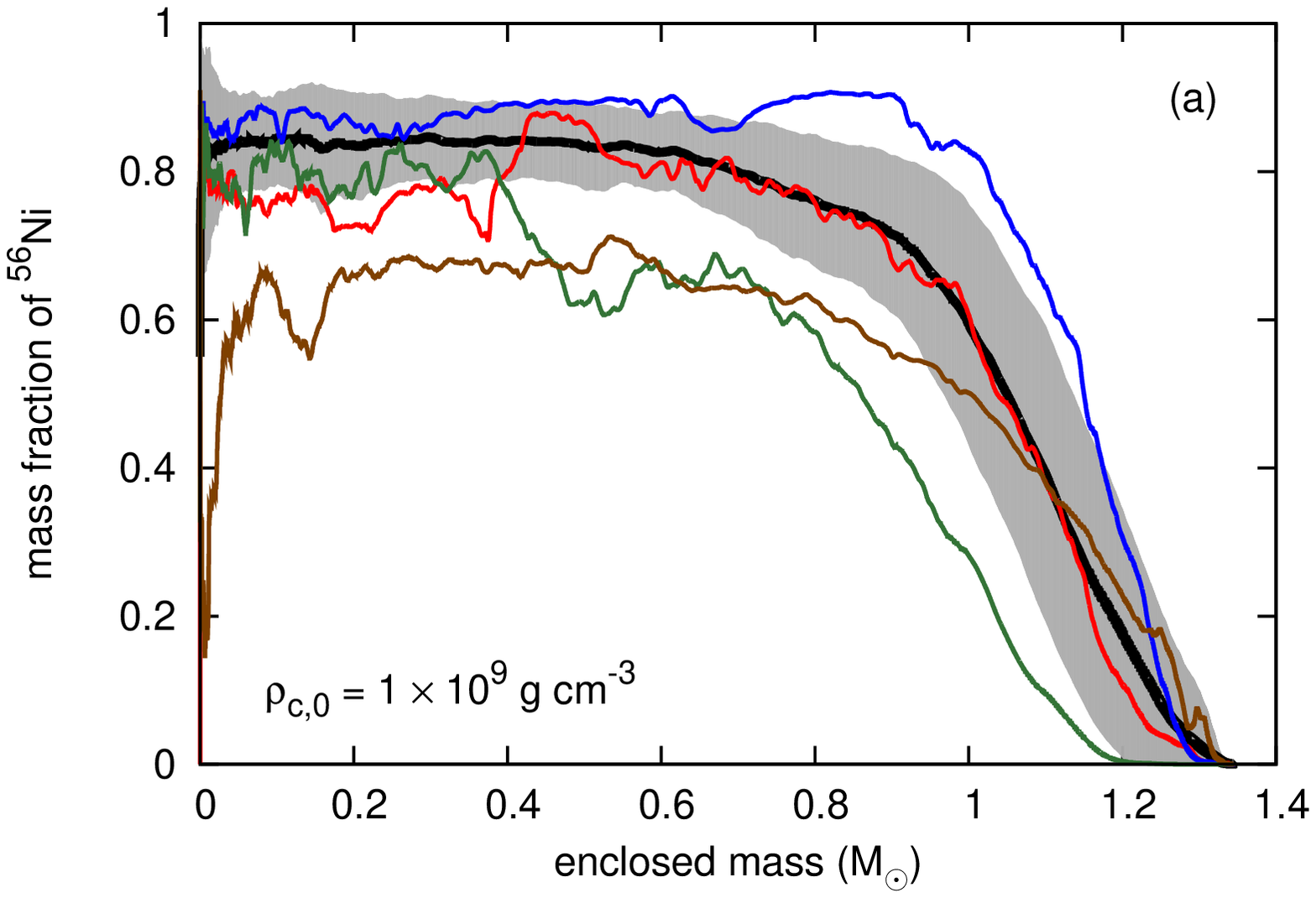}
   } \\
   \subfloat{\label{fig:Ni56_profile3}
      \includegraphics[angle=270,width=\columnwidth]{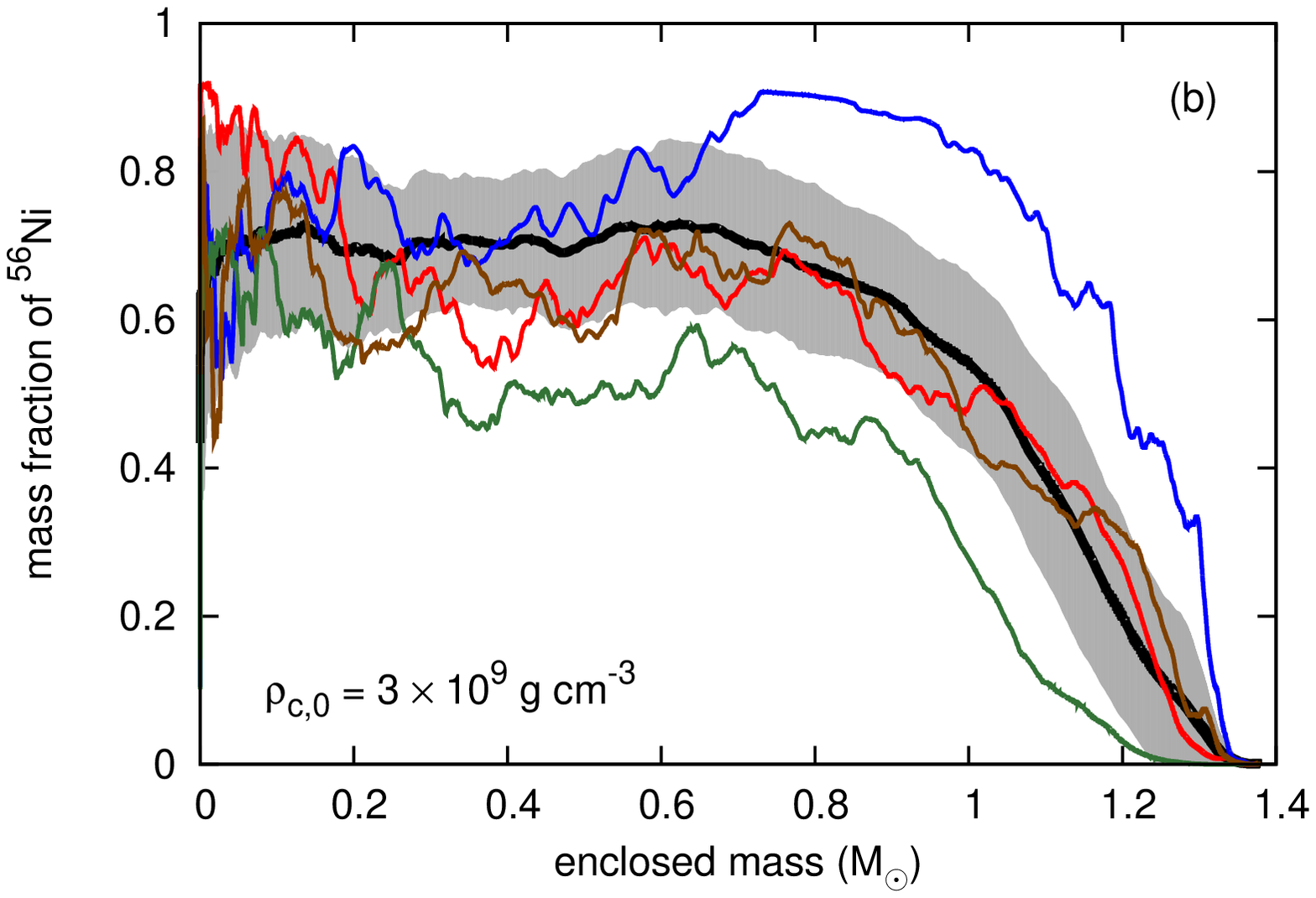}
   } \\
   \subfloat{\label{fig:Ni56_profile5}
      \includegraphics[angle=270,width=\columnwidth]{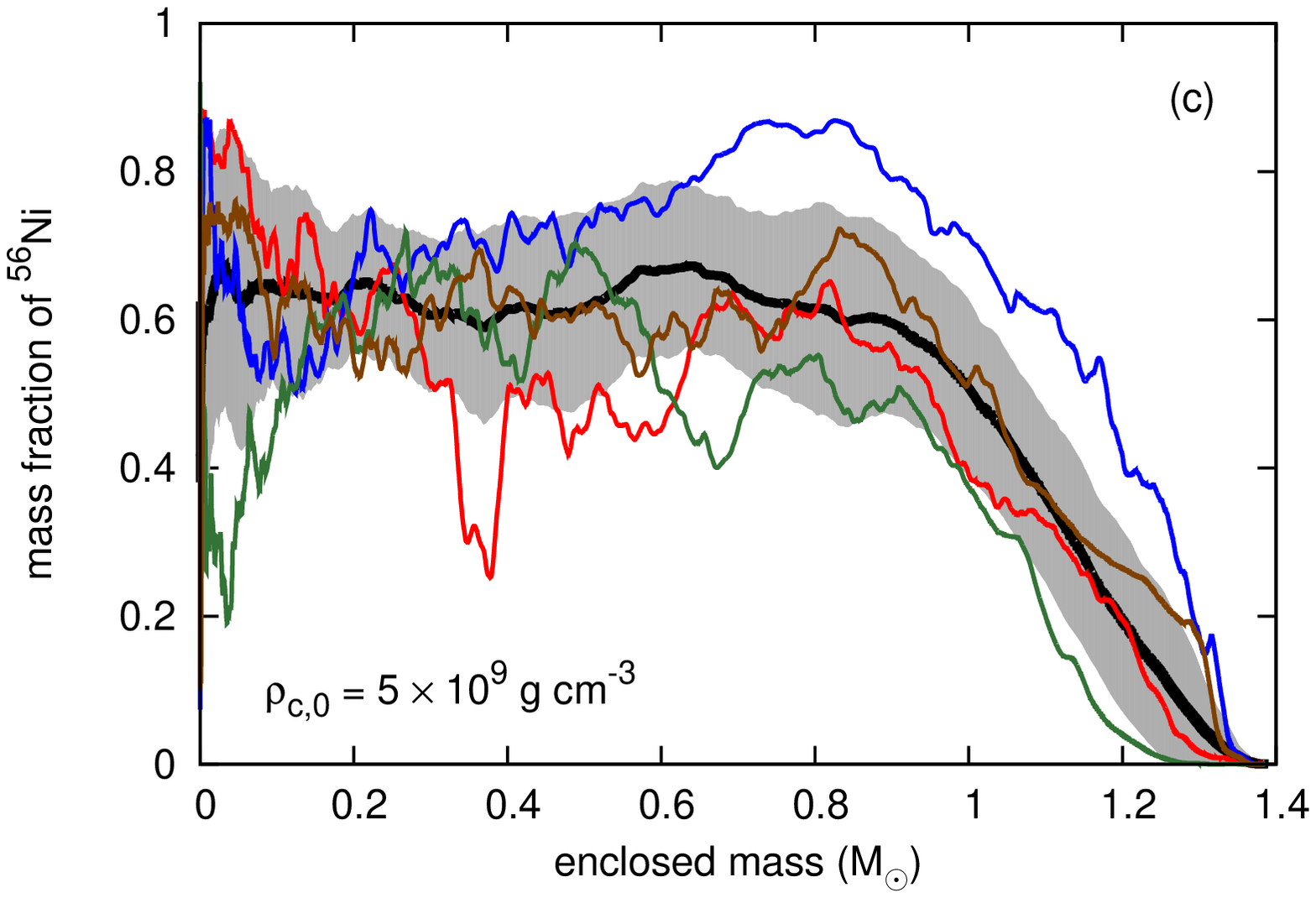}
   }
   \caption{Radially-averaged profiles of \Ni{56} mass fraction at three
      central densities ($1 \times 10^9$, $3 \times 10^9$, $5 \times
      10^9$~g~cm$^{-3}$ in Panels~\ref{fig:Ni56_profile1},
      \ref{fig:Ni56_profile3}, and \ref{fig:Ni56_profile5} respectively).  The
      black curves show the mean profile of all simulations at a given \cdens,
      with the grey band showing the standard deviation.  The four colored
      curves represent four different realizations: 4 (red), 12 (blue), 13
      (green), and 29 (brown).}
   \label{fig:Ni56_profile}
\end{figure}

The principal result from a simulation is the mass of \Ni{56} synthesized in
the explosion, which directly sets the brightness of an event.  The synthesized
\Ni{56} masses are listed in \appref{app:data}.  The next question to be
answered in the analysis concerns the distribution of \Ni{56} in the remnant.
\figref{fig:Ni56_profile} presents radially-averaged profiles of \Ni{56} for
three of the five central densities ($\cdens = 1 \times 10^9$, $3 \times 10^9$,
and $5 \times 10^9$).  Also shown are representative results from realizations
4, 12, 13, and 29 for comparison.  The radial profile of \Ni{56} mass fraction
varies between simulations, but certain details are consistent across the
ensemble.  The inner 0.8~--~1.0~\Msol\ has a high \Ni{56} mass fraction; we
refer to this region as the ``plateau'', although there can be significant
variation within this region.  Outside of the plateau, there is a smooth
decrease to a \Ni{56} mass fraction at or near zero at the surface of the WD;
we refer to this as the ``decline'' region.  It appears that this decline
region may not be fully relaxed to the final profile at time \tIGE, but may
experience some steepening prior to entering the free-expansion phase.
Lower-\cdens\ simulations have less neutronization and therefore generate more
\Ni{56}.  For these simulations the \Ni{56} mass fraction in the plateau tends
to cluster near the maximum value ($\sim\!0.9$), with some deviations down as
far as $\sim\!0.6$.  The higher-\cdens\ simulations have more neutronization
and therefore generate less \Ni{56}, so the typical \Ni{56} mass fraction in
the plateau shows more variation and there exist larger deviations from the
mean (down to $\sim\!0.2$).

The representative realizations presented in \figref{fig:Ni56_profile}
demonstrate the deviation from mean behavior.  In particular, the
representative realizations show ``typical'' behavior (the \Ni{56} profile
nearly matches the mean profile), over- and under-luminous
models, and
\Ni{56} holes (patches within the plateau with significantly less \Ni{56} than
the surrounding regions).  Previous studies have reported the presence of a
\Ni{56} hole in the inner region~\citep{hoeetal2010}.  The green and red curves
in the bottom panel of \figref{fig:Ni56_profile} (realization 4) illustrate
such a \Ni{56} deficit, although the deficit is offset from the center of the
WD in the case of the red curve.  Most of our simulations do not show evidence
of this \Ni{56} hole in the inner region.  Only a few simulations have such a
feature, with \Ni{56} holes being more common in simulations with a higher
\cdens.  Turbulent mixing caused by the burning processes breaks the symmetries
that give rise to a consistent central \Ni{56} hole in 1-d simulations.

\begin{figure*}
   \centering
   \begin{tabular}{cc}
   \subfloat{
      \label{fig:NSE_D5_R04}
      \includegraphics[height=2.8in]{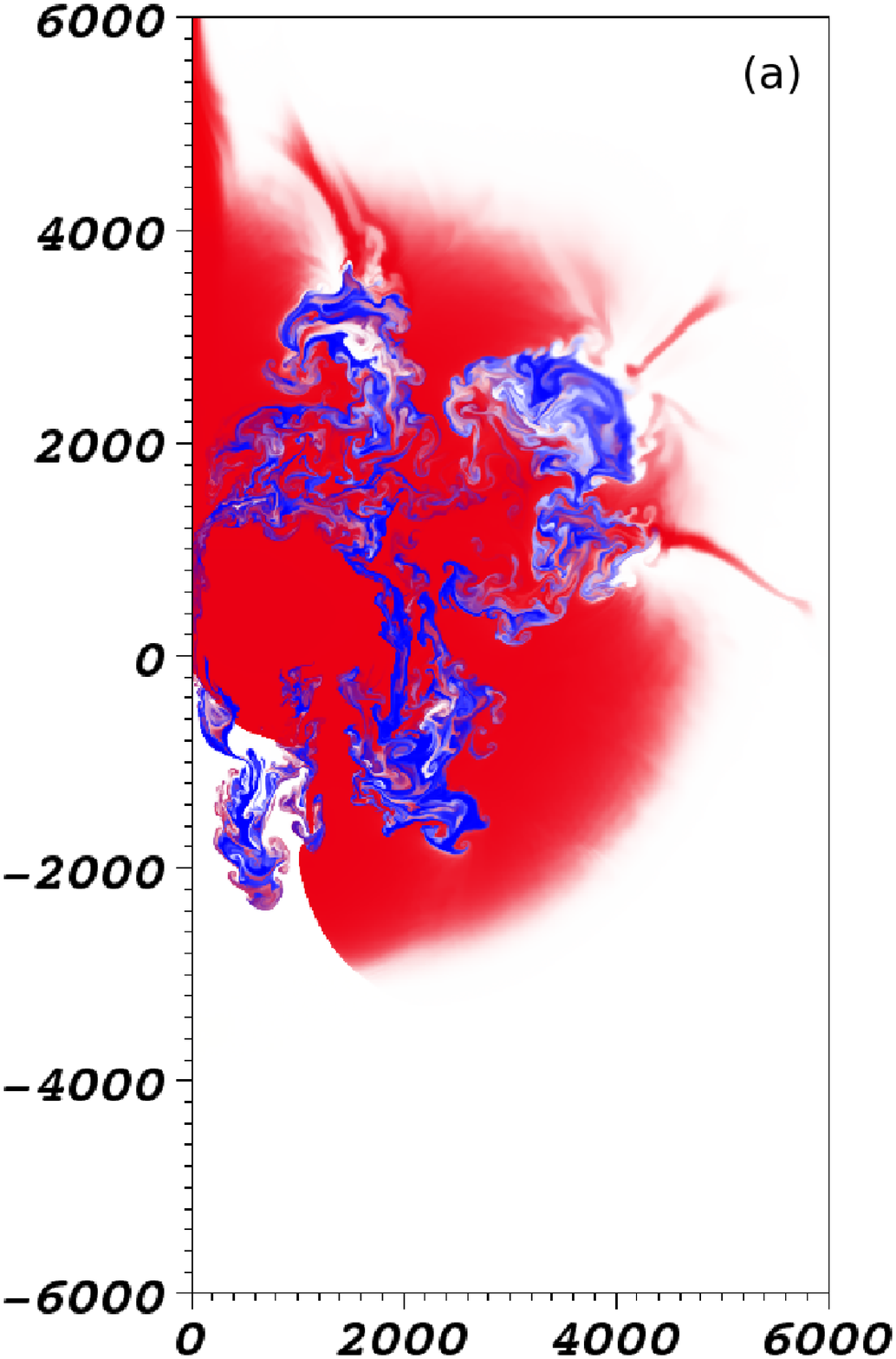}
   }
   \subfloat{
      \label{fig:NSE_D5_R12}
      \includegraphics[height=2.8in]{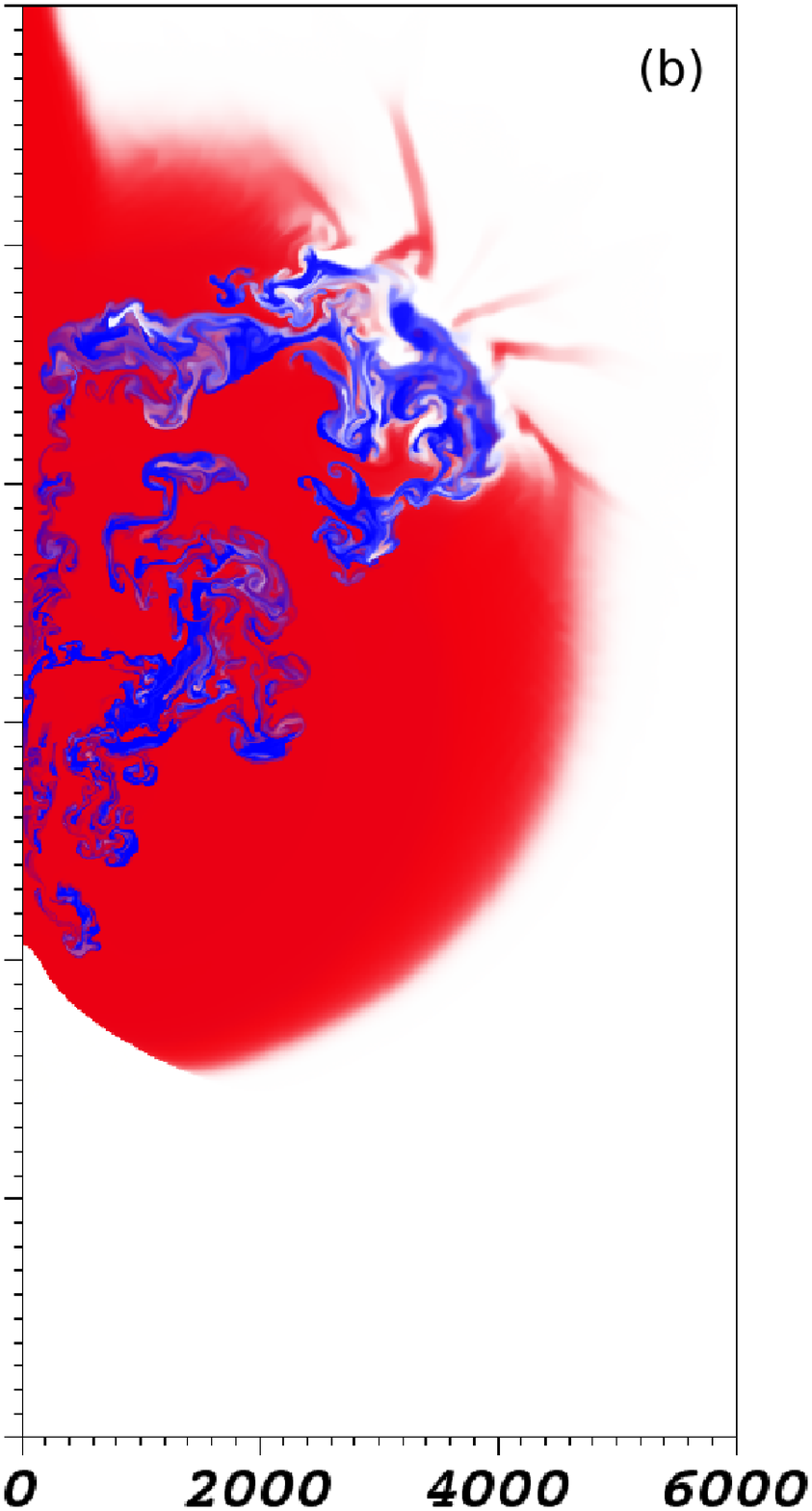}
   }
   \subfloat{
      \label{fig:NSE_D5_R13}
      \includegraphics[height=2.8in]{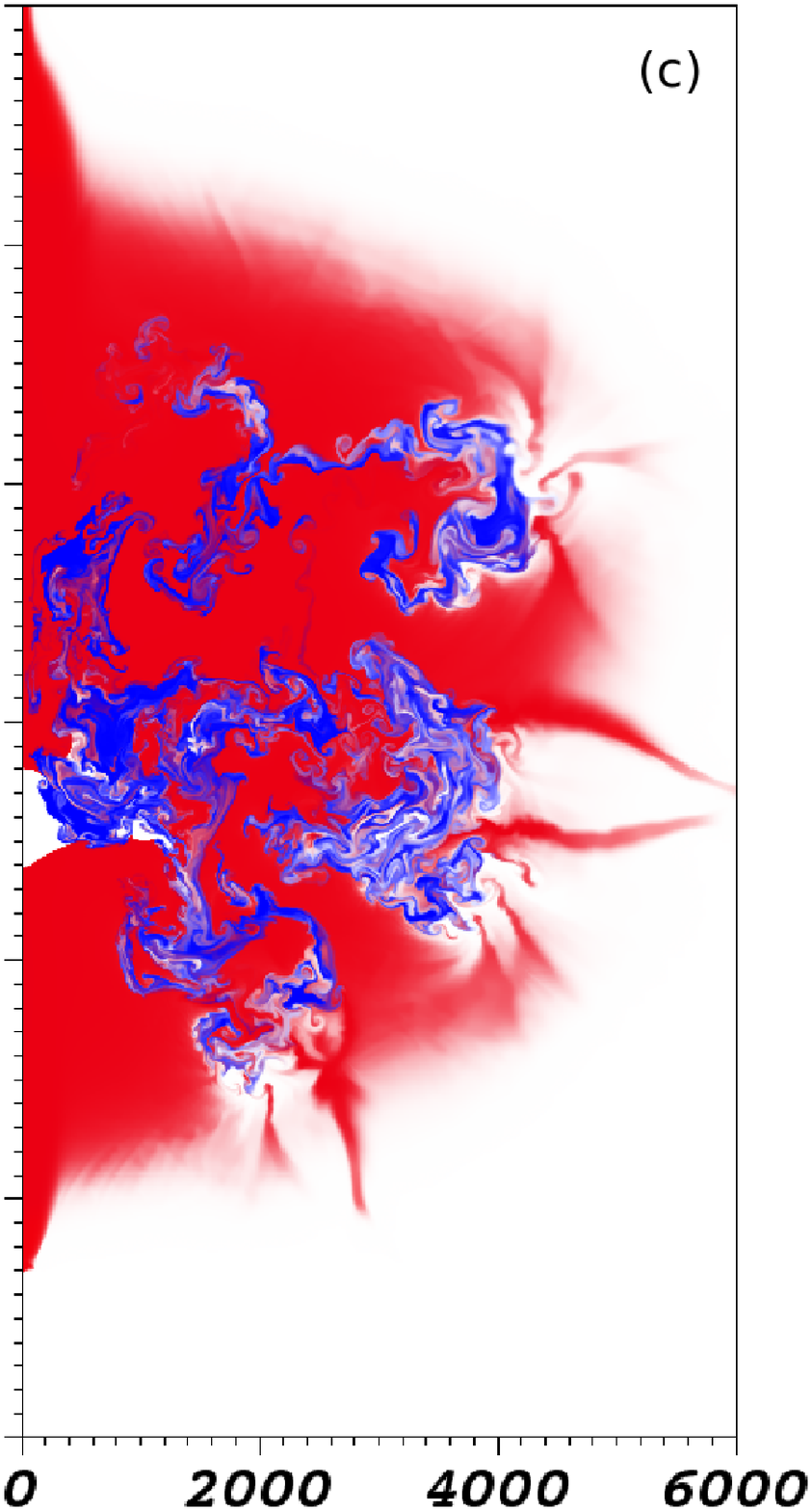}
   }
   \subfloat{
      \label{fig:NSE_D5_R29}
      \includegraphics[height=2.8in]{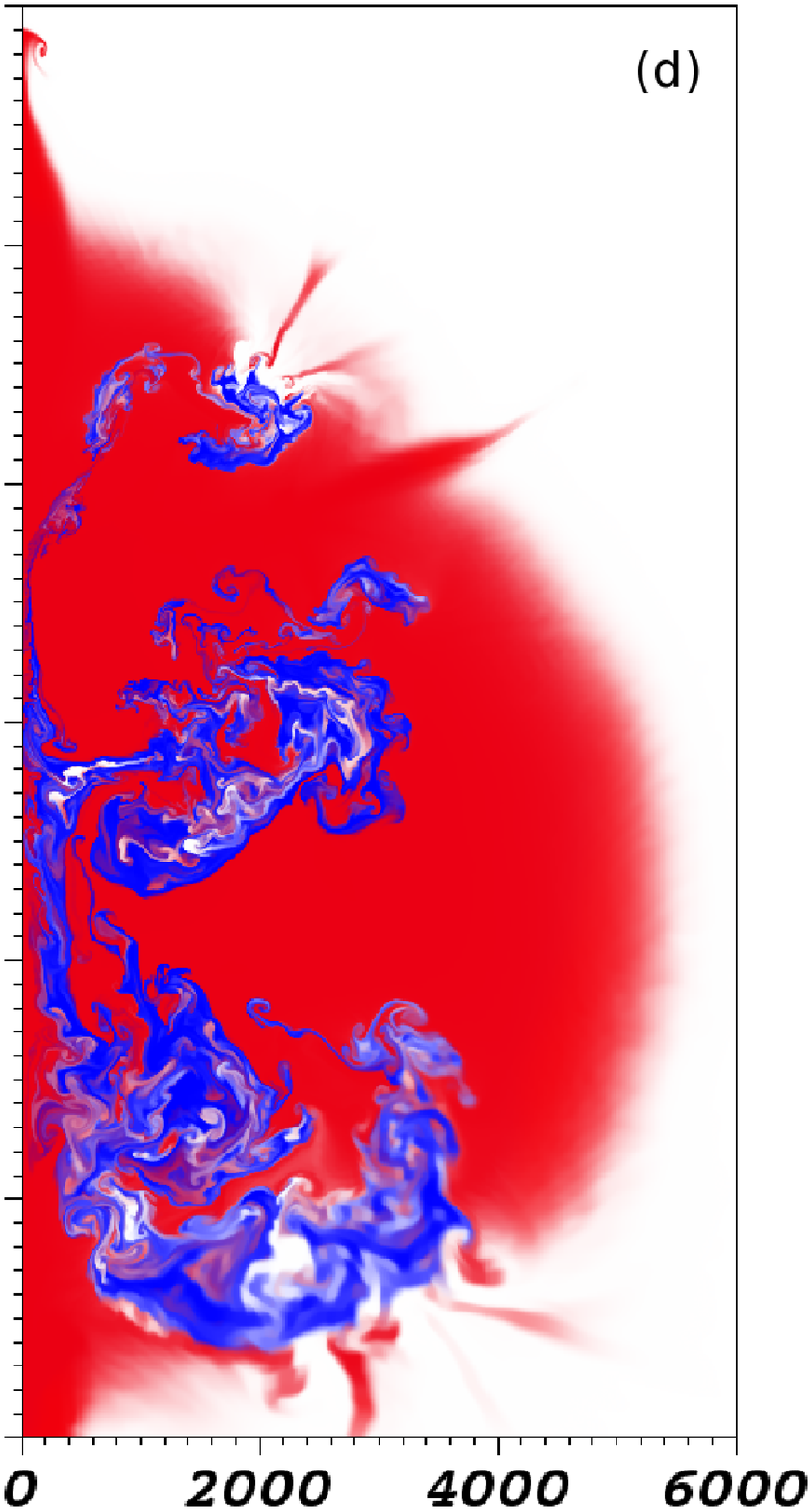}
   }
   \end{tabular}
   \caption{Plots of the inner 6,000~km of four simulations at \tIGE, showing
      mass fractions of
      \Ni{56} (red), stable IGEs (blue), and non-IGE material (white).  These
      are the same simulations plotted in \figref{fig:Ni56_profile5}, with
      $\cdens = 5 \times 10^9$~g~cm$^{-3}$; from left to right, the panels show
      a non-central \Ni{56} hole (realization 4), an overluminous case
      (realization 12), an underluminous case with a central \Ni{56} hole
      (realization 13), and a typical case (realization 29).  The off-center
      \Ni{56} deficit in realization 4 is due to a band of neutronized, stable
      isotopes caused by multiple plumes that have reached a common radius.
      The central \Ni{56} hole in realization 13 is partially due to a
      neutronized region around the equator, and partially due to an
      incompletely-burned region near the axis, below the equator.}
   \label{fig:NSE_2d}
\end{figure*}

\figref{fig:NSE_2d} shows the distributions of \Ni{56}, stable IGEs, and
non-IGE material for the four sample cases shown in \figref{fig:Ni56_profile5}
($\cdens = 5 \times 10^9$).  For our results, the mass of stable IGEs is
approximated by any IGE material that is not \Ni{56}.  The figures present the
inner 6000~km of the domain, and for all cases the bulk of IGEs appears in the
plotted region.  The distributions indicate that \Ni{56} holes can be caused by
incompletely-burned regions (for example, on the axis of
\figref{fig:NSE_D5_R13} just below the equator) or by neutronized regions (for
example, adjoining the previously-mentioned incompletely-burned region of
\figref{fig:NSE_D5_R13}, or the band at a radius of $\sim\!2000$~km in
\figref{fig:NSE_D5_R04}).  The plume rise, velocity fields, and neutronization
may be asymmetric, which is seen especially in Figures~\ref{fig:NSE_D5_R04} and
\ref{fig:NSE_D5_R29}.  The degree of asymmetry observed in our models suggests
that there may be noticeable line-of-sight effects for SNeIa.

\subsection{Distinguishing Age Among SNeIa of Equal Brightness}
\label{sec:EqualBrightness}

\begin{figure}[t]
   \includegraphics[angle=270, width=\columnwidth]{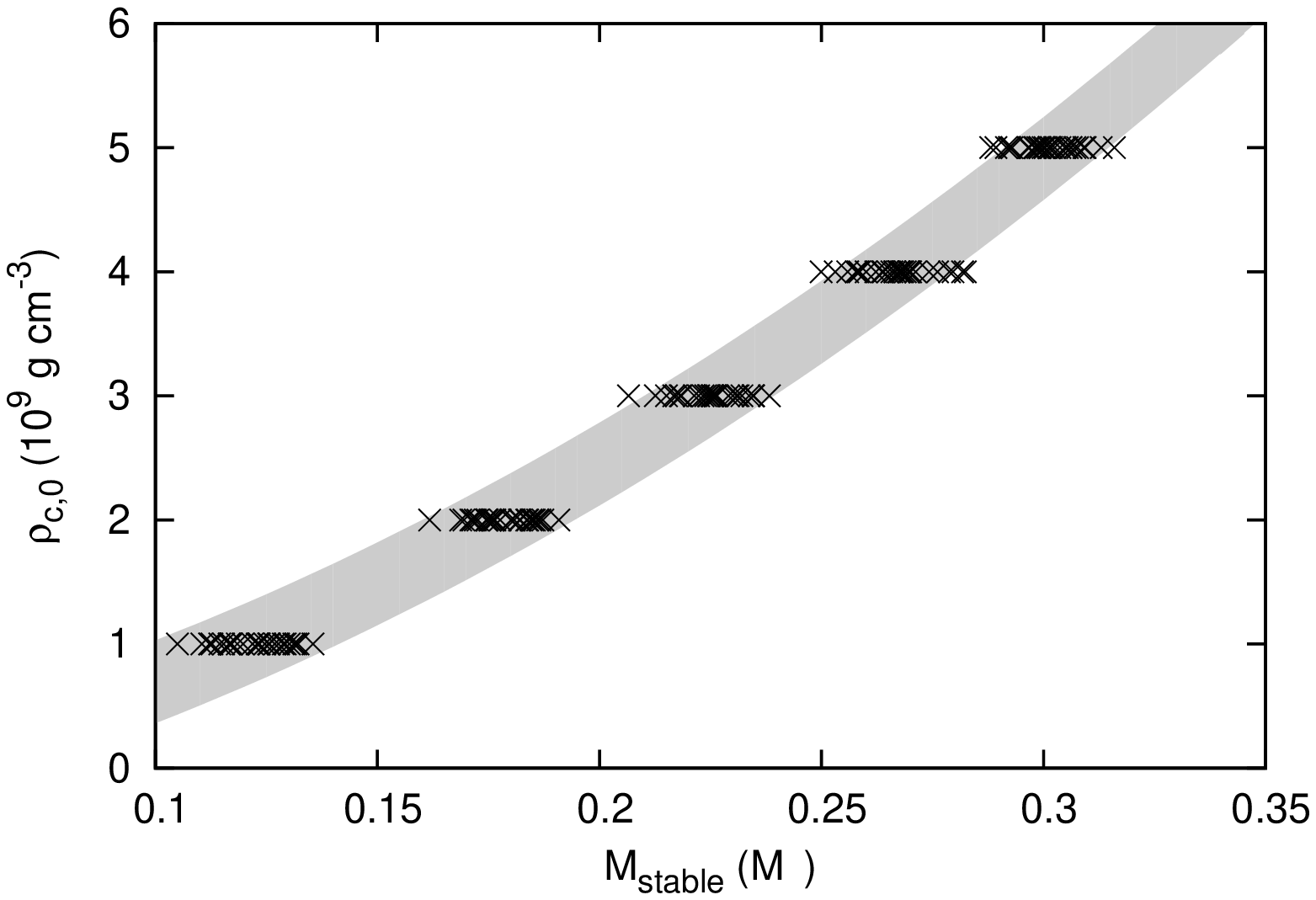}
   \caption{ Plot of \cdens\ vs.\ $M_{\rm stable}$.  The shaded region displays
      two standard deviations around the model (\eqnref{eqn:rho_v_Mstable}).}
   \label{fig:rho_v_Mstable}
\end{figure}

Can we extract from our results insight into the age of a progenitor WD given
the brightness of the SNIa?  From our results, one might conclude that the mass
of stable IGEs ($M_{\rm stable}$) should increase with increasing \cdens, and
that the relation should have a large scatter (akin to the scatter in the
\Ni{56} and IGE relations).  \figref{fig:rho_v_Mstable} illustrates the
relationship between \cdens\ and $M_{\rm stable}$, along with the best-fit
trend and the scatter (the shaded region shows two standard deviations in each
direction around the best fit curve through the data).  The data show an
increasing trend as expected, with the best fit being
\begin{equation}
   \left(\frac{\cdens}{10^9{\rm ~g~cm}^{-3}}\right)
      = 35.4 \left(\frac{M_{\rm stable}}{\Msol}\right)^2
      + 6.92 \left(\frac{M_{\rm stable}}{\Msol}\right)
      - 0.349 \; .
   \label{eqn:rho_v_Mstable}
\end{equation}
Unlike the \Ni{56} and IGE relations, however, there is a very small scatter in
this relation for $M_{\rm stable}$, with a standard deviation of only $0.167
\times 10^9$~g~cm$^{-3}$.  This result is the tightest relation in our data, to
the point where this relation can be meaningfully applied to a single
event.
This tight relationship is unlike all of our other relations, which only apply
to the statistics of large ensembles.

Given observations of multiple SNeIa of the same brightness and a reliable
measure of the mass of stable IGEs from the observations, our
models predict that we can use
\eqnref{eqn:rho_v_Mstable} to determine the relative ages of the progenitors.
The relation provides the central density from observed masses of stable IGEs.
\citet{LesaffreEtAl06} presents relations between the cooling time of the
progenitor and the central density of the progenitor at the ignition of the
thermonuclear runaway.  Applying the relations of \citet{LesaffreEtAl06} to the
central densities, our result allows determination of  the cooling times the
progenitor WDs experienced.  Assuming the progenitors had the same main
sequence mass, we thus obtain a measure of the relative ages of the
progenitors.  Because this result is derived from varying the central density
of the progenitor, we are implicitly assuming that such a variation of \cdens\ 
is the dominant effect on the mass of stable IGEs.  Future work in
three dimensions will consider central density variation combined with other
effects that may be related to age.

\begin{figure}
   \centering
   \subfloat{\label{fig:Ni069}
      \includegraphics[angle=270, width=\columnwidth]{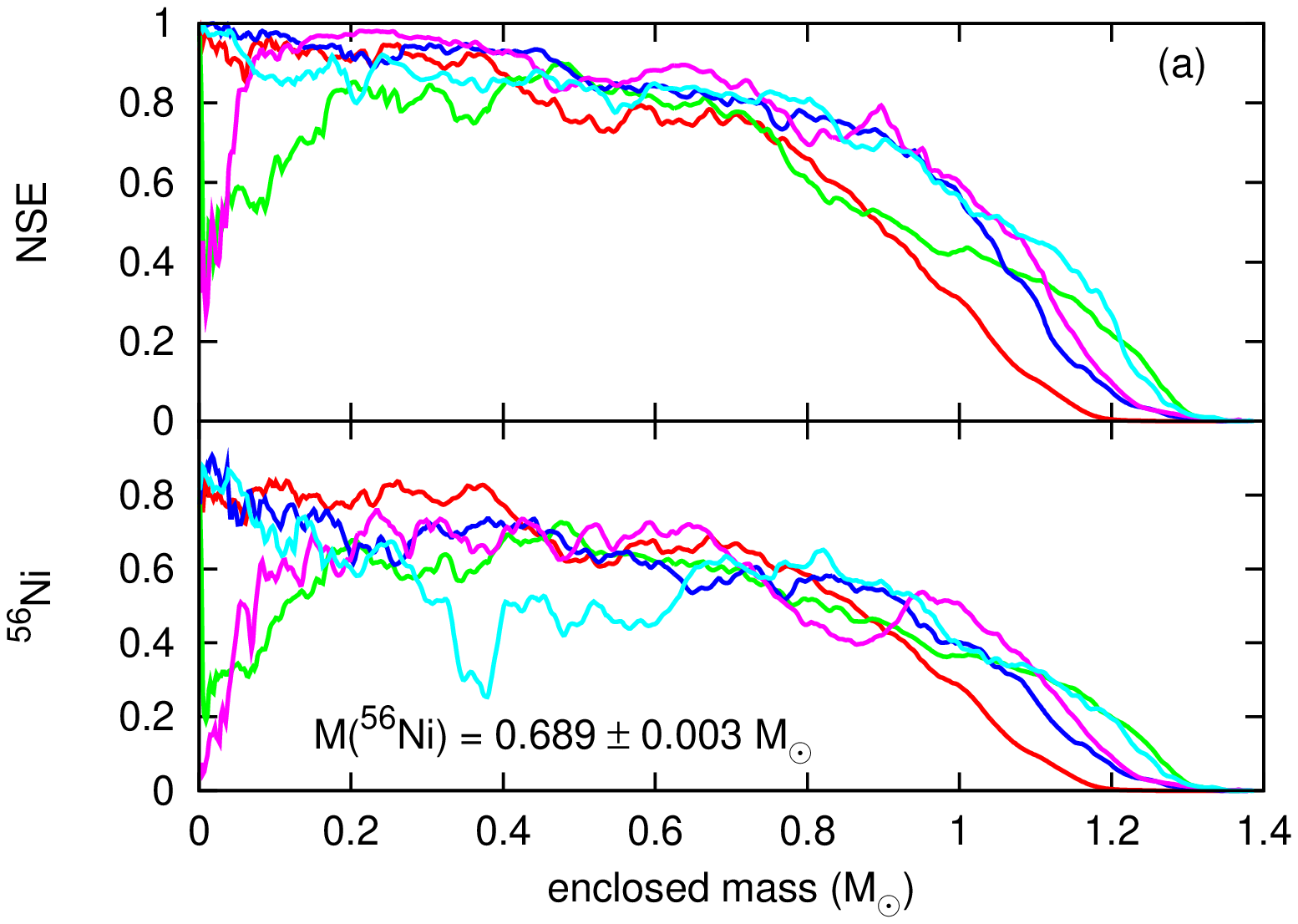}
   } \\
   \subfloat{\label{fig:Ni076}
      \includegraphics[angle=270, width=\columnwidth]{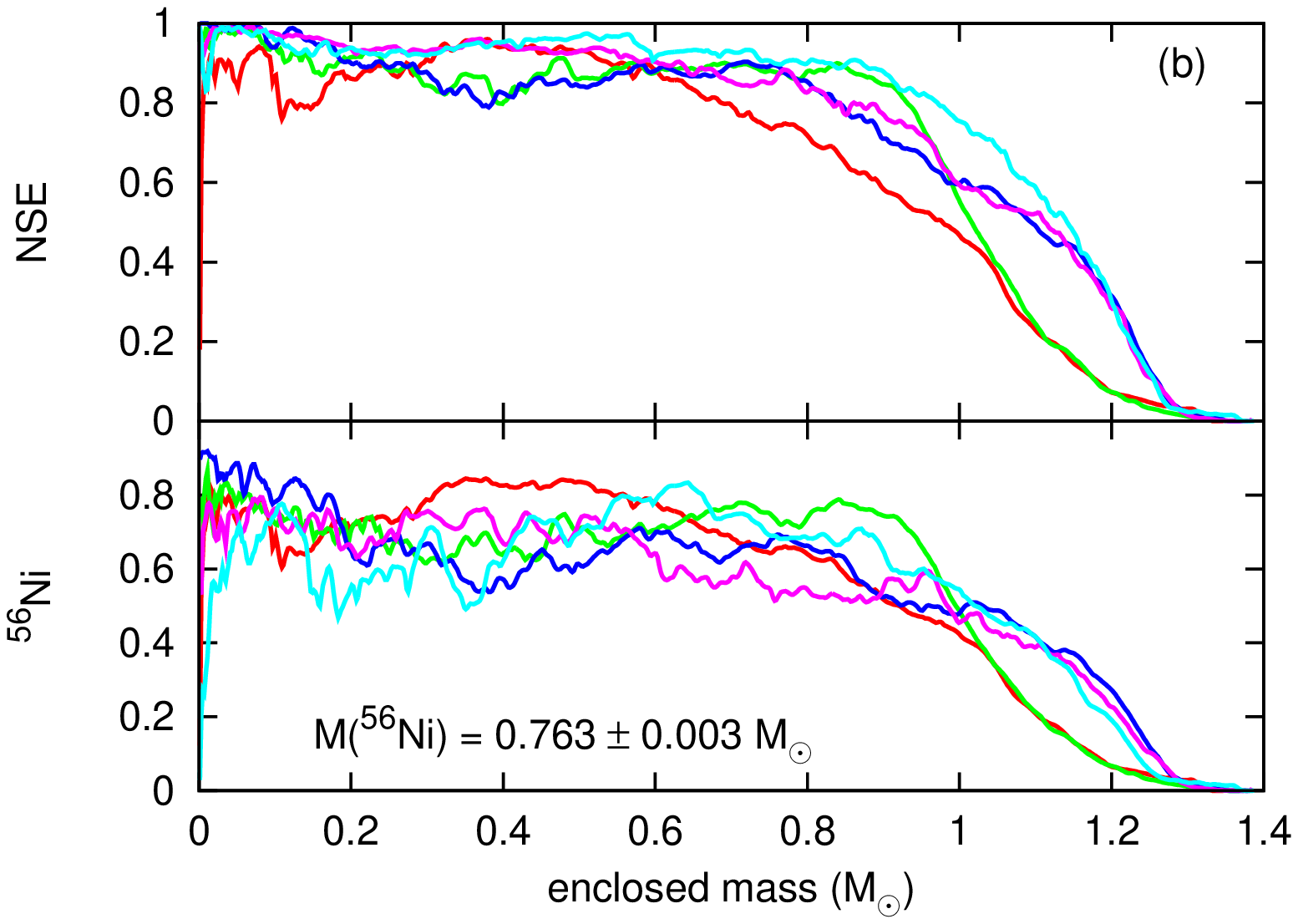}
   } \\
   \subfloat{\label{fig:Ni089}
      \includegraphics[angle=270, width=\columnwidth]{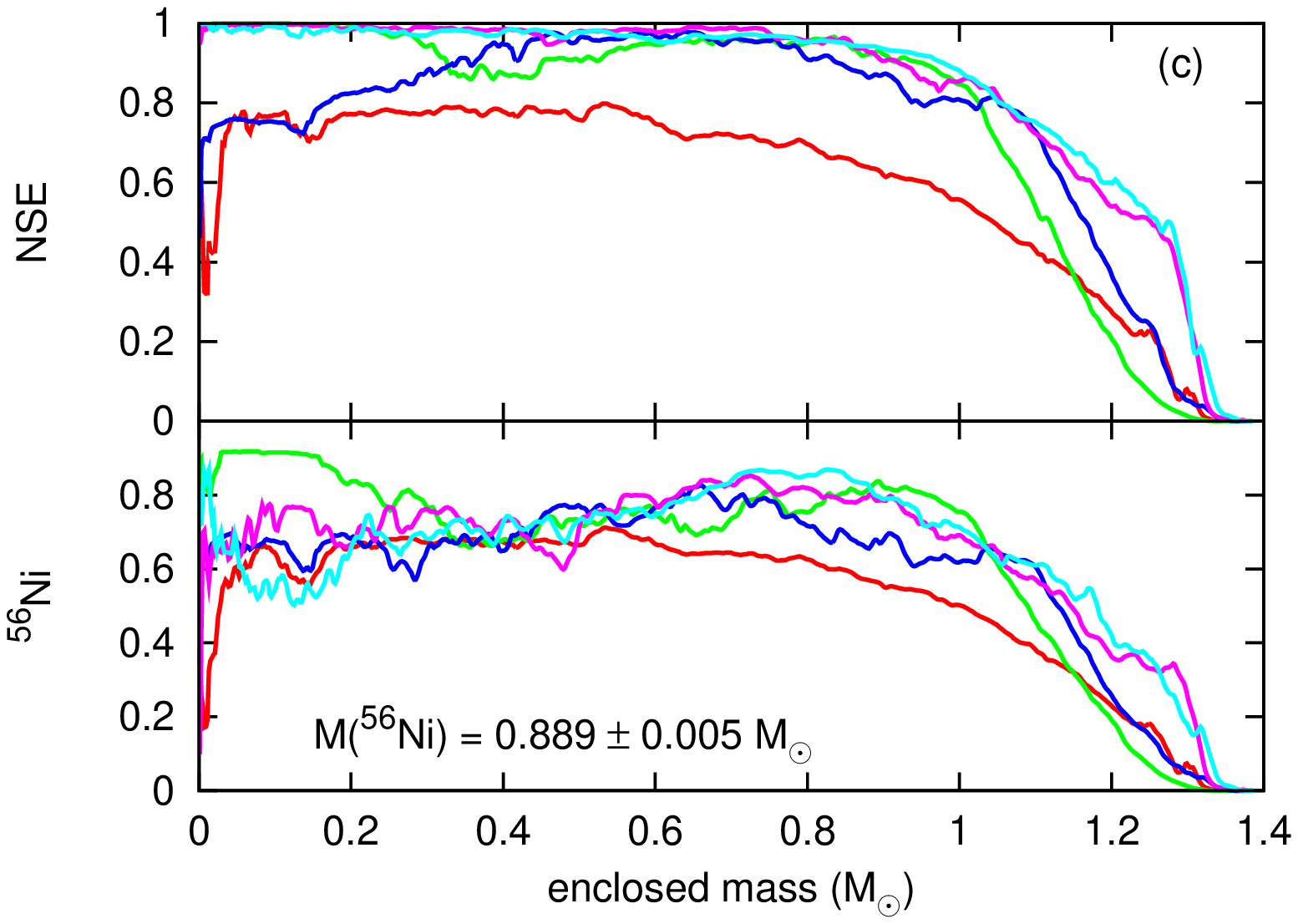}
   }
   \caption{Comparisons of radial profiles of \Ni{56} and IGE mass fractions
      for models with nearly the same total mass of \Ni{56}, but different
      values of \cdens.  The upper set of profiles has a mass of $0.689 \pm
      0.003$~$M_\odot$, the central set of profiles has a mass of $0.763 \pm
      0.003$~$M_\odot$, and the lower set of profiles has a mass of $0.889 \pm
      0.005$~$M_\odot$.  Lines are colored by \cdens: $1.0 \times
      10^9$~g~cm$^{-3}$ (red), $2.0 \times 10^9$~g~cm$^{-3}$ (green), $3.0
      \times 10^9$~g~cm$^{-3}$ (blue), $4.0 \times 10^9$~g~cm$^{-3}$ (magenta),
      $5.0 \times 10^9$~g~cm$^{-3}$ (cyan).}
   \label{fig:NiNSEProfiles}
\end{figure}

Our models, specifically the distribution of IGEs in the expanding remnant,
offer insight into what would be required for observing the mass of stable
IGEs.  While our models exhibit a mild degree of asymmetry, we find systematic
behavior of the distribution of heavy nuclei for events of a given brightness
(or given mass of \Ni{56}).  \figref{fig:NiNSEProfiles} shows profiles of
\Ni{56} and IGE mass fractions in three sets.  Each set has a simulation at
each of the five central densities, but the simulations within a set are chosen
to have approximately the same integrated \Ni{56} mass.  Accordingly, the
simulations within a set may not be from the same realization.  These profiles
are generated at \tIGE, when the burning has essentially ceased, but not yet
into the free-expansion phase.

The plateau (central region) exhibits a slight systematic behavior in
\figref{fig:NiNSEProfiles}.  Considering both the curves of \Ni{56} and IGEs,
we note a tendency of a less-well-defined plateau in the lowest-\Ni{56}-mass
set (top panel).  The start of the decline region occurs at a higher enclosed
mass for simulations with a higher mass of \Ni{56}; the plateau extends farther
out.  Also, there appears to be a mild trend in the plateau region of
simulations within a set: the higher \cdens\ simulations (magenta and cyan
curves) tend to show a wider plateau in the IGE mass fraction.  In addition, we
note that within these results, some simulations exhibit the \Ni{56} hole while
some do not.  The hole may be observed in the drastic decreases of some curves
at the lowest enclosed masses.

\begin{figure*}
   \centering
   \subfloat{\label{fig:NSE_D10}
      \includegraphics[height=2.8in]{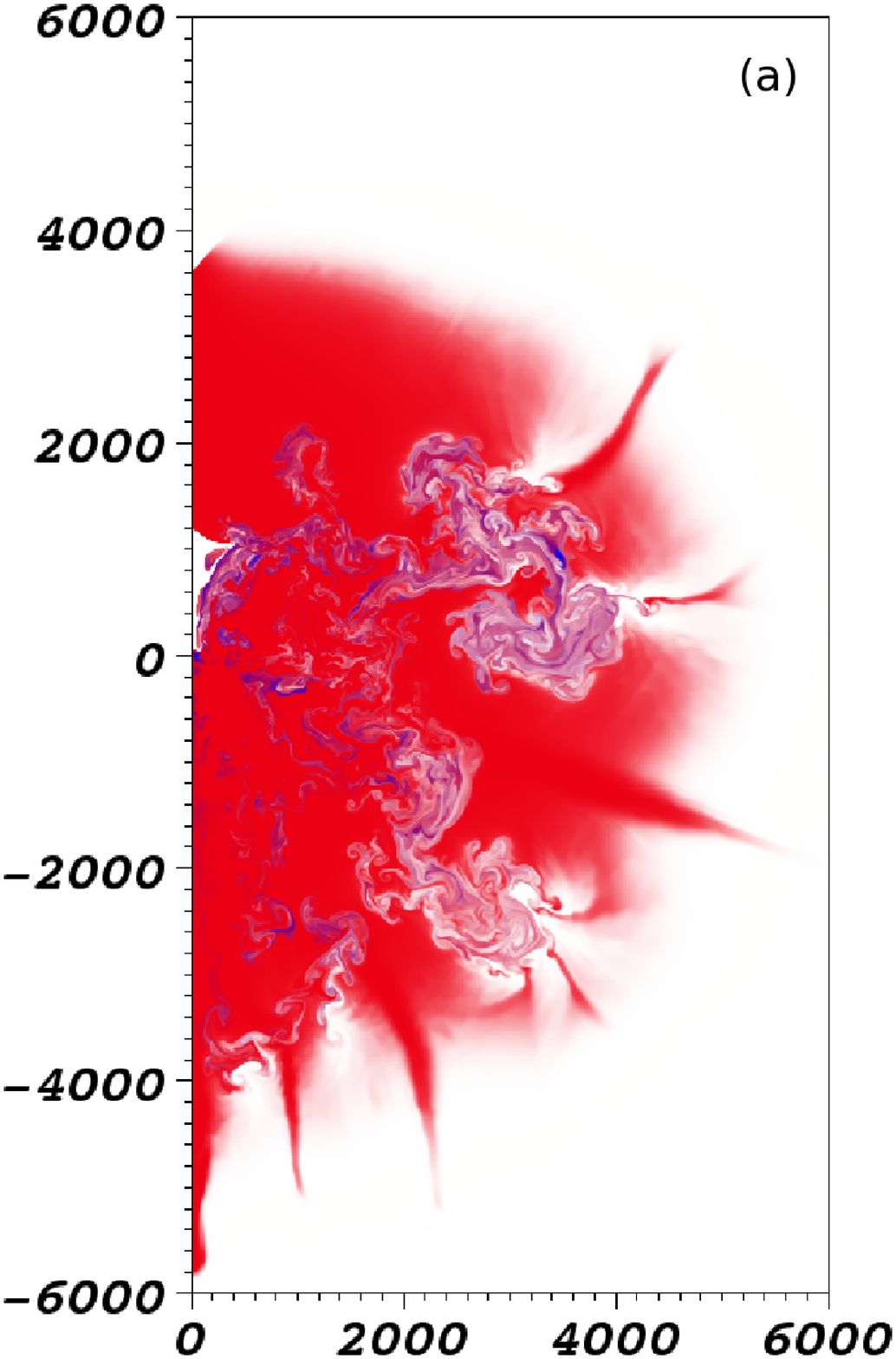}
   } \quad\quad\quad
   \subfloat{\label{fig:NSE_D30}
      \includegraphics[height=2.8in]{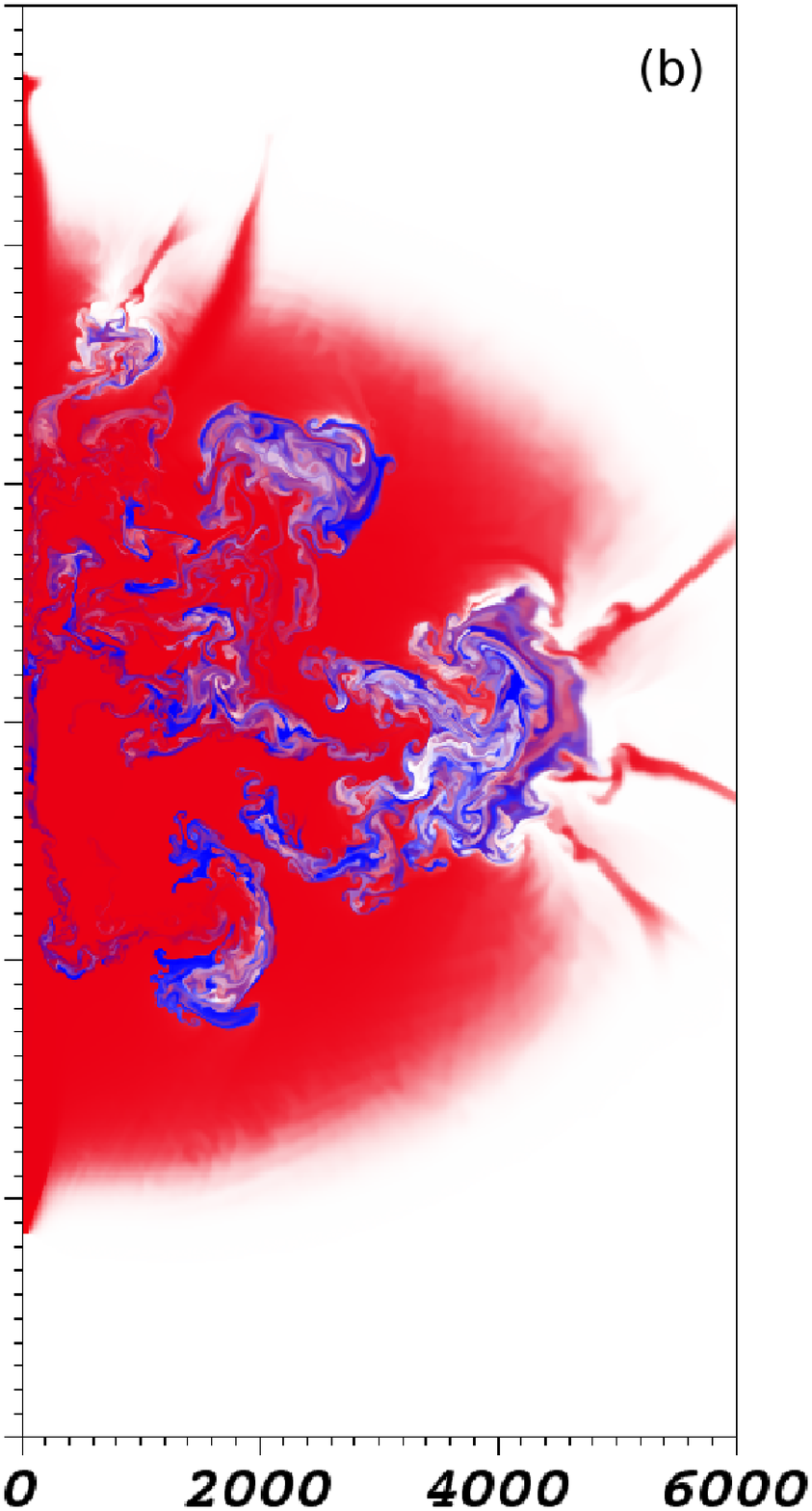}
   } \quad\quad\quad
   \subfloat{\label{fig:NSE_D50}
      \includegraphics[height=2.8in]{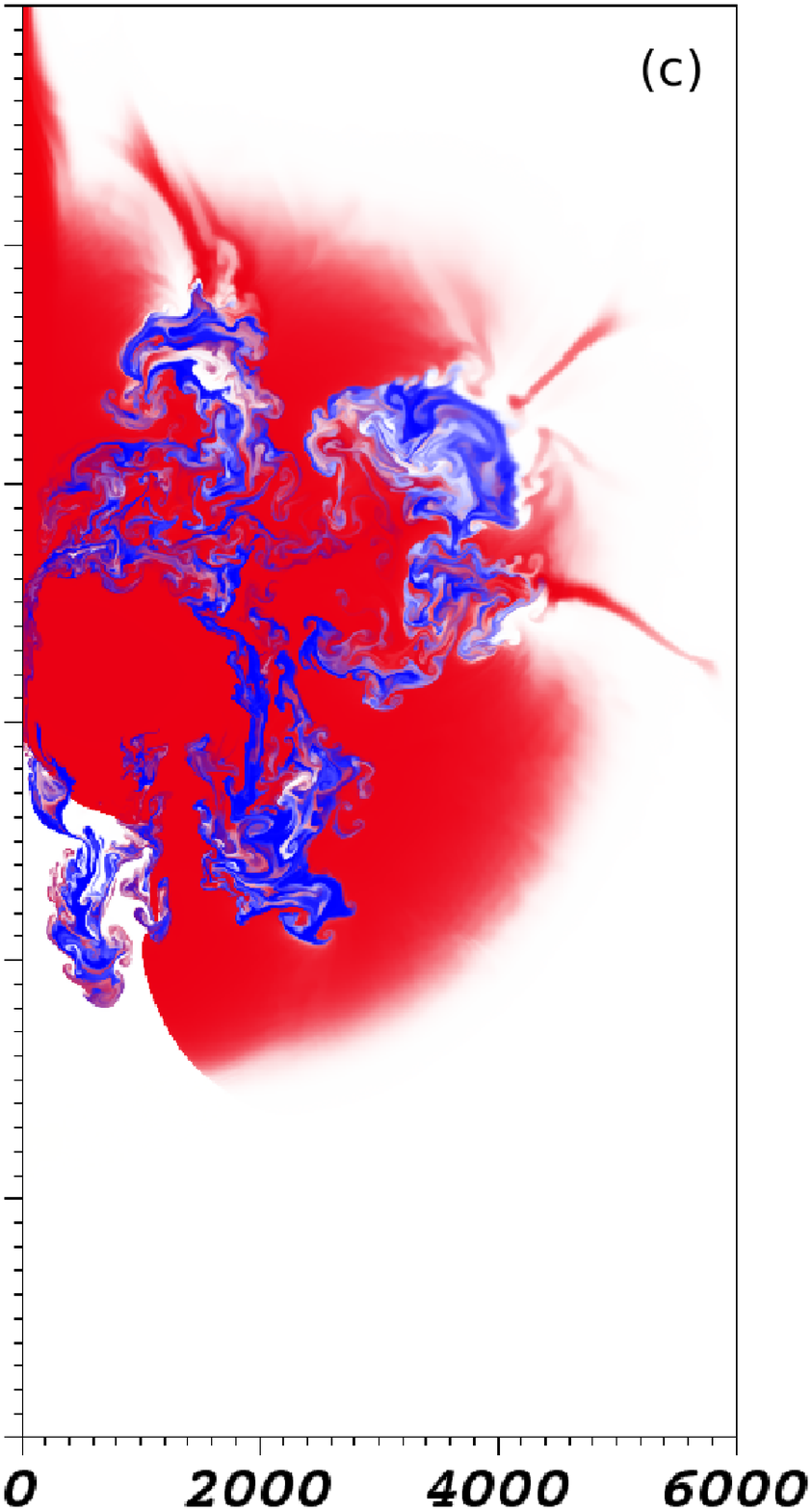}
   }
   \caption{Comparison of \Ni{56} and stable IGE distributions at
      \tIGE\ for simulations with
      the same total \Ni{56} mass.  These images are generated from the same
      simulations used in \figref{fig:Ni069}.  From left to right, the central
      densities are 1.0, 3.0, and 5.0 $\times 10^9$~g~cm$^{-3}$.  Red
      represents the mass fraction of \Ni{56}, blue represents
      stable IGEs, and white represents
      non-IGEs.  Axes are in kilometers.  As \cdens\ increases (with constant
      \Ni{56}), more IGEs are produced and less material is mixed (seen by the
      purple regions).}
   \label{fig:NSEpseudocolor}
\end{figure*}

We also may consider the distribution of the synthesized heavy elements within
the models.  \figref{fig:NSEpseudocolor} shows the distribution of \Ni{56}
vs.\ stable IGEs for 3 of the simulations from \figref{fig:Ni069}.  The
densities of the three simulations ($\cdens = 1 \times 10^9$, $3 \times 10^9$,
and $5 \times 10^9$~g~cm$^{-3}$) span the range of central density.  The trend
for more stable IGEs at higher \cdens\ can clearly be seen by the more
extensive regions colored in blue, indicating higher mass fractions of stable
IGEs.  These 2-d plots also show that as \cdens\ increases, the \Ni{56} and
stable IGEs are more sharply segregated; there are more mixed regions (in
shades of purple) in the lowest-\cdens\ simulation (\figref{fig:NSE_D10}). 

\section{Discussion}
\label{sec:discussion}

The results presented above in~\secref{sec:results} follow directly from 
our simulations and are as rigorous as can be within the limitations of our
models (as described in~\secref{sec:method}). We may extend our results some in order to 
investigate the implications of our trends, particularly the decreasing
proportion of \Ni{56} with increasing central density. 
The principle result from
one of our simulations is the mass of \Ni{56}, and there are well-established 
relationships between the brightness of an observed thermonuclear supernova
and the mass of \Ni{56} synthesized during the explosion~\citep[see][and references
therein]{WoosleyEtAl07:LightCurves}. Thus our trends may have implications for the brightness 
of events.  In this section, we explore these connections by comparing our
trends in \Ni{56} mass to \Ni{56} masses inferred from observations.

In \citet{KruegerEtAl10} (see especially 
Figure~4 of \citet{KruegerEtAl10}, an updated version of which is shown in
\figref{fig:neill} of this paper), we showed that our results agree with the
general trend of dimmer SNeIa from older stellar populations by comparing with
observations from \citet{NeillEtAl09}.  We improve upon that 
result with two additions: recalibration of the \Ni{56} mass produced by our 
models and clarifying the definition of ``age'' by including a range of main 
sequence lifetimes.  Additionally, this section discusses how our results 
compare to other
studies considering the effect of \cdens\ on \Ni{56}.

\subsection{Recalibration of \Ni{56} Yield}
\label{sec:YieldFitting}

As discussed above, on
average our suite of simulations exhibited an overproduction of IGEs and
\Ni{56} when compared to masses estimated from observations of 
remnants~\citep{WoosleyEtAl07:LightCurves}. We estimate a correction to our 
\Ni{56} masses, the details of which are given in \appref{app:recal}
but which we summarize here, that allows us to more
directly compare our results to \Ni{56} masses inferred from observations. 

The recalibration is based on relationships between model parameters 
and explosion yields found in previous studies. \citet{jacketal10} 
found that higher values of \rhoDDT\ led to increased yields due to
less expansion of the star during the deflagration phase. 
\citet{townetal09} found a correspondence between the mass at
densities above $2 \times 10^{7}$~g~cm$^{-3}$ at \tDDT\ and
the mass of IGEs synthesized in the yield (Figure 5). Although
the correspondence found by \citet{townetal09} was not perfect,
these two relationships form the basis for our re-scaling. We note
that the rescaling assumes that even for lower values of \rhoDDT, 
the mass of IGEs is independent of \cdens; this assumption is 
discussed in greater detail in Appendix B based on the expansion
characteristics present in our simulations.  However, this is a critical
assumption in our analysis and is based on a somewhat restricted parameter
space and therefore is subject to future investigation to confirm it.
Recent work by \citet{seitenzahletal11}, who performed full three-dimensional
simulations covering the range of observed \Ni{56} masses, suggests
that this assumption does not hold. See section~\secref{sec:theory}.

\begin{figure}
   \includegraphics[angle=270,width=\columnwidth]{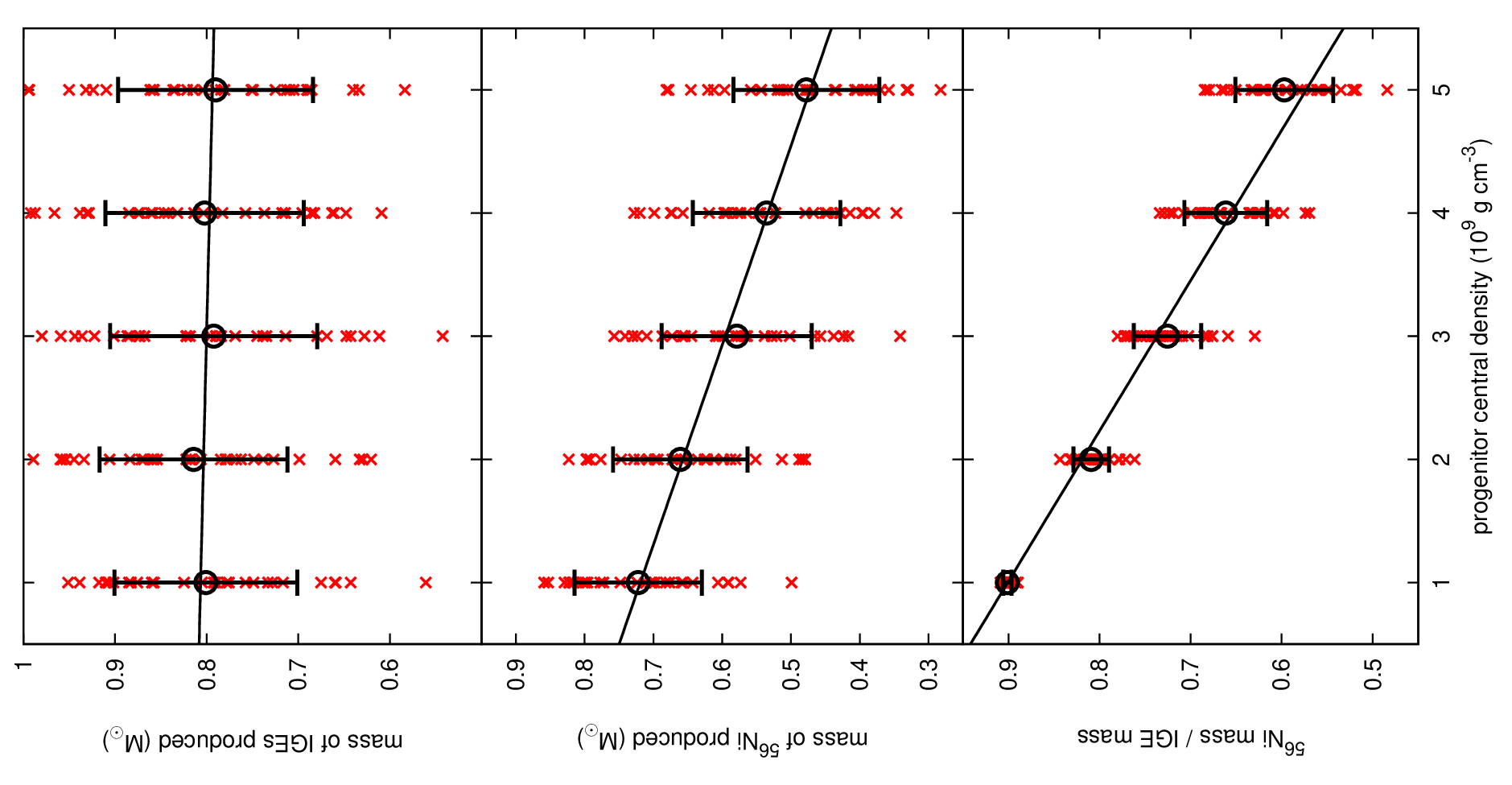}
   \caption{Final yields of our simulations after recalibration using
      Equations~(\ref{eqn:yieldVtDDT}) and (\ref{eqn:slopeVrho}), the parameter
      values given in \tabref{tab:Chi4Params}, and the time shifts given in
      \tabref{tab:TimeShift}.  The black lines are the best-fit trend lines,
      with the averages and standard deviations marked by the circles and the
      vertical error bars.  These plots are directly analogous to those shown
      in \figref{fig:yield_by_rho}.}
   \label{fig:yield_by_rho_extrapolated}
\end{figure}

Performing this recalibration lowers the \Ni{56} masses (see
\figref{fig:yield_by_rho_extrapolated}), giving yields 
that are consistent with yields inferred from observations.  
A similar trend of decreasing \Ni{56} mass with increasing central
density is also found in these extrapolated results.
Future models will extend to three dimensions and may include 
more physically-motivated initial conditions or
more detailed burning models that capture interactions between the turbulent
velocity field and the flame structure.  Such changes will require that we
re-evaluate our choice of model parameters, such as \rhoDDT.

\subsection{Comparison to Observations}
\label{sec:observe}

\begin{figure}[t]
   \includegraphics[angle=270,width=\columnwidth]{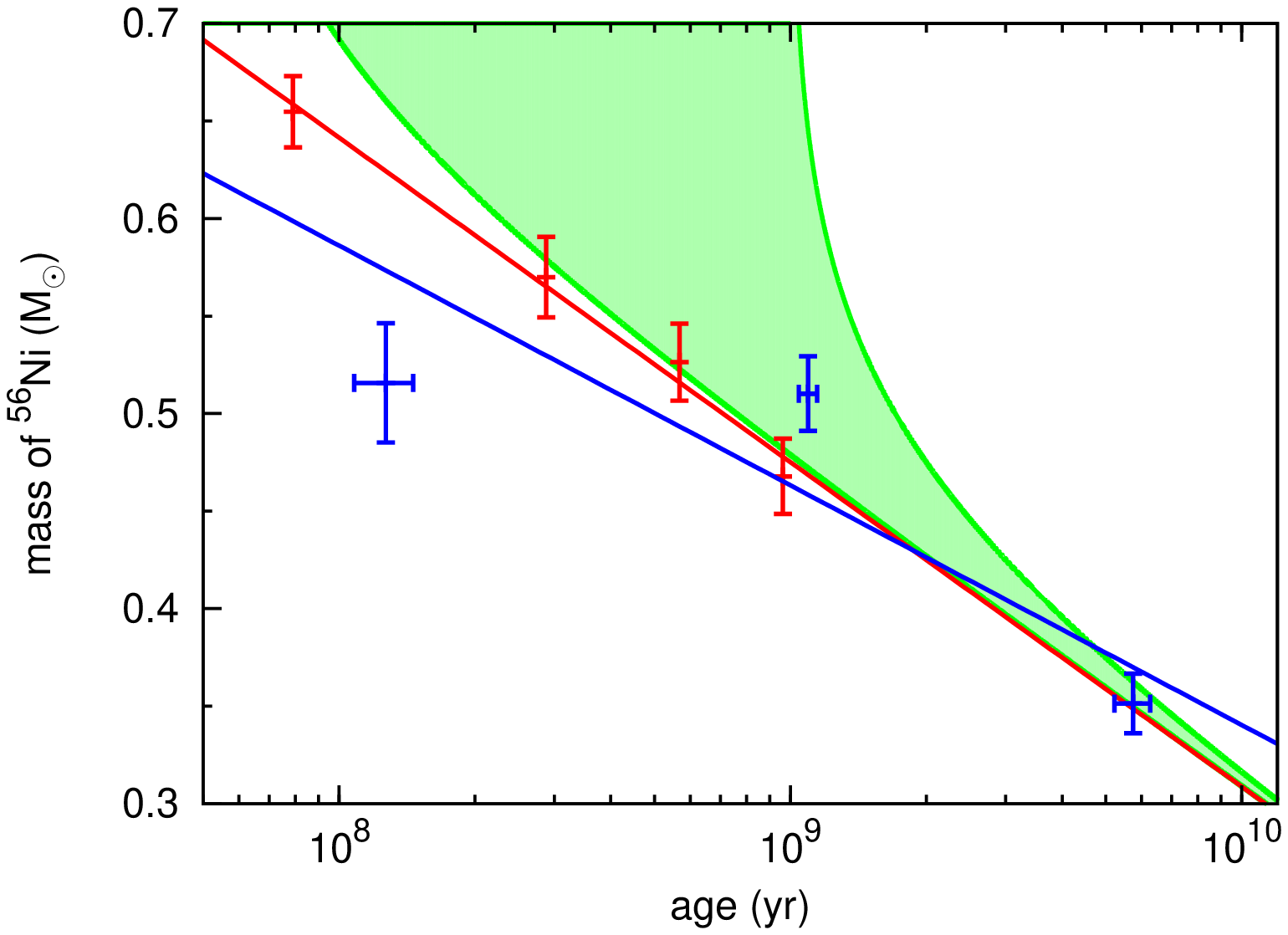}
   \caption{Plot of $M_{\Ni{56}}$ vs.\ age comparing the scaled results 
      of this study to the \Ni{56} masses inferred from the 
      observations of \citet{NeillEtAl09}.  In red are the points from this
      study with no shift (i.e., $\tau_{\rm MS} = 0$~Gyr), along with the
      standard error of the mean and a best-fit trend line following the form
      of \eqnref{eqn:mass_age}.  The green shaded region shows our best-fit
      line with $\tau_{\rm MS} = 0.05 - 1.0$~Gyr.  In blue are the binned and
      averaged points from Figure~5 of \citet{NeillEtAl09}, along with their
      best-fit trend line.}
   \label{fig:neill}
\end{figure}

One of the principal metrics that we extract from each simulation is the mass
of \Ni{56} produced, which is directly related to the brightness of an event.
Using the method outlined in \citet{howelletal+09}, we converted stretch
reported in observational results to \Ni{56} masses to compare to the \Ni{56} masses
(derived in \secref{sec:YieldFitting}) obtained from our simulations.
We then combined our
central density values with the results of \citet{LesaffreEtAl06}, which
correlate the central density at the time of the ignition of the flame front to
the cooling time of the progenitor WD, allowing us to express our results as
ages.  The results of \citet{LesaffreEtAl06} suggest that a WD with a central
density of $1\times10^9$~g~cm$^{-3}$ will not ignite without further accretion,
so for this comparison we neglect our simulations with that value of \cdens.
Our ``age'' was previously defined as the delay time (\tcool), while the
observational results used the delay time, which includes the main-sequence
lifetime ($\tau_{\rm MS}$).  We can improve our comparison by applying a shift
to our data to account for $\tau_{\rm MS}$.  We assume that our progenitors
differ solely in their cooling times in order to separate out other effects;
therefore we take $\tau_{\rm MS}$ to be constant across our results.  Our
best-fit line would now become
\begin{equation}
   M_{\Ni{56}} = \alpha \log_{10} \left( \tcool + \tau_{\rm MS} \right) + \beta,
   \label{eqn:mass_age}
\end{equation}
where $\alpha$ and $\beta$ are fitting parameters.
Because the addition of $\tau_{\rm MS}$ is inside the logarithm, this
corresponds not only to a shift but also an increasing slope.  We have selected
two estimated limiting values for $\tau_{\rm MS}$ (0.05 and 1.0~Gyr,
corresponding to main-sequence masses of approximately 8.0 and 1.5~\Msol
respectively; see \citealt{HansenEtAl04}) and included them with the original
$\tau_{\rm MS} = 0$ result in \figref{fig:neill}.  Adding in a $\tau_{\rm MS}$
consistent with our C/O progenitors brings our results into better agreement
with the two right-most points of \citet{NeillEtAl09}.

As seen in \figref{fig:neill}, our theoretical results are not in complete
agreement with the observed data.  Observationally, the age-brightness
correlation may flatten at young ages, while our data do not, resulting in
our data being overluminous relative to young SNeIa.  This study
isolated the effects
of central density and related that to age assuming a constant main-sequence
mass, but there are other effects that may be correlated.  Examples of such
potentially correlated effects are: main-sequence mass and it's correlation
with central density, metallicity of the progenitor, core $^{12}$C fraction
prior to ignition of the deflagration, sedimentation, and others.  Inclusion of
such effects may modify the results presented here and are the subjects of
future work.

\subsection{Comparison to Other Theoretical Efforts}
\label{sec:theory}

Recent theoretical work by other research groups also addresses the role of
central density in the single-degenerate picture of SNeIa.
\citet{fisheretal10} note that in the gravitationally confined detonation model
(GCD; \citealp{PlewCaldLamb04}), a higher central density leads to increased
energy release during the deflagration phase, which leads to increased
expansion of the star and an increase in the production of intermediate-mass
elements and the corresponding decrease in IGEs.  These results are consistent
with our findings concerning the choice of \rhoDDT\ influencing the production
of IGEs discussed above in \secref{sec:ddt}.  These results are also consistent
with some of our realizations, but \citet{fisheretal10} do not consider the
role of neutronization, preventing comparison to our principal result that the
relative proportion of \Ni{56} decreases with higher central density, producing
a dimmer event.

\citet{seitenzahletal11} performed a study of the DDT scenario with
three-dimensional simulations and a description of the flame energetics that
accounts for neutronization.  \citet{seitenzahletal11} similarly find that the
\Ni{56} fraction of IGEs decreases in WD models with higher central densities,
but they also find that the mass of IGEs increases at higher densities.  The
net effect is a roughly constant mass of \Ni{56} synthesized during the
explosion, implying that central density influences the brightness of an event
only as a secondary parameter.  \citet{seitenzahletal11} comment on differences
between their results and our earlier results \citep{KruegerEtAl10} and,
likewise, we offer discussion here.

First, we note that our results are more similar than they might at first
appear.  Our trend of decreasing \Ni{56} follows from the increased rate of
neutronization at high densities of a roughly constant mass of IGEs synthesized
during the explosion.  In a sense, our trend is consistent with that of
\citet{seitenzahletal11}.  For a given mass of IGEs, the fraction that is
\Ni{56} is lower in WD models with a high central density.  The difference
between our results and \citet{seitenzahletal11} follows from the the increase
in the mass of IGEs synthesized in their models.  If our explosions produced
more IGEs for higher central density progenitors, then we may well see a
roughly constant mass of \Ni{56}.  This observation should be readily apparent
by considering \figref{fig:yield_by_rho}.  If the flat slope of the IGE yield
were instead rising enough, the decreasing slope of the \Ni{56} yield would be
instead rising.  Accordingly, the key to the differences in our results is
understanding the reason for the difference in the production of IGEs. 

While there are many differences between the methodology of
\citet{seitenzahletal11} and ours, including (as they mention) differences in
the flame model (level set vs.\ ADR), differences in the energy release scheme,
and structure of the computational mesh,  perhaps the most substantive
difference is the dimensionality of the simulations.  Three-dimensional
simulations meaningfully describe turbulent flow, which enables use of
turbulence-flame interaction (TFI) models.  Turbulence in two-dimensional
simulations, however, has very different properties; particularly, it has an
inverse-cascade of energy from small to large scales~\citep[see Chapter 10
of][and references therein]{davidson2004}.  Because of the large scope of this
study, only two-dimensional simulations were possible.  As described above, our
models use a flame speed that compensates for buoyancy effects to prevent the
flame from being torn apart by Rayleigh-Taylor-induced turbulence.  Because of
the issue of turbulence in two-dimensional simulations, we chose not to include
models for unresolved turbulence and the turbulence-flame interaction in our
models for this study.

In their three-dimensional simulations, \citet{seitenzahletal11} similarly
account for buoyancy effects, but also implement a method for accounting for
turbulent energy on unresolved scales and the corresponding enhancement to the
flame speed.  Originally proposed by \citet{NiemHill95} and developed in detail
by \citet{Schmetal06a, Schmetal06b}, the method consists of a dynamic measure
of the local turbulent energy on sub-grid scales and sets the flame speed to $s
= \sqrt{s_\ell^2 + C_tq^2}$, where $q$ is a velocity that characterizes the
sub-grid turbulence energy content and $C_t$ is a constant taken to be 4/3.
This addition should boost the burning rate during the deflagration phase,
which will change the density profile of the star when the detonation occurs
and thus change the final yield.  
It may be our omission of turbulence and
flame-turbulence interaction models that accounts for the systematic difference
we see in the yield of IGEs compared to \citet{seitenzahletal11}.
While the difference in treatment of the turbulent flame is a very important 
one, it is difficult to decouple from the dimensionality of the simulations 
performed. The dimensionality appears to influence the expansion rate, which 
may or may not be related to the turbulent flame model. Future three-dimensional
simulations with different turbulent flame models will be able to address these 
uncertainties.

\citet{seitenzahletal11} also mention the choice of \rhoDDT\ as a possible
source of the difference between their study and our earlier results.  As they
describe, \citet{seitenzahletal11} use a dynamic measure that calculates the
probability of a DDT based on the turbulent intensity, density, and fraction of
fuel.  As they note, this prescription is significantly different from just
fixing \rhoDDT\ and results in fewer DDTs.  Considering Table~1 of
\citet{seitenzahletal11}, we can see that the density at which the DDT occurred
appears to be a function of central density due to their DDT criteria: other
than their three models with only five DDT points, their effective DDT density
increases with central density.  A higher DDT density implies an earlier
detonation, which means that the star has expanded less prior to the
detonation, allowing the detonation to burn a larger fraction of the star.  So
it may be possible that the trend of increasing IGEs seen in
\citet{seitenzahletal11} comes from the change in their effective DDT density,
instead of directly from the changing \cdens.  Regardless of whether this
difference in methodology fully explains the difference in results, we can
state that the differences in methodologies make direct comparison between the
two results difficult.

Finally, the relatively small number of simulations performed by
\citet{seitenzahletal11} may not allow them to determine statistically
meaningful trends.  Unfortunately, the expense of three-dimensional simulations
make extensive studies difficult.  The twelve simulations they performed are a
remarkable achievement, but our results suggest that it is necessary to have a
distribution of ignition conditions that reproduces the distribution of
observed yields.  Our statistical ensemble does this in two dimensions, but
demonstrating such a distribution in three dimensions would require significant
computational resources.

As \citet{seitenzahletal11} mention, \citet{Meaketal09} explored the effect of
varying the offset from the center of the initially burned region (what we call
the ``match head'') in the GCD scenario.  Their Figure~12 plots mass of NSE
elements (which we have been referring to as IGEs), mass at $\rho >
10^7$~g~cm$^{-3}$, and mass of \Ni{56} as functions of central density {\em at
the time of the detonation;} all of their models have the same central density
at the ignition of the deflagration.  We note that the GCD scenario is
significantly different from the DDT scenario that we and
\citet{seitenzahletal11} investigated, with the principal difference arising in
the expansion of the star during the deflagration phase of the evolution.  In
the GCD scenario, the star expands during the deflagration as the bubbles rise
because the displaced mass softens the gravitational potential.  The effect is
that GCD models typically experience considerably less expansion than DDT
models.  Also, unlike central ignitions, there is very little material burned
until the detonation occurs.  A direct comparison, therefore, between the
results of \citet{Meaketal09}, with yields presented as functions of the
central density at the ignition of the detonation, and our work, considering
yields as a function of central density at the initial ignition of the
deflagration, is at best uncertain, especially given that our results show that
the effect of neutronization is greatest during the deflagration phase as
illustrated in~\figref{fig:defdet}.

\citet{Roepetal06_2} performed a set of simulations of SNeIa assuming the pure
deflagration model.  They found that as the central density of the progenitor
is increased, the mass of \Ni{56} produced also increases.  Their discussion of
why they produce more iron-group material also applies to our results.  The
yields during the deflagration and detonation phases were presented in
\figref{fig:defdet}.  We find the yield of IGEs increases with central density
during the deflagration phase.  We do not, however, find the same increase in
\Ni{56} mass due to increased neutronization.  On this point our results
disagree with the pure deflagration model as studied by \citet{Roepetal06_2}.
\citet{Bravo_etal_90}, however, studied the effects of neutronization and found
that at higher central densities the yield of \Ni{56} decreases, in agreement
with our results.

Similarly to the \citet{seitenzahletal11} work, \citet{Roepetal06_2} and
\citet{Bravo_etal_90} only performed a modest number of simulations.  In light
of our findings in \secref{sec:Statistics}, we note that it may be  difficult
to draw statistically significant results from a single realization.
\citet{Roepetal06_2} and \citet{Bravo_etal_90} both appear to use only a single
morphology for each parameter set, which our results suggest could result in
deriving an incorrect trend.

\section{Summary and Conclusions}
\label{sec:conclusions}

This paper builds on the results presented in \citet{KruegerEtAl10}, giving
more detail of the study and extending the analysis.  In that paper we showed
that, in our 2-d simulations, a higher central density of the progenitor star
does not impact the production of IGEs, but leads to greater neutronization,
resulting in the production of less \Ni{56}.  We also discussed the relation
between the age of the progenitor and the central density (see, e.g.,
\citealt{LesaffreEtAl06}), and the relation between the brightness and the mass
of \Ni{56} produced.  Thus the statement that a higher density leads to less
\Ni{56} is equivalent to the statement that an older progenitor will produce a
dimmer SNIa.  In this work we expand on the discussion of
\citet{KruegerEtAl10} to give more detail of our models and to improve upon the
age-brightness relation predicted by our simulations.
In particular, we show that by adding a main-sequence lifetime to the
cooling time our brightness-age relation is steeper, more closely matching the
observed behavior of older SNeIa.

In comparing with other theoretical work, we see that the variation of \Ni{56}
mass with progenitor central density is not a settled question.  In this paper
we further developed the idea that, due to the strong nonlinearities of the
processes in SNeIa, a statistical study of an ensemble of SNIa
simulations may be
necessary to determine the true trends.  For our simulations, we find that 15
realizations (morphologies of the initial flame surface) are sufficient to
characterize the mean trends from our models.

We find that the inner region (out to an enclosed mass of 0.8~--~1.0~$M_\odot$)
of the remnant is dominated by \Ni{56}.  However, the stable
(non-\Ni{56}) IGEs tend to be in ``clumps'', instead of well-mixed throughout
this region.  This may give rise to \Ni{56} holes with little or no \Ni{56},
depending on the line of sight through a SNIa remnant.  The outer region of the
remnant will have more intermediate- and low-mass elements, as the burning
becomes less efficient for lower densities.  As the central density increases,
the mean \Ni{56} mass fraction in the inner region drops (roughly 0.8 for
$\cdens = 1 \times 10^9$~g~cm$^{-3}$ to roughly 0.6 for $\cdens = 5 \times
10^9$~g~cm$^{-3}$).  However, the extent of this region (in enclosed-mass
space) does not significantly change.  Variations in the central
density affect
the sharpness of the edges of the stable IGE clumps: a higher
central density
leads to clumps of stable elements that are more sharply defined,
as well as less mixing between
the \Ni{56} and the stable IGEs.

To better connect to observations, we discussed how to distinguish the relative
ages of SNeIa with the same brightness (in other words, the relative initial
central densities of SNeIa that produce the same mass of \Ni{56}).  We found
that, in our models, the best measure of the central density is
the mass of
stable IGEs, where higher central density progenitors produce more stable IGEs
due to their greater rate of neutronization during the subsonic deflagration
phase.

We found that a higher central density leads to a shorter deflagration phase.
Since the rate of neutronization is significantly boosted, the total
neutronization is greater at a higher central density despite there being less
time to neutronize.  The time between the ignition of the first detonation and
the cessation of burning is independent of central density.

As noted in \citet{KruegerEtAl10} and described above, our choice for the DDT
transition density led to an overproduction of \Ni{56}.  In
\secref{sec:YieldFitting} we provided a recalibration of this overall
brightness normalization to extrapolate our results to an expected average
brightness.  Future work will be improved by a better choice of DDT density and
we will report any quantitative changes to the trends reported here.  Due to
the fundamentally 3-d nature of some of the phenomena in a SNIa (such as the
turbulent velocities), we plan to extend this work by performing 3-d
simulations.  Because 3-d simulations are much more computationally expensive
than the corresponding 2-d simulations, a study with 3-d simulations will by
necessity be constrained to a smaller number of simulations.  The choices will
be motivated by this work and seek to span the parameter space explored here.

\acknowledgements

This work was supported by the Department of Energy through grants
DE-FG02-07ER41516, DE-FG02-08ER41570, and DE-FG02-08ER41565, and by NASA
through grant NNX09AD19G.  ACC also acknowledges support from the Department of
Energy under grant DE-FG02-87ER40317.  APJ is currently supported by a National
Research Council Research Associateship.  The authors gratefully acknowledge
the generous assistance of Pierre Lesaffre, fruitful discussions with Mike
Zingale, and the use of weak reaction tables developed by Ivo Seitenzahl.  The
authors also acknowledge the hospitality of the KITP, which is supported by NSF
grant PHY05-51164, during the programs ``Accretion and Explosion: the
Astrophysics of Degenerate Stars'' and ``Stellar Death and Supernovae.''  The
software used in this work was in part developed by the DOE-supported
ASC/Alliances Center for Astrophysical Thermonuclear Flashes at the University
of Chicago.  We thank Nathan Hearn for making his QuickFlash analysis tools
publicly available at http://quickflash.sourceforge.net.  This research
utilized resources at the New York Center for Computational Sciences at Stony
Brook University/Brookhaven National Laboratory which is supported by the
U.S.\ Department of Energy under Contract No.\ DE-AC02-98CH10886 and by the
State of New York.


\begin{appendix}

\section{Randomized Initial Conditions}
\label{app:init}

The initial flame surfaces for each realization are determined by spherical
harmonics.  The formula used is:
\begin{equation}
r = 150{\rm km} + \sum_\ell A_\ell Y_\ell^m,
\end{equation}
where $Y_\ell^m$ are the spherical harmonics and the coefficients $A_\ell$ are
given in \tabref{tab:FlameSpectrum}.  A reference implementation is available
online at \url{http://astronomy.ua.edu/townsley/code/}.  This suite of
simulations used an initial seed value of 1866936915.

\begin{table}[h!]
\centering
\caption{Amplitudes of spherical harmonic perturbations by realization.}
\begin{tabular}{c r r r r r}
\hline \hline
\multirow{2}{*}{real.\ \#}
   & \multicolumn{5}{c}{amplitude of perturbation (km)} \\
   & $A_{12}$ & $A_{13}$ & $A_{14}$ & $A_{15}$ & $A_{16}$ \\
\hline
 1 & -50.88 & -51.22 &  22.23 &  9.979 & -56.23 \\
 2 & -31.03 &  51.57 & -43.21 & -55.76 & -30.22 \\
 3 & -39.89 & -28.06 &  31.48 &  6.848 & -29.80 \\
 4 & -48.89 & -44.21 &  1.509 &  52.44 &  29.00 \\
 5 &  29.67 & -74.12 & -53.44 &  13.83 & -51.24 \\
 6 & -26.95 &  5.169 & -45.79 & -46.34 &  14.93 \\
 7 &  12.67 & -19.76 & -90.22 & -2.312 & -20.83 \\
 8 & -10.23 &  1.206 & -43.74 & -25.57 &  10.26 \\
 9 &  30.25 &  69.91 & -29.37 & -74.17 & -29.50 \\
10 & -15.76 & -4.818 &  14.53 &  1.979 & -15.23 \\
11 & -55.91 &  47.08 & -15.86 & -34.44 & -43.00 \\
12 &  1.021 & -0.685 & -33.13 & -48.32 & -22.22 \\
13 & -10.41 & -32.27 &  5.236 &  27.06 & -12.46 \\
14 & -73.60 & -38.16 & -22.76 &  41.81 &  25.76 \\
15 & -74.20 & -3.179 & -30.20 & -36.56 & -21.16 \\
16 & -86.84 &  16.28 &  19.75 &  14.43 &  4.924 \\
17 &  12.87 &  44.39 & -12.17 & -34.34 & -67.18 \\
18 & -13.88 & -62.62 & -35.93 & -41.57 & -81.32 \\
19 & -31.09 & -49.91 &  2.953 &  57.76 & -5.397 \\
20 &  51.43 & -44.30 & -41.00 &  26.83 & -42.83 \\
21 & -28.15 &  11.27 & -37.43 &  85.20 & -46.84 \\
22 &  53.67 &  4.730 & -52.71 & -46.66 & -65.99 \\
23 & -21.66 & -12.31 &  53.21 &  14.14 & -34.18 \\
24 & -16.25 & -68.30 & -43.63 &  111.6 & -33.86 \\
25 & -56.50 & -31.25 &  44.51 &  0.098 & -40.66 \\
26 & -98.62 & -31.80 & -19.40 &  45.75 & -3.815 \\
27 & -39.29 &  2.525 & -32.95 &  27.10 & -14.95 \\
28 &  29.24 &  79.73 & -30.80 & -19.18 & -96.98 \\
29 & -14.08 &  46.98 & -89.82 & -11.95 &  28.21 \\
30 & -24.91 &  33.12 & -46.75 & -10.40 & -3.940 \\
\hline
\end{tabular}
\label{tab:FlameSpectrum}
\end{table}

\section{Recalibration of \Ni{56} Yields}
\label{app:recal}

As discussed above near the beginning of \secref{sec:results}, on average our
suite of simulations exhibited an overproduction of IGEs and \Ni{56}.  The
yields of IGEs and \Ni{56} are related to the expansion of the WD prior to the
ignition of the detonation: as the star expands during the deflagration phase,
the density decreases and, therefore, less fuel (C and O) remains at
sufficiently high density to burn to IGEs during the detonation.  Thus, the
amount of expansion of the WD prior to the DDT principally determines the yield
of IGEs and \Ni{56}~\citep{Nomo84, Khokhlov1991Delayed-detonat}.  The amount of
expansion is determined by DDT conditions~\citep{jacketal10} along with the
energy deposition history, which depends on initial conditions, dimensionality,
and the burning model.

Given our models and the simulations we have performed, the most
straightforward way to systematically correct the overproduction of IGEs is to
modify the DDT conditions and repeat the simulations.  The value of \rhoDDT\ 
should be determined by the physics of the DDT, but, as the physics of DDTs is
incompletely understood and most likely could not be resolved in these
simulations, we treat \rhoDDT\ as a parameter. The DDT transition is
implemented by burning small regions ahead of rising plumes, when the tip of
the plume reaches \rhoDDT. As demonstrated in \citet{jacketal10}, lowering the
value of \rhoDDT\ is tantamount to increasing \tDDT\ because the material that
undergoes a DDT is at a higher radius and the time required for a rising bubble
of burning material to reach that radius is longer.  This longer \tDDT\ results
in more expansion of the WD prior to detonation and, because the amount of
material at higher density decreases, the detonation produces a lower yield of
IGEs.  In order to address the impact of \rhoDDT\ on our results, the ideal
solution would be to repeat our simulations with \rhoDDT\ set to a lower value
and therefore obtain a longer \tDDT, more expansion, and more realistic yields.
Using the data available from this study and the study of \citet{jacketal10},
we can estimate the results of performing a new suite of simulations using a
different value of \rhoDDT\ and use these findings to inform future studies.
We consider two issues here.  First we attempt to extrapolate to lower
$\rho_{\rm DDT}$ based on the expansion characteristics of our models to
confirm that for a \emph{single value} of $\rho_{\rm DDT}$, the $M_{\rm
IGE}$ appears to continue to be independent of $\rho_{c,0}$.  Second, we
use this justified assumption to extrapolate our yields based on a delay of
the DDT transition.

\begin{figure*}
   \centering
   \subfloat{\label{fig:proxy_ddt}
      \includegraphics[angle=270,width=0.45\columnwidth]{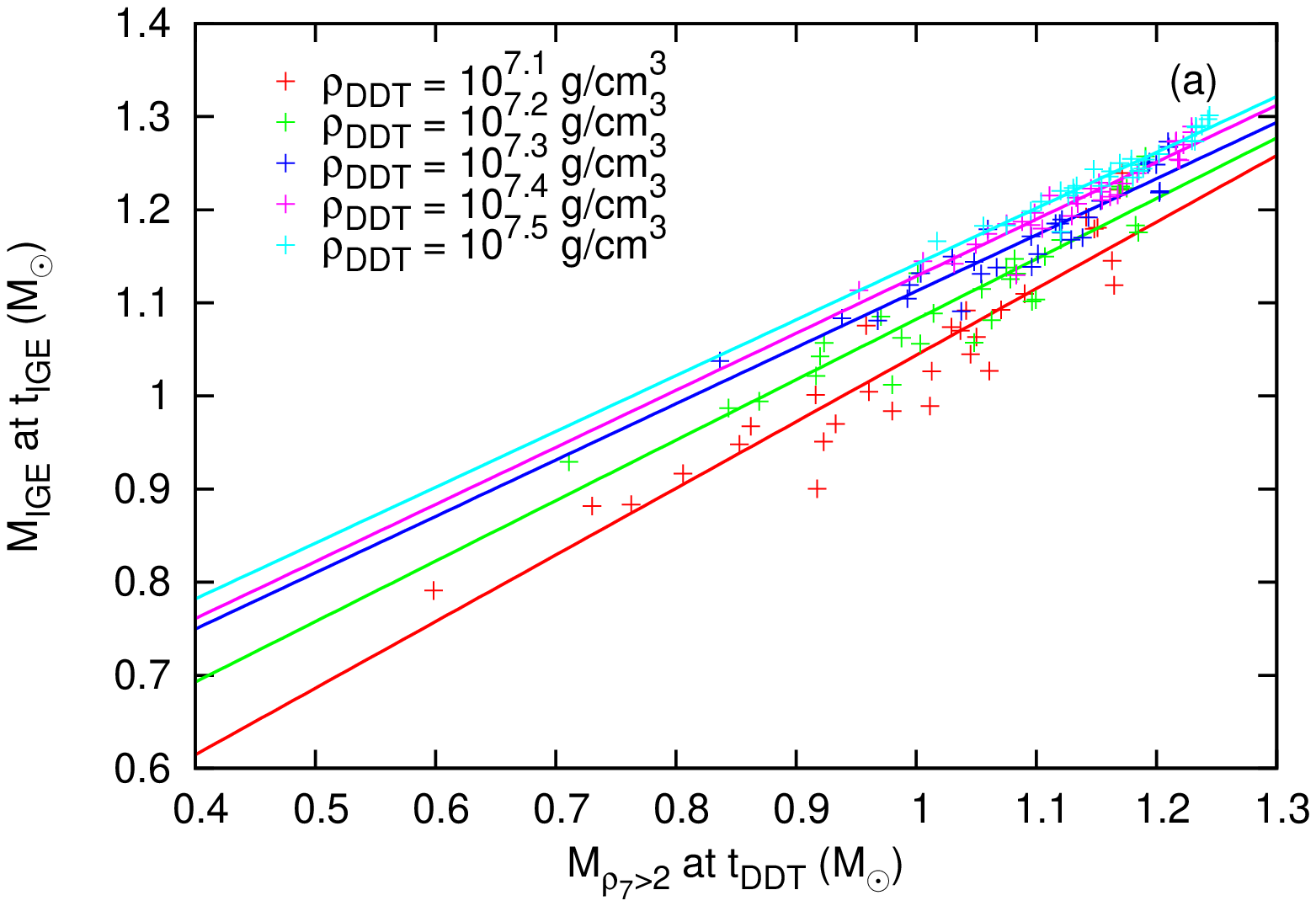}
   }
   \subfloat{\label{fig:proxy_c0}
      \includegraphics[angle=270,width=0.45\columnwidth]{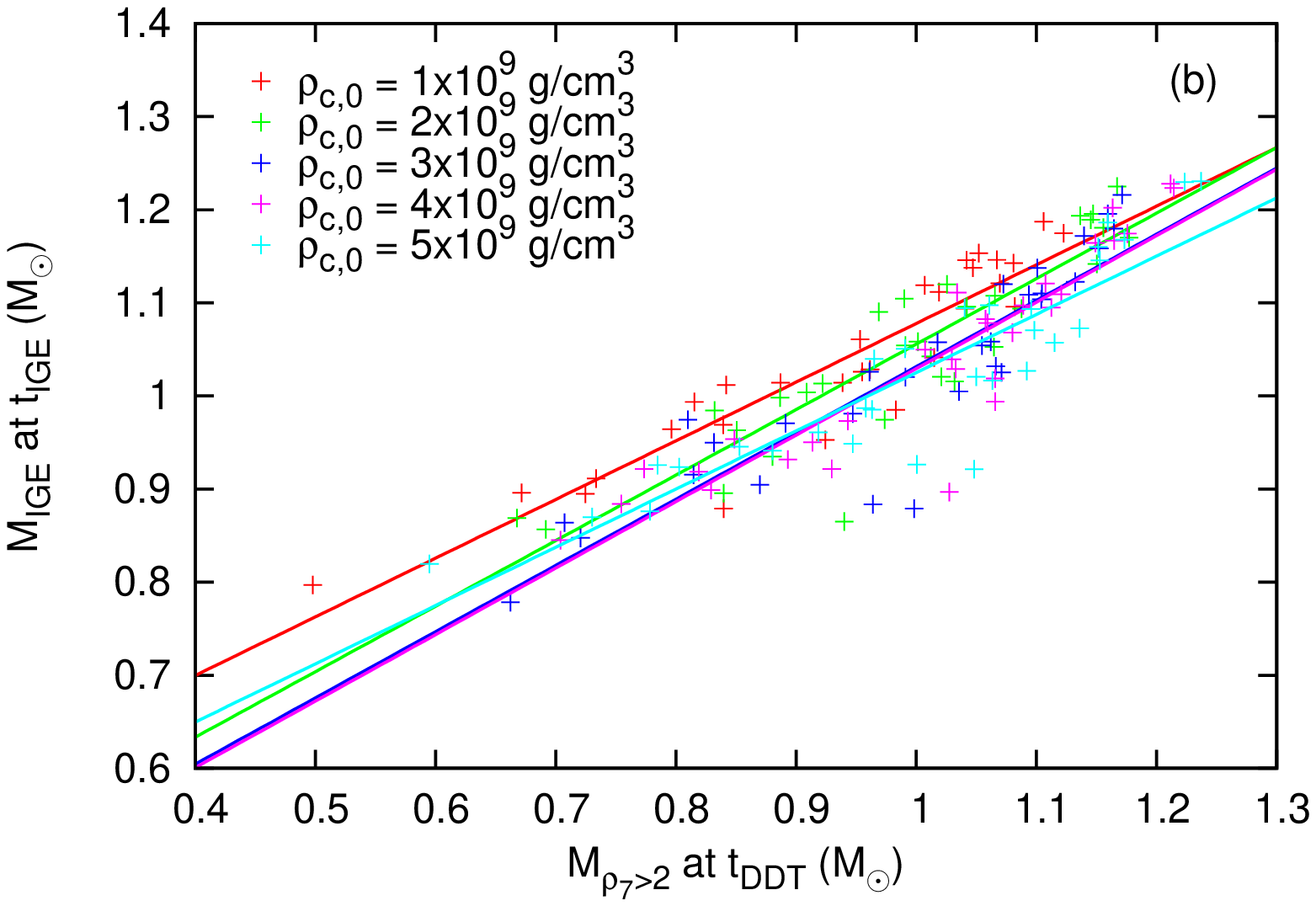}
   }
   \caption{Relationship between \proxy\ and \MIGE, demonstrating
      the dependence of this relationship on \cdens\ and \rhoDDT.  The data in
      the left panel comes from the study of \citet{jacketal10}, which varied
      \rhoDDT\ with a constant \cdens; data points and trend lines are colored
      by \rhoDDT.  The data in the right panel comes from this study; data
      points and trend lines are colored by \cdens.}
   \label{fig:proxy_variation}
\end{figure*}
\begin{figure*}
   \centering
   \includegraphics[angle=270,width=0.45\columnwidth]{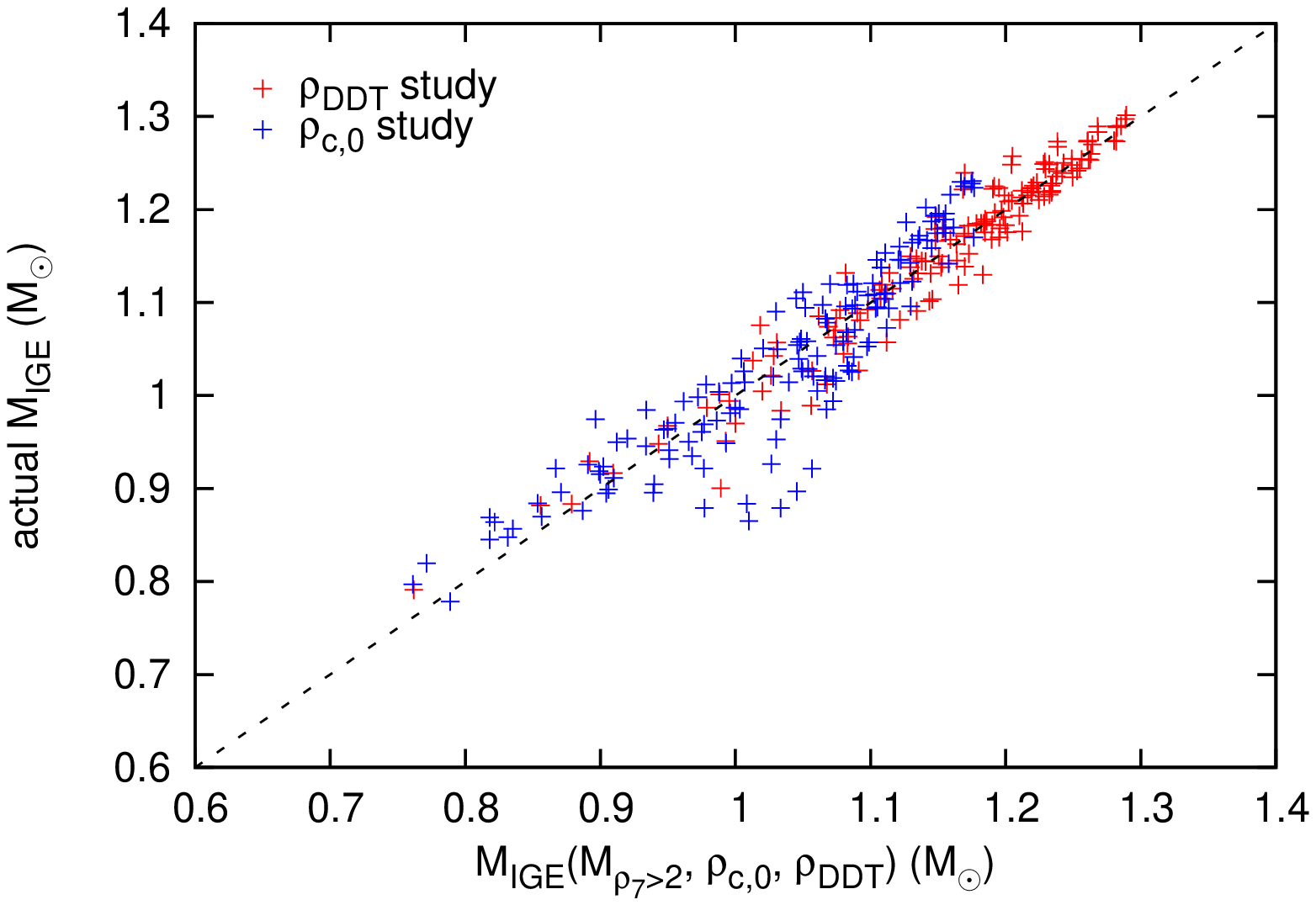}
   \caption{Demonstration of the results of \eqnref{eqn:proxy},
      showing the comparison of the actual \MIGE\ data from simulations to the
      values of \MIGE\ estimated from this relation (assuming $\rhoDDT =
      10^{7.1}$~g~cm$^{-3}$).  Data points are colored by the study they were
      taken from: blue from this study, red from \citet{jacketal10}.  The
      dotted line shows a perfect correspondence between the two measures of
      \MIGE.}
   \label{fig:proxy_v_actual}
\end{figure*}

First we consider how we expect $M_{\rm IGE}$ might change if a lower
$\rho_{\rm DDT}$ were assumed.
\citet{townetal09} demonstrates a correlation between \MIGE\ at \tIGE\ and the
mass of all material with a density greater than $2\times10^7$~g~cm$^{-3}$ at
time \tDDT\ (which we represent by \proxy).  Therefore we can choose a new
value of \rhoDDT\ and use this correlation to estimate the final mass of IGEs
in order to verify that changing \rhoDDT\ does not change the fact that the
mass of IGEs is independent of \cdens.  In order to refine the relation between
\MIGE\ and \proxy, we make use of data from this study and data from the study
presented in \citet{jacketal10}, which varied \rhoDDT.  This will enable us to
calibrate the relationship between \MIGE\ and \proxy\ for differences in
\cdens\ and \rhoDDT.  The relationship between \MIGE\ and \proxy\ appears to be
approximately linear, so we assume that $\MIGE = m \; \proxy + b$.  As can be
seen in \figref{fig:proxy_variation}, the relationship between \proxy\ and
\MIGE\ depends on \cdens\ and \rhoDDT, so the slope and intercept are allowed
to be functions of \cdens\ and \rhoDDT.  We test polynomials of varying degree
and find that the best functions are quadratic in \cdens\ and \rhoDDT\ for both
the slope and the intercept.  The equations for this fitting are
\begin{subequations}
   \begin{align}
      \MIGE
         & = m(\cdens, \rhoDDT) \; \proxy + b(\cdens, \rhoDDT) \\
      m(\cdens, \rhoDDT)
         & = m_0 + \gamma_1      \cdens  + \gamma_2      \cdens^2
                 + \delta_1      \rhoDDT + \delta_2      \rhoDDT^2 \\
      b(\cdens, \rhoDDT)
         & = b_0 + \varepsilon_1 \cdens  + \varepsilon_2 \cdens^2
                 + \zeta_1       \rhoDDT + \zeta_2       \rhoDDT^2.
   \end{align}
   \label{eqn:proxy}
\end{subequations}
Fitting this 10-parameter function to the data from this study and the study of
\citet{jacketal10} yields the parameters given in \tabref{tab:Chi10Params}.  As
a check, we use this relation at the same \rhoDDT\ as was used in our
simulations ($10^{7.1}$~g~cm$^{-3}$) and compare the estimated values of
\MIGE\ to the actual values.  This is shown in \figref{fig:proxy_v_actual}.

\begin{figure*}
   \centering
   \includegraphics[angle=270,width=0.45\columnwidth]{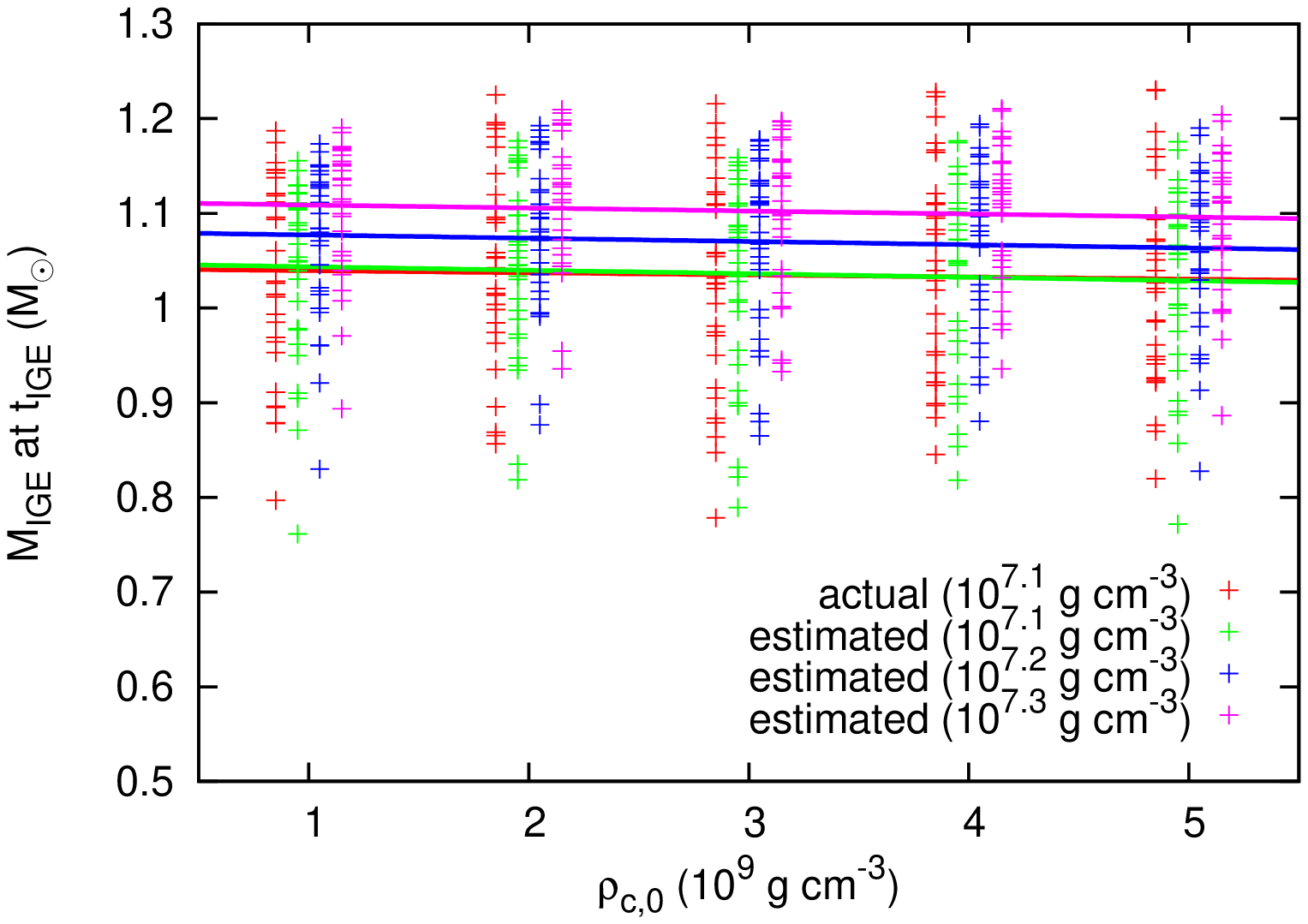}
   \caption{Demonstration of the results of \eqnref{eqn:proxy},
      showing the trend of \MIGE\ with \cdens, using data from this study.  The
      values of \MIGE\ from the simulations are shown in red, with green, blue,
      and magenta showing the estimated \MIGE\ assuming $\rhoDDT = 10^{7.1}$,
      $10^{7.2}$, and $10^{7.3}$~g~cm$^{-3}$ respectively.  The data points are
      slightly offset horizontally for clarity.  This shows that \MIGE\ is
      independent of \cdens\ even if \rhoDDT\ varies.  This figure also
      demonstrates that a lower value of \rhoDDT\ leads to a lower mean value
      of \MIGE.}
   \label{fig:proxy_compare}
\end{figure*}

With this function, we can extract \proxy\ assuming different values of
\rhoDDT.  Based on the data available, we cannot choose a value of \rhoDDT\ 
lower than what used in this study, but we can test higher values of \rhoDDT.
\figref{fig:proxy_compare} shows that for three sample values of \rhoDDT,
\MIGE\ is still independent of \cdens; formally, we say that \MIGE\ is
independent of \cdens\ when the magnitude of the slope of the line relating
these two quantities is less than the uncertainty in the slope.  We therefore
assume that, even for lower values of \rhoDDT, \MIGE\ is independent of \cdens.
This allows us to apply a uniform shift in \MIGE (constant for all simulations)
to correct for the overproduction of IGEs.  Observations tell us that the mass
of IGEs should lie approximately in the range of $0.7 - 0.9$~\Msol\ (see, e.g.,
\citealt{WoosleyEtAl07:LightCurves}), so we can choose to force our mean mass
of IGEs to be in the center of this range (0.8~\Msol).

\begin{table}
   \centering
   \caption{Fit parameters for \eqnref{eqn:proxy}.  These values
      assume that \cdens\ is in $10^9$~g~cm$^{-3}$, \rhoDDT\ is in
      $10^7$~g~cm$^{-3}$, and both masses are in \Msol.}
   \begin{tabular}{c c}
      \hline \hline 
      parameter       & value                 \\[0.7ex]
      \hline 
      $m_0$           & $ 9.099\times10^{-1}$ \\[0.7ex]
      $\gamma_1$      & $ 1.453\times10^{-1}$ \\[0.7ex]
      $\gamma_2$      & $-2.429\times10^{-2}$ \\[0.7ex]
      $\delta_1$      & $-4.225\times10^{-1}$ \\[0.7ex]
      $\delta_2$      & $ 8.555\times10^{-2}$ \\[0.7ex]
      $b_0$           & $ 8.358\times10^{-3}$ \\[0.7ex]
      $\varepsilon_1$ & $-1.847\times10^{-1}$ \\[0.7ex]
      $\varepsilon_2$ & $ 2.869\times10^{-2}$ \\[0.7ex]
      $\zeta_1$       & $ 6.231\times10^{-1}$ \\[0.7ex]
      $\zeta_2$       & $-1.196\times10^{-1}$ \\[0.7ex]
      \hline 
   \end{tabular}
   \label{tab:Chi10Params}
\end{table}

\begin{figure*}
   \centering
   \subfloat{\label{fig:YieldIGE}
     \includegraphics[angle=270, width=0.45\columnwidth]{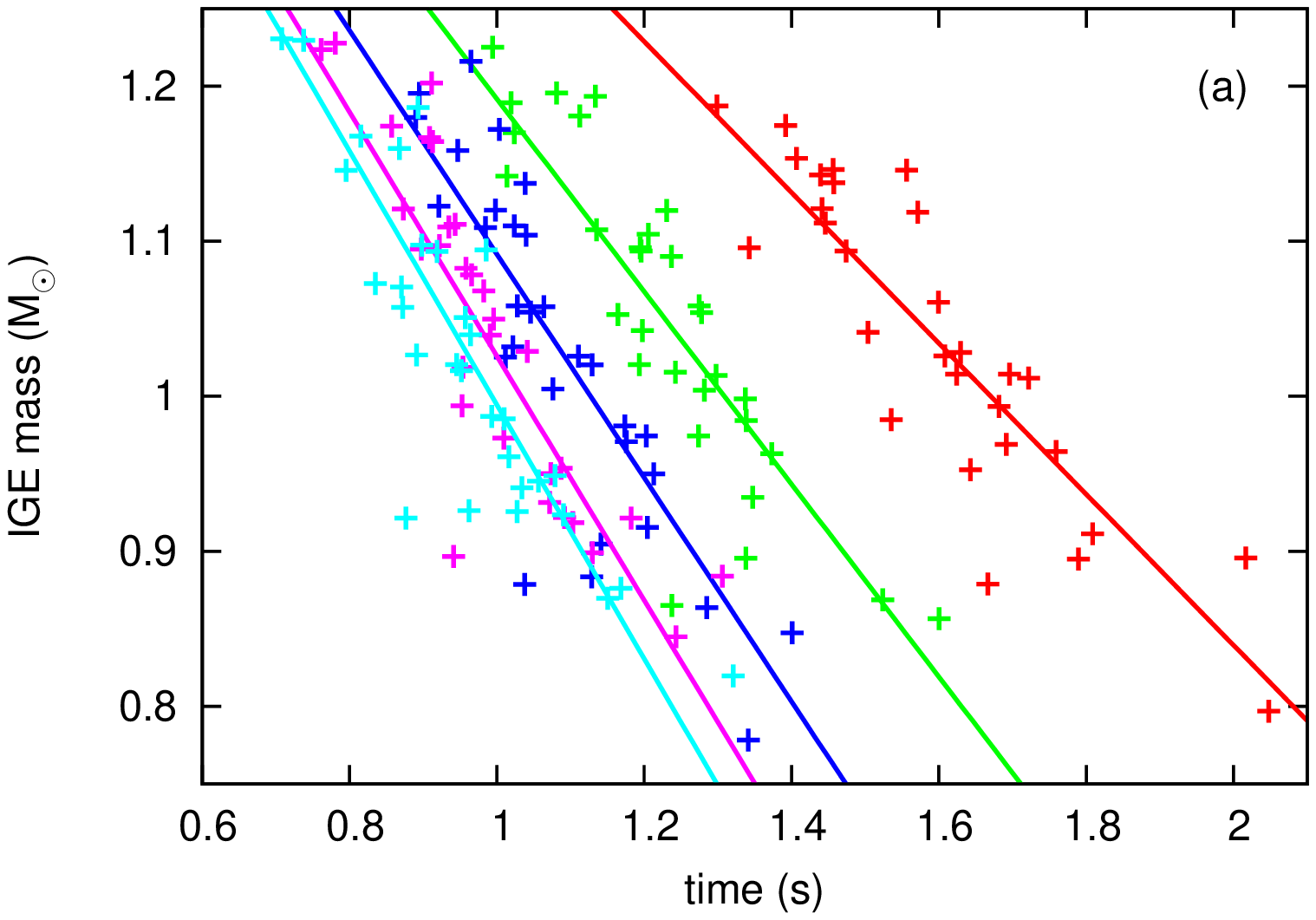}
   }
   \subfloat{\label{fig:YieldNi56}
     \includegraphics[angle=270, width=0.45\columnwidth]{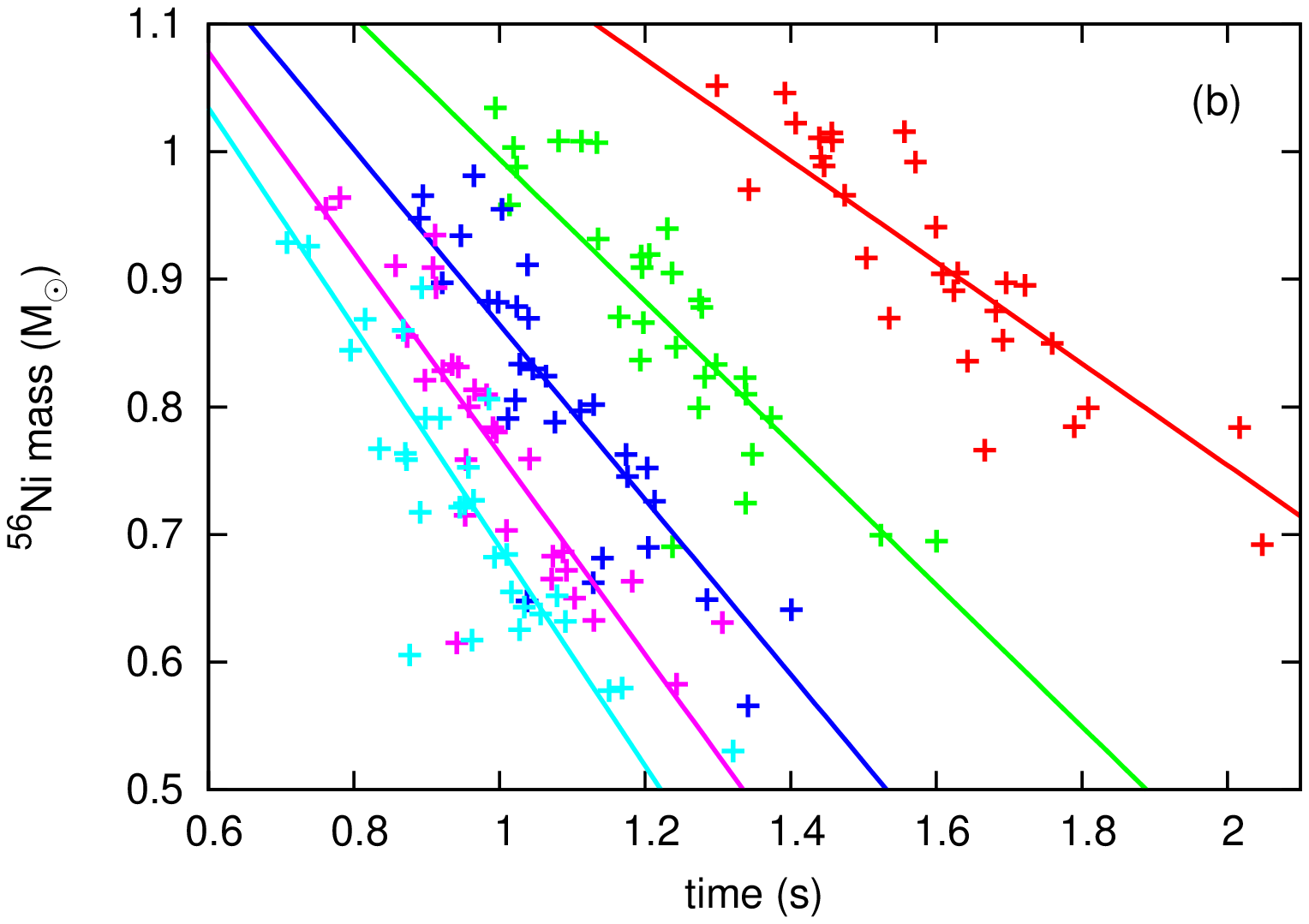}
   }
   \caption{Yield of IGEs (left panel) and \Ni{56} (right panel)
      as functions of \tDDT\ for the 149 simulations.  Also shown are the
      best-fit trend lines in the form given by
      Equations~(\ref{eqn:yieldVtDDT}) and (\ref{eqn:slopeVrho}).  Lines and
      data points are colored by \cdens: $1.0 \times 10^9$~g~cm$^{-3}$ (red),
      $2.0 \times 10^9$~g~cm$^{-3}$ (green), $3.0 \times 10^9$~g~cm$^{-3}$
      (blue), $4.0 \times 10^9$~g~cm$^{-3}$ (magenta), $5.0 \times
      10^9$~g~cm$^{-3}$ (cyan).}
   \label{fig:Yield}
\end{figure*}

Now that we have established a reasonable expectation that $M_{\rm IGE}$
will continue to be independent of $\rho_{c,0}$ even at other $\rho_{\rm
DDT}$ values, we proceed to extrapolate our yields based around this premise.
We can relate this change in \MIGE\ to a change in the mass of \Ni{56} through
adjustments to \tDDT.  The value of \tDDT, which depends on \rhoDDT,  is not a
free parameter that can be independently adjusted in our models; however, in
order to correct the mass of \Ni{56} we can treat \tDDT\ as a parameter.
\figref{fig:Yield} shows the masses of IGEs and \Ni{56} plotted as functions of
\tDDT, where each point is a single simulation colored by \cdens.  Also shown
is a trend line based on \eqnref{eqn:yieldVtDDT}, explained below.  Each of
these quantities shows five separate trends, one for each value of \cdens.
Performing independent linear fits for each \cdens\ gives lines that appear to
be correlated: the five lines intersect each other near $\tDDT = 0$, and the
slopes appear to be a function of \cdens.  Based on these correlations, we
derived a new fit using the function
\begin{equation}
   y(\cdens, \tDDT) = y_0 + s(\cdens) \; \tDDT, \label{eqn:yieldVtDDT}
\end{equation}
where $y$ is the mass of either IGEs or \Ni{56}.  We found that the best form
for the slope $s(\cdens)$ is:
\begin{equation}
   s(\cdens) = a \; \cdens^2 + b \; \cdens + c.  \label{eqn:slopeVrho}
\end{equation}
Minimizing the \chisq\ for this 4-parameter function results in the parameter
values shown in \tabref{tab:Chi4Params}.

\begin{table}
   \centering
   \caption{Fit parameters for Equations~(\ref{eqn:yieldVtDDT})
      and (\ref{eqn:slopeVrho}).  These values assume that \tDDT\ is in
      seconds, \cdens\ is in $10^9$~g~cm$^{-3}$, and mass is in \Msol.}
   \begin{tabular}{c c c}
      \hline \hline 
      parameter & \MIGE                   & $M_{\Ni{56}}$           \\[0.7ex]
      \hline 
      $y_0$     & 1.813                   & 1.550                   \\[0.7ex]
      $a$       & $1.716 \times 10^{-2}$  & $1.422 \times 10^{-2}$  \\[0.7ex]
      $b$       & $-1.859 \times 10^{-1}$ & $-2.007 \times 10^{-1}$ \\[0.7ex]
      $c$       & $-3.180 \times 10^{-1}$ & $-2.114 \times 10^{-1}$ \\[0.7ex]
      \hline 
   \end{tabular}
   \label{tab:Chi4Params}
\end{table}


Given the shift in the mass of IGEs, we can then use the $\MIGE(\cdens, \tDDT)$
relation from Equations~(\ref{eqn:yieldVtDDT}) and (\ref{eqn:slopeVrho}) to
compute a corresponding change in \tDDT.  However, since the relation (in
particular the slope) depends on \cdens, the time shift will also depend on
\cdens; i.e., we will have $\Delta \tDDT (\cdens)$.  The five values of $\Delta
\tDDT$ are given in \tabref{tab:TimeShift}.  Using this density-dependent time
shift and the $M_{\Ni{56}}(\cdens, \tDDT)$ relation, we can then derive a
consistent adjustment to the mass of \Ni{56}.  This adjusted mass of \Ni{56} is
used in \secref{sec:YieldFitting}.

\begin{table}
   \centering
   \caption{Time shifts for the recalibration of the \Ni{56} and
      IGE masses.}
   \begin{tabular}{c c}
      \hline \hline 
      \cdens\ (g~cm$^{-3}$) & $\Delta \tDDT$ (s) \\[0.7ex]
      \hline 
      $1 \times 10^9$       & 0.4850             \\[0.7ex]
      $2 \times 10^9$       & 0.3800             \\[0.7ex]
      $3 \times 10^9$       & 0.3273             \\[0.7ex]
      $4 \times 10^9$       & 0.2999             \\[0.7ex]
      $5 \times 10^9$       & 0.2884             \\[0.7ex]
      \hline 
   \end{tabular}
   \label{tab:TimeShift}
\end{table}

\section{Simulation Results}
\label{app:data}

The structure and composition of the progenitor models are discussed in detail
in \secref{sec:init}, along with the creation of a statistical ensemble.  The
simulations and the code are discussed in \secref{sec:method}.  
As we have stressed, the principal result from our simulations is a mass
of \Ni{56}. Our models reproduce the qualitative description given in
\citet{WoosleyEtAl07:LightCurves} that produces a type Ia supernova 
light curve. The ideal solution for comparing our results to observed
light would be for us to post-process our results with a radiation transfer
method to actually calculate light curves. In the absence of that technology,
we may infer observed properties of supernova events suggested by our results.  
We report some of those here, including stretch, the scaling of brightness
with rapidity of decline that is reported in observational 
results~\citep{howelletal+09,jha2007}.
The tabulated data are
\begin{enumerate}
   \item \cdens: central density of pre-supernova white dwarf (g~cm$^{-3}$);
   \item $r$: realization number\footnotemark[10];
   \item \tDDT: time of the first deflagration-to-detonation transition (s);
   \item $M_{\rm IGE}(\tDDT)$: mass of Fe-group elements synthesized by time
      \tDDT\ (\Msol);
   \item $M_{\rm \Ni{56}}(\tDDT)$: mass of \Ni{56} synthesized by time
      \tDDT\ (\Msol);
   \item \tIGE: time that production of Fe-group elements ceases (s);
   \item $M_{\rm IGE}(\tIGE)$: mass of Fe-group elements synthesized by time
      \tIGE\  (\Msol);
   \item $M_{\rm \Ni{56}}(\tIGE)$: mass of \Ni{56} synthesized by time
      \tIGE\ (\Msol);
   \item $t^*_{\rm DDT}$: recalibrated\footnotemark[11] value of \tDDT\ (s);
   \item $M^*_{\rm IGE}(t^*_{\rm IGE})$: recalibrated\footnotemark[11] value of
      $M_{\rm IGE}(\tIGE)$ (\Msol);
   \item $M^*_{\rm \Ni{56}}(t^*_{\rm IGE})$: recalibrated\footnotemark[11]
      value of $M_{\rm \Ni{56}}(\tIGE)$ (\Msol);
   \item $s$: recalibrated\footnotemark[11] value of stretch\footnotemark[12];
   \item \tcool: cooling time\footnotemark[12] (yr).
\end{enumerate}
\footnotetext[10]{See \appref{app:init} and \secref{sec:method} for more
   details.}
\footnotetext[11]{See \appref{app:recal} and \secref{sec:YieldFitting} for
   details of the recalibration.}
\footnotetext[12]{See \secref{sec:observe} for details of the derivation of $s$
   and \tcool.}
The initial composition for the progenitor is shown in
\tabref{tab:ProgComposition}.  The total mass of the star and the total mass of
the initial convective, isentropic core are both shown for each progenitor in
\tabref{tab:DensityMass}.  Both of these tables are shown in \secref{sec:init}.

\begin{center}
\begin{longtable}{ccccccccccccc}
\caption{Data extracted from simulations.} \label{tab:data} \\[1ex]

\hline \hline \\[1ex]
\cdens &
$r$ &
\tDDT &
$M_{\rm IGE}(\tDDT)$ &
$M_{\rm \Ni{56}}(\tDDT)$ &
\tIGE &
$M_{\rm IGE}(\tIGE)$ &
$M_{\rm \Ni{56}}(\tIGE)$ &
$t^*_{\rm DDT}$ &
$M^*_{\rm IGE}(t^*_{\rm IGE})$ &
$M^*_{\rm \Ni{56}}(t^*_{\rm IGE})$ &
$s$ &
\tcool
 \\[1ex]
(g~cm$^{-3}$)&
 &
(s) &
(\Msol) &
(\Msol) &
(s) &
(\Msol) &
(\Msol) &
(s) &
(\Msol) &
(\Msol) &
 &
(yr)
 \\[1ex]
\hline \\[1ex]
\endfirsthead

\multicolumn{3}{c}{{\tablename} \thetable{} -- Continued} \\[1ex]
  \hline \hline \\[1ex]
\cdens &
$r$ &
\tDDT &
$M_{\rm IGE}(\tDDT)$ &
$M_{\rm \Ni{56}}(\tDDT)$ &
\tIGE &
$M_{\rm IGE}(\tIGE)$ &
$M_{\rm \Ni{56}}(\tIGE)$ &
$t^*_{\rm DDT}$ &
$M^*_{\rm IGE}(t^*_{\rm IGE})$ &
$M^*_{\rm \Ni{56}}(t^*_{\rm IGE})$ &
$s$ &
\tcool
 \\[1ex]
(g~cm$^{-3}$)&
~ &
(s) &
(\Msol) &
(\Msol) &
(s) &
(\Msol) &
(\Msol) &
(s) &
(\Msol) &
(\Msol) &
~ &
(yr)
 \\[1ex]
\hline \\[1ex]
\endhead

\\[2ex]
  \multicolumn{3}{l}{{Continued on Next Page\ldots}} \\[1ex]
\endfoot

  \hline
\endlastfoot

1e+9 & 1  & 1.439 & 1.256e-1 & 7.744e-2 & 2.046 & 1.143e+0 & 1.011e+0 & 1.924 & 9.067e-1 & 8.177e-1 & 1.203 & \ldots\footnotemark[b] \\[1ex]
1e+9 & 2  & 1.608 & 1.756e-1 & 1.210e-1 & 2.040 & 1.026e+0 & 9.042e-1 & 2.093 & 7.900e-1 & 7.112e-1 & 1.138 & \ldots\footnotemark[b] \\[1ex]
1e+9 & 3  & 1.809 & 1.773e-1 & 1.229e-1 & 2.203 & 9.113e-1 & 7.991e-1 & 2.294 & 6.752e-1 & 6.061e-1 & 1.066 & \ldots\footnotemark[b] \\[1ex]
1e+9 & 4  & 1.696 & 1.766e-1 & 1.256e-1 & 2.083 & 1.014e+0 & 8.972e-1 & 2.181 & 7.783e-1 & 7.042e-1 & 1.134 & \ldots\footnotemark[b] \\[1ex]
1e+9 & 5  & 1.504 & 1.490e-1 & 9.677e-2 & 2.039 & 1.041e+0 & 9.166e-1 & 1.989 & 8.052e-1 & 7.237e-1 & 1.146 & \ldots\footnotemark[b] \\[1ex]
1e+9 & 6  & 1.456 & 1.176e-1 & 7.158e-2 & 1.959 & 1.146e+0 & 1.015e+0 & 1.941 & 9.102e-1 & 8.217e-1 & 1.205 & \ldots\footnotemark[b] \\[1ex]
1e+9 & 7  & 1.474 & 1.313e-1 & 8.181e-2 & 2.008 & 1.094e+0 & 9.659e-1 & 1.959 & 8.575e-1 & 7.729e-1 & 1.177 & \ldots\footnotemark[b] \\[1ex]
1e+9 & 8  & 1.477 & 1.164e-1 & 6.855e-2 & 2.400 & 8.780e-1 & 7.687e-1 & \ldots\footnotemark[a] & \ldots\footnotemark[a] & \ldots\footnotemark[a] & \ldots\footnotemark[a] & \ldots\footnotemark[a] \\[1ex]
1e+9 & 9  & 1.682 & 1.637e-1 & 1.110e-1 & 2.095 & 9.935e-1 & 8.751e-1 & 2.167 & 7.575e-1 & 6.822e-1 & 1.119 & \ldots\footnotemark[b] \\[1ex]
1e+9 & 10 & 2.017 & 2.081e-1 & 1.495e-1 & 2.430 & 8.959e-1 & 7.837e-1 & 2.502 & 6.599e-1 & 5.907e-1 & 1.054 & \ldots\footnotemark[b] \\[1ex]
1e+9 & 11 & 1.457 & 1.310e-1 & 8.431e-2 & 1.878 & 1.138e+0 & 1.008e+0 & 1.942 & 9.016e-1 & 8.155e-1 & 1.202 & \ldots\footnotemark[b] \\[1ex]
1e+9 & 12 & 1.556 & 1.099e-1 & 6.654e-2 & 2.025 & 1.146e+0 & 1.016e+0 & 2.041 & 9.098e-1 & 8.227e-1 & 1.206 & \ldots\footnotemark[b] \\[1ex]
1e+9 & 13 & 2.048 & 2.017e-1 & 1.429e-1 & 2.473 & 7.969e-1 & 6.920e-1 & 2.533 & 5.608e-1 & 4.990e-1 & 0.980 & \ldots\footnotemark[b] \\[1ex]
1e+9 & 14 & 1.600 & 1.616e-1 & 1.135e-1 & 2.015 & 1.061e+0 & 9.410e-1 & 2.085 & 8.246e-1 & 7.480e-1 & 1.161 & \ldots\footnotemark[b] \\[1ex]
1e+9 & 15 & 1.441 & 1.169e-1 & 7.452e-2 & 1.863 & 1.121e+0 & 9.955e-1 & 1.927 & 8.850e-1 & 8.026e-1 & 1.194 & \ldots\footnotemark[b] \\[1ex]
1e+9 & 16 & 1.629 & 1.756e-1 & 1.198e-1 & 2.106 & 1.028e+0 & 9.049e-1 & 2.114 & 7.923e-1 & 7.120e-1 & 1.139 & \ldots\footnotemark[b] \\[1ex]
1e+9 & 17 & 1.721 & 1.701e-1 & 1.204e-1 & 2.136 & 1.012e+0 & 8.951e-1 & 2.206 & 7.757e-1 & 7.021e-1 & 1.132 & \ldots\footnotemark[b] \\[1ex]
1e+9 & 18 & 1.407 & 1.160e-1 & 7.173e-2 & 1.906 & 1.154e+0 & 1.022e+0 & 1.892 & 9.175e-1 & 8.295e-1 & 1.210 & \ldots\footnotemark[b] \\[1ex]
1e+9 & 19 & 1.691 & 1.690e-1 & 1.157e-1 & 2.068 & 9.690e-1 & 8.524e-1 & 2.176 & 7.330e-1 & 6.594e-1 & 1.103 & \ldots\footnotemark[b] \\[1ex]
1e+9 & 20 & 1.643 & 1.910e-1 & 1.341e-1 & 2.166 & 9.528e-1 & 8.358e-1 & 2.128 & 7.168e-1 & 6.428e-1 & 1.092 & \ldots\footnotemark[b] \\[1ex]
1e+9 & 21 & 1.299 & 9.447e-2 & 5.357e-2 & 1.824 & 1.187e+0 & 1.052e+0 & 1.784 & 9.513e-1 & 8.589e-1 & 1.226 & \ldots\footnotemark[b] \\[1ex]
1e+9 & 22 & 1.535 & 1.459e-1 & 9.739e-2 & 2.032 & 9.850e-1 & 8.697e-1 & 2.020 & 7.489e-1 & 6.767e-1 & 1.115 & \ldots\footnotemark[b] \\[1ex]
1e+9 & 23 & 1.759 & 1.648e-1 & 1.131e-1 & 2.234 & 9.643e-1 & 8.498e-1 & 2.244 & 7.283e-1 & 6.568e-1 & 1.102 & \ldots\footnotemark[b] \\[1ex]
1e+9 & 24 & 1.666 & 1.851e-1 & 1.265e-1 & 2.179 & 8.789e-1 & 7.662e-1 & 2.151 & 6.428e-1 & 5.732e-1 & 1.041 & \ldots\footnotemark[b] \\[1ex]
1e+9 & 25 & 1.789 & 1.865e-1 & 1.315e-1 & 2.243 & 8.950e-1 & 7.846e-1 & 2.274 & 6.590e-1 & 5.916e-1 & 1.055 & \ldots\footnotemark[b] \\[1ex]
1e+9 & 26 & 1.446 & 1.348e-1 & 9.107e-2 & 1.844 & 1.112e+0 & 9.889e-1 & 1.931 & 8.758e-1 & 7.959e-1 & 1.190 & \ldots\footnotemark[b] \\[1ex]
1e+9 & 27 & 1.392 & 9.724e-2 & 6.110e-2 & 1.845 & 1.175e+0 & 1.046e+0 & 1.877 & 9.387e-1 & 8.530e-1 & 1.223 & \ldots\footnotemark[b] \\[1ex]
1e+9 & 28 & 1.342 & 1.132e-1 & 7.015e-2 & 1.828 & 1.096e+0 & 9.702e-1 & 1.827 & 8.598e-1 & 7.772e-1 & 1.179 & \ldots\footnotemark[b] \\[1ex]
1e+9 & 29 & 1.624 & 1.738e-1 & 1.169e-1 & 2.176 & 1.014e+0 & 8.911e-1 & 2.109 & 7.783e-1 & 6.981e-1 & 1.130 & \ldots\footnotemark[b] \\[1ex]
1e+9 & 30 & 1.571 & 1.277e-1 & 8.167e-2 & 2.000 & 1.119e+0 & 9.920e-1 & 2.056 & 8.828e-1 & 7.990e-1 & 1.192 & \ldots\footnotemark[b] \\[1ex]
2e+9 & 1  & 1.274 & 2.186e-1 & 1.056e-1 & 1.772 & 9.744e-1 & 7.991e-1 & 1.654 & 7.384e-1 & 5.879e-1 & 1.052 & 7.908e+7 \\[1ex]
2e+9 & 2  & 1.338 & 2.363e-1 & 1.183e-1 & 1.827 & 8.956e-1 & 7.246e-1 & 1.718 & 6.596e-1 & 5.133e-1 & 0.992 & 7.908e+7 \\[1ex]
2e+9 & 3  & 1.373 & 2.242e-1 & 1.122e-1 & 1.867 & 9.631e-1 & 7.917e-1 & 1.753 & 7.270e-1 & 5.804e-1 & 1.046 & 7.908e+7 \\[1ex]
2e+9 & 4  & 1.231 & 2.044e-1 & 9.844e-2 & 1.722 & 1.120e+0 & 9.398e-1 & 1.611 & 8.838e-1 & 7.285e-1 & 1.149 & 7.908e+7 \\[1ex]
2e+9 & 5  & 1.164 & 2.030e-1 & 9.237e-2 & 1.672 & 1.053e+0 & 8.705e-1 & 1.544 & 8.166e-1 & 6.592e-1 & 1.103 & 7.908e+7 \\[1ex]
2e+9 & 6  & 1.278 & 1.909e-1 & 8.627e-2 & 1.779 & 1.054e+0 & 8.780e-1 & 1.658 & 8.180e-1 & 6.668e-1 & 1.109 & 7.908e+7 \\[1ex]
2e+9 & 7  & 1.237 & 1.942e-1 & 8.247e-2 & 1.628 & 1.090e+0 & 9.049e-1 & 1.617 & 8.541e-1 & 6.936e-1 & 1.127 & 7.908e+7 \\[1ex]
2e+9 & 8  & 1.196 & 1.727e-1 & 6.563e-2 & 1.789 & 1.094e+0 & 9.092e-1 & 1.576 & 8.577e-1 & 6.979e-1 & 1.129 & 7.908e+7 \\[1ex]
2e+9 & 9  & 1.281 & 2.196e-1 & 1.018e-1 & 1.742 & 1.004e+0 & 8.231e-1 & 1.661 & 7.678e-1 & 6.118e-1 & 1.070 & 7.908e+7 \\[1ex]
2e+9 & 10 & 1.600 & 2.540e-1 & 1.408e-1 & 2.141 & 8.565e-1 & 6.948e-1 & 1.980 & 6.205e-1 & 4.835e-1 & 0.967 & 7.908e+7 \\[1ex]
2e+9 & 11 & 1.195 & 1.883e-1 & 8.399e-2 & 1.598 & 1.096e+0 & 9.183e-1 & 1.575 & 8.598e-1 & 7.070e-1 & 1.135 & 7.908e+7 \\[1ex]
2e+9 & 12 & 1.134 & 1.356e-1 & 4.169e-2 & 1.604 & 1.194e+0 & 1.007e+0 & 1.514 & 9.575e-1 & 7.957e-1 & 1.190 & 7.908e+7 \\[1ex]
2e+9 & 13 & 1.524 & 2.363e-1 & 1.168e-1 & 1.957 & 8.689e-1 & 6.994e-1 & 1.904 & 6.328e-1 & 4.882e-1 & 0.971 & 7.908e+7 \\[1ex]
2e+9 & 14 & 1.242 & 2.062e-1 & 1.029e-1 & 1.727 & 1.016e+0 & 8.468e-1 & 1.622 & 7.795e-1 & 6.356e-1 & 1.087 & 7.908e+7 \\[1ex]
2e+9 & 15 & 1.024 & 1.429e-1 & 5.202e-2 & 1.569 & 1.170e+0 & 9.880e-1 & 1.404 & 9.340e-1 & 7.768e-1 & 1.179 & 7.908e+7 \\[1ex]
2e+9 & 16 & 1.347 & 2.412e-1 & 1.257e-1 & 1.816 & 9.349e-1 & 7.629e-1 & 1.727 & 6.988e-1 & 5.516e-1 & 1.024 & 7.908e+7 \\[1ex]
2e+9 & 17 & 1.275 & 2.046e-1 & 9.945e-2 & 1.704 & 1.058e+0 & 8.838e-1 & 1.655 & 8.223e-1 & 6.726e-1 & 1.112 & 7.908e+7 \\[1ex]
2e+9 & 18 & 1.019 & 1.490e-1 & 5.379e-2 & 1.586 & 1.189e+0 & 1.003e+0 & 1.399 & 9.533e-1 & 7.921e-1 & 1.188 & 7.908e+7 \\[1ex]
2e+9 & 19 & 1.206 & 1.913e-1 & 8.005e-2 & 1.619 & 1.105e+0 & 9.195e-1 & 1.586 & 8.685e-1 & 7.083e-1 & 1.136 & 7.908e+7 \\[1ex]
2e+9 & 20 & 1.298 & 2.323e-1 & 1.153e-1 & 1.716 & 1.013e+0 & 8.333e-1 & 1.678 & 7.774e-1 & 6.220e-1 & 1.077 & 7.908e+7 \\[1ex]
2e+9 & 21 & 0.994 & 1.383e-1 & 4.546e-2 & 1.503 & 1.225e+0 & 1.034e+0 & 1.374 & 9.892e-1 & 8.231e-1 & 1.206 & 7.908e+7 \\[1ex]
2e+9 & 22 & 1.197 & 1.861e-1 & 8.048e-2 & 1.792 & 1.043e+0 & 8.660e-1 & 1.577 & 8.065e-1 & 6.548e-1 & 1.100 & 7.908e+7 \\[1ex]
2e+9 & 23 & 1.337 & 2.027e-1 & 9.030e-2 & 1.780 & 9.984e-1 & 8.230e-1 & 1.717 & 7.624e-1 & 6.117e-1 & 1.070 & 7.908e+7 \\[1ex]
2e+9 & 24 & 1.237 & 2.299e-1 & 1.069e-1 & 1.870 & 8.652e-1 & 6.905e-1 & 1.617 & 6.291e-1 & 4.792e-1 & 0.963 & 7.908e+7 \\[1ex]
2e+9 & 25 & 1.338 & 2.190e-1 & 1.059e-1 & 1.736 & 9.844e-1 & 8.099e-1 & 1.718 & 7.484e-1 & 5.986e-1 & 1.060 & 7.908e+7 \\[1ex]
2e+9 & 26 & 1.135 & 1.792e-1 & 7.881e-2 & 1.557 & 1.107e+0 & 9.315e-1 & 1.515 & 8.714e-1 & 7.202e-1 & 1.144 & 7.908e+7 \\[1ex]
2e+9 & 27 & 1.112 & 1.329e-1 & 4.918e-2 & 1.539 & 1.181e+0 & 1.008e+0 & 1.492 & 9.448e-1 & 7.970e-1 & 1.191 & 7.908e+7 \\[1ex]
2e+9 & 28 & 1.014 & 1.529e-1 & 5.527e-2 & 1.451 & 1.142e+0 & 9.583e-1 & 1.394 & 9.059e-1 & 7.471e-1 & 1.161 & 7.908e+7 \\[1ex]
2e+9 & 29 & 1.193 & 2.064e-1 & 8.958e-2 & 1.766 & 1.021e+0 & 8.368e-1 & 1.573 & 7.845e-1 & 6.255e-1 & 1.080 & 7.908e+7 \\[1ex]
2e+9 & 30 & 1.081 & 1.509e-1 & 5.572e-2 & 1.529 & 1.196e+0 & 1.008e+0 & 1.461 & 9.596e-1 & 7.971e-1 & 1.191 & 7.908e+7 \\[1ex]
3e+9 & 1  & 1.111 & 2.392e-1 & 8.266e-2 & 1.533 & 1.026e+0 & 7.969e-1 & 1.438 & 7.899e-1 & 5.726e-1 & 1.040 & 2.879e+8 \\[1ex]
3e+9 & 2  & 1.064 & 2.459e-1 & 8.436e-2 & 1.503 & 1.058e+0 & 8.240e-1 & 1.391 & 8.217e-1 & 5.997e-1 & 1.061 & 2.879e+8 \\[1ex]
3e+9 & 3  & 1.285 & 2.507e-1 & 8.898e-2 & 1.723 & 8.638e-1 & 6.488e-1 & 1.612 & 6.278e-1 & 4.244e-1 & 0.911 & 2.879e+8 \\[1ex]
3e+9 & 4  & 1.174 & 2.375e-1 & 8.843e-2 & 1.751 & 9.808e-1 & 7.627e-1 & 1.501 & 7.448e-1 & 5.383e-1 & 1.013 & 2.879e+8 \\[1ex]
3e+9 & 5  & 1.022 & 2.249e-1 & 7.251e-2 & 1.556 & 1.032e+0 & 8.054e-1 & 1.349 & 7.958e-1 & 5.810e-1 & 1.047 & 2.879e+8 \\[1ex]
3e+9 & 6  & 1.024 & 1.950e-1 & 4.817e-2 & 1.531 & 1.110e+0 & 8.787e-1 & 1.351 & 8.739e-1 & 6.544e-1 & 1.100 & 2.879e+8 \\[1ex]
3e+9 & 7  & 0.998 & 2.130e-1 & 5.642e-2 & 1.414 & 1.120e+0 & 8.818e-1 & 1.325 & 8.841e-1 & 6.575e-1 & 1.102 & 2.879e+8 \\[1ex]
3e+9 & 8  & 1.039 & 2.016e-1 & 5.049e-2 & 1.538 & 1.104e+0 & 8.693e-1 & 1.367 & 8.679e-1 & 6.449e-1 & 1.093 & 2.879e+8 \\[1ex]
3e+9 & 9  & 1.141 & 2.510e-1 & 8.595e-2 & 1.652 & 9.048e-1 & 6.815e-1 & 1.469 & 6.687e-1 & 4.571e-1 & 0.943 & 2.879e+8 \\[1ex]
3e+9 & 10 & 1.401 & 2.650e-1 & 1.099e-1 & 1.881 & 8.474e-1 & 6.409e-1 & 1.728 & 6.114e-1 & 4.165e-1 & 0.903 & 2.879e+8 \\[1ex]
3e+9 & 11 & 1.046 & 2.204e-1 & 6.880e-2 & 1.484 & 1.054e+0 & 8.297e-1 & 1.373 & 8.182e-1 & 6.054e-1 & 1.065 & 2.879e+8 \\[1ex]
3e+9 & 12 & 0.965 & 1.622e-1 & 3.069e-2 & 1.478 & 1.216e+0 & 9.813e-1 & 1.292 & 9.799e-1 & 7.569e-1 & 1.167 & 2.879e+8 \\[1ex]
3e+9 & 13 & 1.341 & 2.612e-1 & 9.459e-2 & 1.777 & 7.783e-1 & 5.657e-1 & 1.669 & 5.423e-1 & 3.414e-1 & 0.821 & 2.879e+8 \\[1ex]
3e+9 & 14 & 1.076 & 2.186e-1 & 7.401e-2 & 1.578 & 1.005e+0 & 7.881e-1 & 1.403 & 7.688e-1 & 5.637e-1 & 1.033 & 2.879e+8 \\[1ex]
3e+9 & 15 & 0.947 & 1.748e-1 & 4.569e-2 & 1.413 & 1.159e+0 & 9.341e-1 & 1.274 & 9.226e-1 & 7.097e-1 & 1.137 & 2.879e+8 \\[1ex]
3e+9 & 16 & 1.129 & 2.631e-1 & 9.851e-2 & 1.664 & 8.836e-1 & 6.622e-1 & 1.456 & 6.475e-1 & 4.378e-1 & 0.924 & 2.879e+8 \\[1ex]
3e+9 & 17 & 1.129 & 2.272e-1 & 7.756e-2 & 1.589 & 1.020e+0 & 8.019e-1 & 1.456 & 7.843e-1 & 5.775e-1 & 1.044 & 2.879e+8 \\[1ex]
3e+9 & 18 & 0.890 & 1.738e-1 & 4.342e-2 & 1.426 & 1.180e+0 & 9.478e-1 & 1.217 & 9.438e-1 & 7.235e-1 & 1.146 & 2.879e+8 \\[1ex]
3e+9 & 19 & 1.204 & 2.529e-1 & 8.665e-2 & 1.639 & 9.155e-1 & 6.899e-1 & 1.531 & 6.794e-1 & 4.655e-1 & 0.950 & 2.879e+8 \\[1ex]
3e+9 & 20 & 1.176 & 2.544e-1 & 9.120e-2 & 1.608 & 9.708e-1 & 7.453e-1 & 1.503 & 7.347e-1 & 5.210e-1 & 0.999 & 2.879e+8 \\[1ex]
3e+9 & 21 & 0.895 & 1.713e-1 & 4.005e-2 & 1.403 & 1.195e+0 & 9.655e-1 & 1.222 & 9.594e-1 & 7.411e-1 & 1.157 & 2.879e+8 \\[1ex]
3e+9 & 22 & 1.028 & 2.095e-1 & 6.057e-2 & 1.509 & 1.058e+0 & 8.335e-1 & 1.355 & 8.223e-1 & 6.092e-1 & 1.068 & 2.879e+8 \\[1ex]
3e+9 & 23 & 1.213 & 2.392e-1 & 7.568e-2 & 1.626 & 9.500e-1 & 7.260e-1 & 1.540 & 7.139e-1 & 5.017e-1 & 0.982 & 2.879e+8 \\[1ex]
3e+9 & 24 & 1.038 & 2.525e-1 & 8.053e-2 & 1.647 & 8.788e-1 & 6.479e-1 & 1.365 & 6.428e-1 & 4.236e-1 & 0.910 & 2.879e+8 \\[1ex]
3e+9 & 25 & 1.203 & 2.463e-1 & 9.050e-2 & 1.629 & 9.745e-1 & 7.522e-1 & 1.530 & 7.384e-1 & 5.278e-1 & 1.004 & 2.879e+8 \\[1ex]
3e+9 & 26 & 0.984 & 2.020e-1 & 5.757e-2 & 1.385 & 1.109e+0 & 8.828e-1 & 1.312 & 8.728e-1 & 6.584e-1 & 1.103 & 2.879e+8 \\[1ex]
3e+9 & 27 & 1.003 & 1.638e-1 & 3.844e-2 & 1.433 & 1.172e+0 & 9.547e-1 & 1.331 & 9.360e-1 & 7.304e-1 & 1.150 & 2.879e+8 \\[1ex]
3e+9 & 28 & 0.921 & 1.909e-1 & 5.010e-2 & 1.375 & 1.123e+0 & 8.973e-1 & 1.249 & 8.866e-1 & 6.729e-1 & 1.113 & 2.879e+8 \\[1ex]
3e+9 & 29 & 1.011 & 2.293e-1 & 6.818e-2 & 1.490 & 1.025e+0 & 7.908e-1 & 1.339 & 7.894e-1 & 5.665e-1 & 1.036 & 2.879e+8 \\[1ex]
3e+9 & 30 & 1.038 & 1.857e-1 & 4.406e-2 & 1.473 & 1.137e+0 & 9.113e-1 & 1.366 & 9.013e-1 & 6.869e-1 & 1.122 & 2.879e+8 \\[1ex]
4e+9 & 1  & 0.996 & 2.532e-1 & 6.652e-2 & 1.449 & 1.050e+0 & 7.802e-1 & 1.295 & 8.138e-1 & 5.443e-1 & 1.018 & 5.682e+8 \\[1ex]
4e+9 & 2  & 0.958 & 2.501e-1 & 5.462e-2 & 1.376 & 1.083e+0 & 8.002e-1 & 1.258 & 8.465e-1 & 5.642e-1 & 1.034 & 5.682e+8 \\[1ex]
4e+9 & 3  & 1.182 & 2.528e-1 & 5.884e-2 & 1.581 & 9.215e-1 & 6.633e-1 & 1.482 & 6.855e-1 & 4.274e-1 & 0.914 & 5.682e+8 \\[1ex]
4e+9 & 4  & 1.092 & 2.416e-1 & 6.524e-2 & 1.683 & 9.218e-1 & 6.719e-1 & 1.392 & 6.857e-1 & 4.359e-1 & 0.922 & 5.682e+8 \\[1ex]
4e+9 & 5  & 0.898 & 2.296e-1 & 4.362e-2 & 1.344 & 1.095e+0 & 8.208e-1 & 1.198 & 8.589e-1 & 5.848e-1 & 1.050 & 5.682e+8 \\[1ex]
4e+9 & 6  & 0.935 & 2.141e-1 & 3.618e-2 & 1.436 & 1.109e+0 & 8.331e-1 & 1.235 & 8.732e-1 & 5.972e-1 & 1.059 & 5.682e+8 \\[1ex]
4e+9 & 7  & 0.943 & 2.383e-1 & 4.479e-2 & 1.390 & 1.111e+0 & 8.314e-1 & 1.243 & 8.749e-1 & 5.954e-1 & 1.058 & 5.682e+8 \\[1ex]
4e+9 & 8  & 1.041 & 2.287e-1 & 3.968e-2 & 1.524 & 1.029e+0 & 7.590e-1 & 1.341 & 7.930e-1 & 5.231e-1 & 1.001 & 5.682e+8 \\[1ex]
4e+9 & 9  & 1.009 & 2.551e-1 & 5.758e-2 & 1.513 & 9.731e-1 & 7.033e-1 & 1.309 & 7.370e-1 & 4.673e-1 & 0.952 & 5.682e+8 \\[1ex]
4e+9 & 10 & 1.306 & 2.604e-1 & 6.975e-2 & 1.739 & 8.840e-1 & 6.309e-1 & 1.606 & 6.479e-1 & 3.950e-1 & 0.881 & 5.682e+8 \\[1ex]
4e+9 & 11 & 0.966 & 2.429e-1 & 5.893e-2 & 1.385 & 1.078e+0 & 8.134e-1 & 1.266 & 8.423e-1 & 5.775e-1 & 1.044 & 5.682e+8 \\[1ex]
4e+9 & 12 & 0.911 & 1.863e-1 & 2.445e-2 & 1.411 & 1.202e+0 & 9.346e-1 & 1.211 & 9.661e-1 & 6.986e-1 & 1.130 & 5.682e+8 \\[1ex]
4e+9 & 13 & 1.243 & 2.776e-1 & 7.328e-2 & 1.693 & 8.450e-1 & 5.828e-1 & 1.543 & 6.090e-1 & 3.468e-1 & 0.828 & 5.682e+8 \\[1ex]
4e+9 & 14 & 0.991 & 2.307e-1 & 5.550e-2 & 1.432 & 1.040e+0 & 7.836e-1 & 1.291 & 8.035e-1 & 5.477e-1 & 1.021 & 5.682e+8 \\[1ex]
4e+9 & 15 & 0.857 & 1.973e-1 & 3.404e-2 & 1.295 & 1.174e+0 & 9.107e-1 & 1.157 & 9.383e-1 & 6.747e-1 & 1.114 & 5.682e+8 \\[1ex]
4e+9 & 16 & 1.072 & 2.842e-1 & 8.135e-2 & 1.520 & 9.316e-1 & 6.651e-1 & 1.372 & 6.956e-1 & 4.291e-1 & 0.916 & 5.682e+8 \\[1ex]
4e+9 & 17 & 0.982 & 2.230e-1 & 5.106e-2 & 1.475 & 1.068e+0 & 8.094e-1 & 1.282 & 8.320e-1 & 5.735e-1 & 1.041 & 5.682e+8 \\[1ex]
4e+9 & 18 & 0.761 & 1.836e-1 & 3.043e-2 & 1.272 & 1.224e+0 & 9.556e-1 & 1.061 & 9.875e-1 & 7.196e-1 & 1.143 & 5.682e+8 \\[1ex]
4e+9 & 19 & 1.103 & 2.649e-1 & 6.228e-2 & 1.518 & 9.185e-1 & 6.502e-1 & 1.403 & 6.824e-1 & 4.142e-1 & 0.901 & 5.682e+8 \\[1ex]
4e+9 & 20 & 1.087 & 2.577e-1 & 6.216e-2 & 1.549 & 9.537e-1 & 6.863e-1 & 1.387 & 7.177e-1 & 4.504e-1 & 0.936 & 5.682e+8 \\[1ex]
4e+9 & 21 & 0.780 & 1.877e-1 & 2.909e-2 & 1.286 & 1.228e+0 & 9.640e-1 & 1.080 & 9.919e-1 & 7.281e-1 & 1.149 & 5.682e+8 \\[1ex]
4e+9 & 22 & 0.954 & 2.283e-1 & 4.964e-2 & 1.456 & 1.019e+0 & 7.587e-1 & 1.254 & 7.827e-1 & 5.227e-1 & 1.000 & 5.682e+8 \\[1ex]
4e+9 & 23 & 1.130 & 2.532e-1 & 5.409e-2 & 1.637 & 8.991e-1 & 6.326e-1 & 1.430 & 6.631e-1 & 3.966e-1 & 0.883 & 5.682e+8 \\[1ex]
4e+9 & 24 & 0.941 & 2.751e-1 & 5.869e-2 & 1.506 & 8.969e-1 & 6.151e-1 & 1.241 & 6.608e-1 & 3.791e-1 & 0.864 & 5.682e+8 \\[1ex]
4e+9 & 25 & 1.073 & 2.644e-1 & 6.909e-2 & 1.538 & 9.503e-1 & 6.829e-1 & 1.373 & 7.142e-1 & 4.470e-1 & 0.933 & 5.682e+8 \\[1ex]
4e+9 & 26 & 0.922 & 2.217e-1 & 4.436e-2 & 1.397 & 1.097e+0 & 8.284e-1 & 1.222 & 8.612e-1 & 5.924e-1 & 1.055 & 5.682e+8 \\[1ex]
4e+9 & 27 & 0.909 & 1.905e-1 & 3.106e-2 & 1.348 & 1.167e+0 & 9.091e-1 & 1.209 & 9.308e-1 & 6.731e-1 & 1.113 & 5.682e+8 \\[1ex]
4e+9 & 28 & 0.873 & 2.169e-1 & 4.070e-2 & 1.321 & 1.121e+0 & 8.550e-1 & 1.173 & 8.848e-1 & 6.191e-1 & 1.075 & 5.682e+8 \\[1ex]
4e+9 & 29 & 0.953 & 2.442e-1 & 4.582e-2 & 1.522 & 9.938e-1 & 7.150e-1 & 1.253 & 7.577e-1 & 4.790e-1 & 0.963 & 5.682e+8 \\[1ex]
4e+9 & 30 & 0.912 & 1.922e-1 & 2.870e-2 & 1.418 & 1.164e+0 & 8.932e-1 & 1.212 & 9.284e-1 & 6.573e-1 & 1.102 & 5.682e+8 \\[1ex]
5e+9 & 1  & 0.993 & 2.659e-1 & 5.307e-2 & 1.441 & 9.871e-1 & 6.823e-1 & 1.282 & 7.510e-1 & 4.344e-1 & 0.921 & 9.625e+8 \\[1ex]
5e+9 & 2  & 0.964 & 2.683e-1 & 4.278e-2 & 1.357 & 1.040e+0 & 7.268e-1 & 1.252 & 8.037e-1 & 4.789e-1 & 0.962 & 9.625e+8 \\[1ex]
5e+9 & 3  & 1.151 & 2.732e-1 & 4.979e-2 & 1.593 & 8.698e-1 & 5.775e-1 & 1.439 & 6.338e-1 & 3.296e-1 & 0.807 & 9.625e+8 \\[1ex]
5e+9 & 4  & 1.010 & 2.522e-1 & 4.409e-2 & 1.589 & 9.855e-1 & 6.843e-1 & 1.298 & 7.495e-1 & 4.364e-1 & 0.923 & 9.625e+8 \\[1ex]
5e+9 & 5  & 0.835 & 2.397e-1 & 3.034e-2 & 1.303 & 1.073e+0 & 7.673e-1 & 1.123 & 8.366e-1 & 5.195e-1 & 0.998 & 9.625e+8 \\[1ex]
5e+9 & 6  & 0.919 & 2.352e-1 & 3.124e-2 & 1.425 & 1.094e+0 & 7.911e-1 & 1.207 & 8.575e-1 & 5.432e-1 & 1.017 & 9.625e+8 \\[1ex]
5e+9 & 7  & 0.871 & 2.459e-1 & 3.091e-2 & 1.368 & 1.071e+0 & 7.635e-1 & 1.159 & 8.346e-1 & 5.157e-1 & 0.994 & 9.625e+8 \\[1ex]
5e+9 & 8  & 1.079 & 2.583e-1 & 4.057e-2 & 1.553 & 9.488e-1 & 6.520e-1 & 1.367 & 7.127e-1 & 4.042e-1 & 0.891 & 9.625e+8 \\[1ex]
5e+9 & 9  & 1.028 & 2.723e-1 & 3.992e-2 & 1.438 & 9.257e-1 & 6.253e-1 & 1.316 & 6.897e-1 & 3.775e-1 & 0.862 & 9.625e+8 \\[1ex]
5e+9 & 10 & 1.321 & 2.654e-1 & 3.999e-2 & 1.744 & 8.197e-1 & 5.304e-1 & 1.609 & 5.836e-1 & 2.825e-1 & 0.746 & 9.625e+8 \\[1ex]
5e+9 & 11 & 0.945 & 2.598e-1 & 4.614e-2 & 1.369 & 1.020e+0 & 7.213e-1 & 1.234 & 7.844e-1 & 4.735e-1 & 0.958 & 9.625e+8 \\[1ex]
5e+9 & 12 & 0.893 & 2.114e-1 & 2.626e-2 & 1.402 & 1.186e+0 & 8.934e-1 & 1.181 & 9.502e-1 & 6.455e-1 & 1.094 & 9.625e+8 \\[1ex]
5e+9 & 13 & 1.168 & 2.718e-1 & 4.903e-2 & 1.658 & 8.762e-1 & 5.797e-1 & 1.457 & 6.401e-1 & 3.318e-1 & 0.810 & 9.625e+8 \\[1ex]
5e+9 & 14 & 0.957 & 2.477e-1 & 4.220e-2 & 1.356 & 1.051e+0 & 7.527e-1 & 1.246 & 8.148e-1 & 5.048e-1 & 0.985 & 9.625e+8 \\[1ex]
5e+9 & 15 & 0.816 & 2.211e-1 & 3.014e-2 & 1.230 & 1.168e+0 & 8.686e-1 & 1.104 & 9.317e-1 & 6.207e-1 & 1.076 & 9.625e+8 \\[1ex]
5e+9 & 16 & 0.962 & 2.841e-1 & 5.486e-2 & 1.476 & 9.264e-1 & 6.173e-1 & 1.250 & 6.903e-1 & 3.694e-1 & 0.854 & 9.625e+8 \\[1ex]
5e+9 & 17 & 0.952 & 2.412e-1 & 3.602e-2 & 1.394 & 1.017e+0 & 7.242e-1 & 1.240 & 7.806e-1 & 4.764e-1 & 0.960 & 9.625e+8 \\[1ex]
5e+9 & 18 & 0.708 & 1.996e-1 & 2.283e-2 & 1.240 & 1.231e+0 & 9.288e-1 & 0.996 & 9.945e-1 & 6.809e-1 & 1.118 & 9.625e+8 \\[1ex]
5e+9 & 19 & 1.056 & 2.749e-1 & 4.684e-2 & 1.514 & 9.454e-1 & 6.377e-1 & 1.345 & 7.093e-1 & 3.898e-1 & 0.876 & 9.625e+8 \\[1ex]
5e+9 & 20 & 1.017 & 2.628e-1 & 3.967e-2 & 1.450 & 9.610e-1 & 6.549e-1 & 1.305 & 7.250e-1 & 4.070e-1 & 0.894 & 9.625e+8 \\[1ex]
5e+9 & 21 & 0.738 & 2.054e-1 & 2.201e-2 & 1.248 & 1.230e+0 & 9.259e-1 & 1.026 & 9.936e-1 & 6.781e-1 & 1.116 & 9.625e+8 \\[1ex]
5e+9 & 22 & 0.872 & 2.439e-1 & 4.205e-2 & 1.420 & 1.057e+0 & 7.587e-1 & 1.160 & 8.213e-1 & 5.108e-1 & 0.990 & 9.625e+8 \\[1ex]
5e+9 & 23 & 1.090 & 2.600e-1 & 3.872e-2 & 1.487 & 9.235e-1 & 6.318e-1 & 1.379 & 6.874e-1 & 3.839e-1 & 0.869 & 9.625e+8 \\[1ex]
5e+9 & 24 & 0.877 & 2.843e-1 & 4.293e-2 & 1.417 & 9.215e-1 & 6.055e-1 & 1.165 & 6.854e-1 & 3.577e-1 & 0.840 & 9.625e+8 \\[1ex]
5e+9 & 25 & 1.034 & 2.672e-1 & 4.927e-2 & 1.469 & 9.410e-1 & 6.431e-1 & 1.323 & 7.050e-1 & 3.952e-1 & 0.881 & 9.625e+8 \\[1ex]
5e+9 & 26 & 0.898 & 2.510e-1 & 4.255e-2 & 1.344 & 1.098e+0 & 7.913e-1 & 1.186 & 8.615e-1 & 5.435e-1 & 1.017 & 9.625e+8 \\[1ex]
5e+9 & 27 & 0.985 & 2.265e-1 & 3.279e-2 & 1.413 & 1.094e+0 & 8.063e-1 & 1.274 & 8.583e-1 & 5.584e-1 & 1.029 & 9.625e+8 \\[1ex]
5e+9 & 28 & 0.795 & 2.264e-1 & 2.744e-2 & 1.272 & 1.146e+0 & 8.445e-1 & 1.084 & 9.096e-1 & 5.966e-1 & 1.059 & 9.625e+8 \\[1ex]
5e+9 & 29 & 0.891 & 2.486e-1 & 3.129e-2 & 1.406 & 1.027e+0 & 7.174e-1 & 1.180 & 7.907e-1 & 4.695e-1 & 0.954 & 9.625e+8 \\[1ex]
5e+9 & 30 & 0.868 & 2.162e-1 & 2.560e-2 & 1.343 & 1.160e+0 & 8.600e-1 & 1.156 & 9.239e-1 & 6.122e-1 & 1.070 & 9.625e+8 \\[1ex]
\end{longtable}
\end{center}
\footnotetext[a]{As discussed in \secref{sec:ddt}, the results of $\cdens =
   1.0\times10^9$~g~cm$^{-3}$, realization 8 were excluded from the analysis.}
\footnotetext[b]{As discussed in \secref{sec:observe}, the results of
   \cite{LesaffreEtAl06} do not allow us to generate cooling times for results
   with $\cdens = 1.0\times 10^9$~g~cm$^{-3}$.}

\end{appendix}

\end{document}